\documentclass[12pt]{article}
\usepackage{amssymb}
\usepackage{amsfonts}
\usepackage{amsmath}
\usepackage{amsthm}
\usepackage{amscd}
\usepackage{epsfig}
\usepackage{graphicx}

\evensidemargin \oddsidemargin

\newcommand{\Lap}{-\frac{1}{2}\Delta}

\newcommand{\grad}{\nabla}
\newcommand{\ep}{\varepsilon}

\renewcommand{\O}{\operatorname{O}}

\renewcommand{\th}{\frac{3}{2}}
\newcommand{\scriptT} { \mathcal{T} }

\def\scriptU{ {\mathcal{U} }}
\def\scriptUt{ { \widetilde{\mathcal{U}} }}
\def\scriptB{ {\mathcal{B} }}

\def\balpha{{\boldsymbol \alpha}}
\def\a{{\alpha}} 
\def\b{{\beta}}
\def\g{{ \gamma}}
\def\k{{\kappa}}
\def\fp{{\mathfrak p}}
\def\bk{{\mathbf k}}
\def\fq{{\mathfrak q}}
\def\fr{{\mathfrak r}}
\def\fu{{\mathfrak u}}
\def\fw{{\mathfrak w}}
\def\({\left(}
\def\[{ \left[}
\def\){ \right)}
\def\]{ \right]}
\def\hV{\hat{V}_0}
\def\bE{{\mathbf E}}
\def\pp{ {p'}}
\def\PP{ {P'}}

\def\rr{ {r'} }

\def\uu{{u'}}
\def\<{\langle}
\def\>{\rangle}
\def\bell{\mathfrak{l}}

\newtheorem{theorem}{Theorem}[section]

\newtheorem{lemma}[theorem]{Lemma}
\newtheorem{proposition}[theorem]{Proposition}
\numberwithin{equation}{section}

\newtheorem*{main}{Main Theorem}

\title{The Linear Boltzmann Equation as the Low Density Limit of a
 Random Schr\"odinger Equation}
\author{David Eng\thanks{Supported by MacCraken Fellowship.
Courant Institute, New York University.  
251 Mercer Street, New York, NY 10012, eng@math.stanford.edu} \;
and L\'aszl\'o Erd\H os\thanks{Partially
supported by NSF grant DMS-0200235 and EU-IHP Network ``Analysis
and Quantum'' HPRN-CT-2002-0027.  Institute of Mathematics, 
University of Munich, Theresienstr.  39, D-80333 Munich Germany, 
 lerdos@mathematik.uni-muenchen.de. On leave from School of Mathematics, 
GeorgiaTech, USA}}

\date{June 12, 2005}

\begin{document}
\maketitle

\begin{abstract}
We study the long time evolution of a quantum particle interacting with a 
random potential in the Boltzmann-Grad low density limit.
We prove that the phase space density of the quantum evolution 
defined through  the Husimi function converges weakly to a linear 
Boltzmann equation. The Boltzmann
 collision kernel is given by the full quantum scattering cross section
of the obstacle potential. 
\end{abstract}

\section{The Model and the Result}

The Schr\"odinger equation with 
a random potential describes the propagation  of
quantum particles in an environment with random impurities. In the first
approximation one neglects the interaction between the particles and
the problem reduces to a one-body Schr\"odinger equation.
 With high concentration
of impurities the particle is localized, in particular no conduction occurs
\cite{Ai, AM, A,  DK, FS, FMSS}. In the low concentration regime 
conduction is expected to
occur but there are no rigorous mathematical proof of the existence of the
extended states except for
the Bethe lattice \cite{Kl, Kl2}. In this paper we study the long time 
evolution in 
the low concentration regime in a specific scaling limit, called the 
low density or Boltzmann-Grad
limit. Our model is the quantum analogue of the low density Lorentz gas.
As the time increases,  the concentration will be scaled down  in such
a way that the total interaction between the particle
and the obstacles remains bounded for a typical configuration.
 Therefore our result is far
from the extended states regime which requires to understand the
behavior of the Schr\"odinger evolution  
for arbitrary long time,
independently of the fixed (low)  concentration of impurities.
We start by defining our model and stating the main result.

Let $\Lambda_L\subset\mathbb{R}^d $ be a cube of width $L$ and 
let $V_0(x)$ be a smooth radial function
with a sufficiently strong decay to be specified later. 
 Denote by $\omega = (x_\alpha), \alpha = 1, \ldots, N$,
 the configuration of uniformly
distributed obstacles and let 
\begin{equation} 
\varrho := \frac{N}{L^d}
\label{def:dens}
\end{equation}
be the density of 
the obstacles.  We are interested in the evolution of a quantum particle 
in the random environment generated by these obstacles.  The Schr\"odinger
 equation governing the quantum particle is given by:
\begin{equation} \label{E:shro}
        i \partial_t \psi_t = H_{N,L} \psi_t \, , \,\,\, \psi_{t=0} 
= \psi_0\; ,
\end{equation} where the Hamiltonian is given by:
\begin{equation} \label{E:hamiltonian}
        H_{N,L} = H := \Lap + V_\omega ,\, \,\qquad  V_\omega = 
\sum\limits_{\alpha=1}^N V_\alpha, \,\,\,   
 V_\alpha(x) := V_0(x-x_\alpha) \; 
\end{equation} 
with periodic boundary conditions on $\Lambda_L$.
We have used lower case letters, $(x,t)$, to denote the space and time
 variables
 in the microscopic (atomic) scale.  

We shall always take first 
the simultaneous $L\to \infty$, $N\to\infty$ limits, with
a fixed density, $\varrho= N/L^d$, before any other limit. 
The finite box $\Lambda_L$
 is just a technical device to avoid infinite 
summation in the potential term. 
 Our method works for any dimension $d\geq 3$, 
but we restrict ourselves here to the case $d=3$.

 As a first step toward a study of conduction, one 
considers certain scaling limits. 
Let $\ep$ be the  scale separation parameter between microscopic and
macroscopic variables. In reality,  $\ep = 1 \AA  / 1 \; \mathrm{cm} =
 10^{-8}$;
 here we always take the idealized $\ep\to0$ limit.  
Define the macroscopic coordinates $(X,T)$ by
\begin{equation*}
        (X,T) := (x\ep, t\ep)\;.
\end{equation*} Note  that the velocity is not rescaled, 
$\frac{X}{T}=\frac{x}{t}$.
In this paper  we will treat the following scaling limit problem:

\vspace{12pt}

\noindent $\bf{Low \,density \,limit \,(L.D.L.):}$  
Let $\varrho = \ep \varrho_0$ for some fixed positive density $\varrho_0$:
\begin{equation} \label{E:ldl}
        i \partial_t \psi_{\omega,t}^\ep =
\Big[\Lap + \sum\limits_{\alpha =1}^N
                 V_{\a}(x)\Big]\psi_{\omega,t}^\ep   \, .
\end{equation}
Another interesting scaling limit which has been studied in the literature is:

\vspace{12pt}

\noindent $\bf{Weak\, coupling\, limit\, (W.C.L.):}$  
Fix the density $\varrho = \varrho_0$ and scale the strength of the external
 potential by $\sqrt{\ep}$:
\begin{equation} \label{E:wcl}
        i \partial_t \psi_{\omega,t}^\ep = \Big[\Lap +
 \sqrt{\ep} \sum\limits_{\alpha =1}^N
                 V_{\a}(x)\Big]\psi_{\omega,t}^\ep\;.
\end{equation} 
In a related model,
the random obstacle potential $\sum_\alpha V_\alpha$
 is replaced with a Gaussian
field $V_\omega(x)$ with a decaying, $\ep$-independent covariance.

It turns out that in both the L.D.L and W.C.L. models
 these are the weakest interaction
strengths that result in a nontrivial (non-free) macroscopic
evolution in the scaling limit $\ep\to0$.

The Wigner transform of a wave function $\psi$ is defined by:
\begin{equation}        \label{E:wigdef}
        W_\psi(x,v) := \int_{\mathbb{R}^3} \overline{\psi\Big(x +
 \frac{z}{2}\Big)} \psi\Big(x - \frac{z}{2}\Big) e^{ivz} dz \; .
\end{equation} The Wigner transform typically has no
 definite sign but the associated Husimi function is 
nonnegative at appropriate scales.  The Husimi function 
at scale $(\ell_1,\ell_2)$ is defined by:
\begin{equation*}
        H_\psi^{\ell_1,\ell_2} := 
W_\psi *_x G^{\ell_1/\sqrt{2}} *_v G^{\ell_2/\sqrt{2}}\;, 
\end{equation*} where $G^\delta$ is the standard Gaussian 
with variance $\delta^2$, i.e.,
\begin{equation} \label{E:gaussian}
        G^\delta(z) := (2\pi \delta^2)^{-\th} e^{- \frac{z^2}{2\delta^2}} \; .
\end{equation} 
The Husimi function at scale $\ell_1=\ell$, $\ell_2 = \ell^{-1}$
 is  the coherent state at scale $\ell $ defined by
$$
       H_\psi^{\ell, \ell^{-1}}(x,v)=  
        C_{\psi, \ell} (x, v): = \langle
     \psi,  \pi_{x, v}^\ell \psi  \rangle \; ,
$$
where 
$\pi_{x, v}^\ell  $ is the projection onto the $L^2$ normalized state 
 $G^{\ell} (x-z) e^{i zv} $.  
Clearly $   C_{\psi, \ell}  $ is positive and 
\begin{equation}
\int   C_{\psi, \ell}
 (x,v)    dx dv = \| \psi \|_2^2= 1\; .
\label{coh}
\end{equation}
Thus $C_{\psi, \ell}$ can be considered as a probability density on the phase 
space at atomic  scale.  The accuracy for the space variable
 in the coherent state $C_{\psi,\ell}$ is of order $\ell$, and 
the accuracy is of order $\ell^{-1}$ for the velocity  variable.
  This is optimal by the uncertainty principle.   Unfortunately we cannot keep
 this accuracy along our proof, and we need a small extra smoothing.

 The basic object we shall study is the Husimi function on scale 
$\ell_1 = \ep^{-1+\mu}$, $\ell_2= \ep^\mu$ with some $0<\mu < 1/2$, 
which can also  be written as
\begin{equation}
\label{E:husimi}
        H_\psi^{\ep^{-1+\mu},  \ep^\mu} 
= H_\psi^{\ep^{-\mu},  \ep^\mu}\ast_x G^{\ell (\ep)/\sqrt{2}} 
=C_{\psi, \ep^{-\mu}}\ast_x G^{\ell (\ep)/\sqrt{2}}\;,
\end{equation}
where $\ell(\ep) = \ep^{\mu}\sqrt {\ep^{-2} - \ep^{-4\eta}}$. 
We can rescale it to the macroscopic scale by defining
\begin{equation}
        H^{(\ep, \mu)}_\psi (X, V): = 
\ep^{-3} H_\psi^{\ep^{-1+\mu},  \ep^\mu}(X/\ep, V) \ge 0 \; .
\label{E:husimi2}
\end{equation}
From (\ref{coh}), (\ref{E:husimi}) and (\ref{E:husimi2}) 
it follows that $H^{(\ep, \mu)}_\psi (X, V)$ defines a probability
density on the macroscopic phase space $\mathbb{R}^6$.

Notice that the velocity in $ H^{(\ep, \mu)}_\psi$
is not rescaled.   The accuracy for both the (macroscopic) space 
and velocity variables in $H^{(\ep, \mu)}_\psi$ is now of order $\ep^\mu$. 
We shall use this nonnegative phase space
density function to represent the true quantum mechanical function $\psi$.
Our goal is to prove that the macroscopic phase space density  of
$\psi^\ep_t$ converges to a solution of the  linear Boltzmann equation as 
in the classical case, except that the classical differential scattering 
cross section  is replaced by the quantum differential scattering cross section. 

We now recall the linear Boltzmann equation for a time dependent phase
space density $F_T(X,V)$ with collision kernel $ \Sigma (U, V) $:
\begin{align}
        \partial_T &F_T(X, V) + V \cdot\nabla_X F_T (X, V) \notag \\
        =& \int  \Big[ \Sigma (U, V) F_T(X,  U)- \Sigma (V, U) F_T(X, V)\Big] dU 
    \notag \\
        =& \int  \Sigma (U, V) F_T(X,  U) dU -  \Sigma F_T(X, V) \; ,
        \label{E:boltz}
\end{align}
where $\Sigma: =\int \Sigma (V, U)dU  $ is the total cross section.  
In our setting $\Sigma(V,U)$ will be defined later on in the Main Theorem.

For any function $f$ on $\mathbb{R}^3$
we introduce the norms
$$
   \| f \|_{M,N}: = \|  \<x\>^{M} \< \grad\>^{N} f \|_2, 
\qquad N, M \in \mathbb{N}\; ,
$$
where $\<x\> := (1+x^2)^{1/2}$.

Suppose  $V_0$ is a smooth, decaying and radially symmetric potential such that
\begin{align}
       \lambda_0:= \| V_0 \|_{50, 50} 
        \label{E:deflambda}
\end{align}
is sufficiently small.
 In particular, the one body Hamiltonian $H_1:= -\frac{1}{ 2}\Delta + V_0 $ 
has no bound states and 
asymptotic completeness holds, i.e. 
both the incoming and outgoing Hilbert spaces are the full space 
$L^2( \mathbb{R}^3)$. 
Recall the wave operators  
$\Omega_{\mp} = \lim_{s \to \infty} e^{\pm isH_0} e^{\mp isH_1} $, 
where $H_0:= -\frac{1}{ 2}\Delta$.  The kernel of the scattering operator 
$ S = \Omega^\ast_- \Omega_+ $ in the Fourier space exists and can be  written as 
$$
S(u, v) = \delta (u-v) -2 \pi i   \delta (u^2-v^2) T_{\mathrm{scat}}(u, v) .
$$
The  differential scattering  cross section can be defined as 
\begin{equation}
        \label{E:cross}
        \sigma(u, v): = 4\pi \, \underbrace {  \delta (u^2-v^2) }_{\hbox{ on-shell}}
 |T_{\mathrm{scat}}(u, v)|^2 .
\end{equation}
We shall choose initial data of the form
\begin{align*}
        \psi_0^{\ep} (x) = \ep^{3/2} h(\ep x) e^{iu_0 \cdot x}\; ,
\end{align*}
where $u_0\in\mathbb{R}^3$, 
$ \| h \|_{30,30} < \infty$ and $h$ is $L^2$-normalized.
  This implies $\|  \widehat{\psi}_0^{\ep} \|_{30,0}  < \infty  $ .
  
We will usually drop the "hat" on the initial wave function as we will be working 
in momentum space. 
 It should be noted that the specific form of our initial 
wave function is used only in the last step  - in the identification
of the limit.  Our result certainly holds for a general
 class of initial conditions which satisfy 
$\|\widehat{\psi}_0^{\ep} \|_{30,0}  < \infty  $ and that have a 
limiting macroscopic phase space density.

It is straightforward to check that
 the rescaled Husimi functions (\ref{E:husimi2})
of the initial data converge 
weakly  as probability measures on $\mathbb{R}^{6}$ as  
$\varepsilon \to 0$, i.e.
$$
 H^{(\ep,\mu)}_{\psi^\ep_0}(X, V) dX dV \to |h(X)|^2\delta(V-u_0) dX dV\; .
$$ 
We define 
$$
     F_0(X,V):= |h(X)|^2\delta(V-u_0) 
$$ 
to be the initial data of the limiting Boltzmann equation.
We can now state our theorem
 concerning the low density limit:

\begin{main}Suppose $d = 3$ and let $\mu>0$ be 
sufficiently small.  Suppose the random environment $\omega$ is uniformly
 distributed with density $\varrho = \varrho_0\ep$ with some fixed $\varrho_0>0$. 
 Let $V_0$ be a radially symmetric potential such that $
\lambda_0:= \| V_0 \|_{50,50}$ is sufficiently small.
  Let  $\psi_{ \omega, t}^\ep $ be the solution to the Schr\"odinger equation
 (\ref{E:ldl}) with 
$L^2$-normalized initial data $\psi^\ep_0$  of the following form
$$
        \psi^\ep_0(x)= \ep^{3/2}h(\ep x)e^{iu_0 \cdot x}
$$
where $ \| h \|_{30,30} < \infty $, $\| h\|_2=1$.
Then for any $T>0$, and any bounded, continuous test function $J$, 
\begin{align*}
        \lim_{\ep\to0} \lim_{L\to \infty} \Big| \int dX dV J(X,V)\,\Big[  \bE 
                H^{(\ep,\mu)}_{\psi_{\omega, 
                T/\ep}^\ep }(X, V) - F_T(X,V) \Big] \Big| = 0  \, ,
\end{align*}
where $F_T(X, V)$ satisfies the linear Boltzmann
equation (\ref{E:boltz}) with initial data given by
$F_0(X,V) = |h(X)|^2\delta(V-u_0)$
 and with effective collision kernel
$\Sigma(U, V) = \varrho_0 \, \sigma(U, V)$. Here $\sigma(U, V)$ 
is the differential scattering cross section given in (\ref{E:cross}). 
\end{main}
    
 Our result holds for a larger class of distributions of obstacles, 
but for simplicity we assume the uniform distribution in this paper.

The analogous result in the W.C.L. model 
was proven by H.~Spohn \cite{Sp} in the case where the obstacles are
 distributed according to a Gaussian law and macroscopic time is small, $T\leq 
T_0$.
 His result was extended to higher order correlation functions by 
Ho-Landau-Wilkins \cite{HLW} under the same assumptions.
 The W.C.L. model with a Gaussian field 
was proven globally in time by Erd\H os and Yau in \cite{EY1}. 
 Later the method was extended to general distributions by Chen in \cite{Chen}. 
 Chen also showed \cite{Chen2} 
that the convergence of the expected Wigner transform to the Boltzmann equation 
holds in $L^r$ for $r\geq 1$.

  The present proof is similar in  spirit to the W.C.L. proof in \cite{EY1}. 
The main difference between the two models and proofs lies in 
the Boltzmann
collision kernel, $\Sigma$. In the L.D.L. model, $\Sigma$ involves summing
up the complete Born series of each individual obstacle scattering
in contrast to the W.C.L. model, where only the first Born approximation
is needed. Unlike in the W.C.L. model, in
the low density environment
once the quantum particle is in the neighborhood of
an obstacle, it can collide with it many times with a non-vanishing
amplitude. Moreover,  if
two obstacles are near to each other, then complicated double recollision
patterns arise with comparable amplitudes.
On a technical level, this difference forced us to completely
reorganize the diagrammatic expasion of \cite{EY1}. Most importantly,
the recollision diagrams have bigger amplitude in the L.D.L model and
their estimate required several new ideas. 

The classical analogue of the L.D.L. model
 is the classical Lorentz gas.  It is proved by G. Gallavotti \cite{Ga}, 
H. Spohn \cite{Sp2}, and Boldrighini, Bunimovich and Sinai \cite{BBS} 
that the evolution of the phase space density of a classical Lorentz gas 
converges to a linear Boltzmann equation.  However, the classical W.C.L. 
behavior is governed by a Brownian motion instead of the Boltzmann equation - 
see Kesten and Papanicolaou \cite{KP} and D\"urr, 
Goldstein and Lebowitz \cite{DGL}.

In principle, one is interested in the behavior of  
$T\to H^{(\ep,\mu)}_{\psi_{\omega, T/\ep}^\ep }$ 
as a process  for typical $\omega$. This means one has to consider 
the joint distributions of
 $(  H^{(\ep,\mu)}_{\psi_{\omega, T_1/\ep}^\ep }, \cdots , 
H^{(\ep,\mu)}_{\psi_{\omega, T_n/\ep}^\ep } )$. 
We believe that there is no intrinsic 
difficulty to extend our method to this setting.
But the proof will certainly be much more involved.

%
%

\section{Preliminaries}

%
%
\subsection{Notation} 
For convenience, we fix a convention to avoid problems with factors of $2 \pi$ 
arising from the Fourier transform.  We define $dx$ to be the Lebesgue measure
 on $\mathbb{R}^3$ {\it divided by} $(2\pi)^{-3/2}$; i.e.,
\begin{equation*}
        \int dx = \frac{1}{(2 \pi)^{3/2}} \int_{\mathbb{R}^3} d^*x ,
\end{equation*} where we reserve the notation $d^*x$ for the genuine $3$-dimensional
 Lebesgue measure.  This convention will apply to any space or momentum variable
 in $3$ dimensions but $\it{not}$ to one-dimensional integration
 (like time variables and their variables),
 where integration will be the standard, unscaled, Lebesgue measure on the line. 
 With this convention, the three
dimensional Fourier transform (which will be usually denoted by a hat) is:
\begin{equation*}
        \widehat{f}(p) = \mathcal{F}f(p) := \int f(x)e^{-ipx} dx
\end{equation*} and its inverse:
\begin{equation*}
        f(x) = \mathcal{F}^{-1}\widehat{f}(x) =
 \int \widehat{f}(p) e^{ipx} dp\;.
\end{equation*}
Wave functions will always be represented in momentum space, $\psi(p)$, hence
we can omit the hat from their notation.

The other convention is related to the fact that we will be considering 
the problem on the torus $\Lambda := L\mathbb{T}^3$ where $\mathbb{T}^3$ is 
the unit torus. 
 Correspondingly, all the momenta in this paper will be on the discrete
 lattice $(\mathbb{Z} / L)^3$. 
The momentum variables will be denoted by letters $p,q,r, u, v$ or $w$.
 The delta function is defined as:
\begin{align}
        \delta(p) &:= |\Lambda| \text{    for $p=0$} \notag \\
        \delta(p) &:= 0 
 \text{    for $p \in (\mathbb{Z}/L)^3\setminus \{ 0\}$ .} \label{E:deltadef}
\end{align} 
Nevertheless we will use the continuous formalism, under the identification 
\begin{equation}        \label{E:contconv}
        \int dp := \frac{1}{(2 \pi)^\th |\Lambda|}
 \sum\limits_{p\in (\mathbb{Z} \setminus L)^3}\; .
\end{equation} 
Again, this will only apply to momenta variables.  The delta functions
 with time variables will remain the usual continuous delta functions.  
The convention                                               
(\ref{E:contconv}) should not cause any confusion, as $L \to \infty$ can 
be taken at any stage of the proof independently of all other limits.

Gothic script will be used to denote a set of variables, in particular, momenta.
  Define:
\begin{align}
        \fp_{m_1, m_2} :=& \{  p_j \}_{j=m_1}^{m_2} = 
(p_{m_1}, p_{m_1+1}, \ldots , p_{m_2})\notag \\
        \fp_{m} :=& \fp_{1,m} \label{E:defngothic}.
\end{align} In some  instances, we will need to single out the first momenta 
and write $(p_0, \fp_m)$ instead of $\fp_{0,m}$. Similar convention applies 
to other momentum variables. Moreover, for any
  $\bell_{0,b} := (\ell_0, \ldots, \ell_b)$,
 where $\ell_j$ are  non-negative integers for $j=0, \ldots, b$, we define
\begin{align}
        \fp_{0,b}^{\bell_{0,b}} := ( \underbrace{p_0, \ldots, p_0}_{\ell_0+1} , 
                \ldots, \underbrace{p_b, \ldots, p_b}_{\ell_b +1})
        \label{E:defptoell}.
\end{align}
If $\operatorname{Log} x$ is the standard natural logarithm function 
(on the positive line), define for $x>0$:
\begin{equation} \label{E:deflog}
        \log x  : = \begin{cases}
                        1 &\text{for $x\leq e$} \\
                        \operatorname{Log} x &\text{for $x>e$ .}
                        \end{cases} 
\end{equation}

If $x\ge 1$, we define $x^{\O(1)}$ to be $x^k$ 
for some positive constant $k$ which is independent of any parameter
 (such as $\ep$).  

Finally, if $A, B$ are fully ordered sets, unordered set operations 
will be denoted by their usual notation (e.g. $\cup, \cap, \in$, etc.). 
 Define $A\oplus B$
 to be the concatenation of $A$ and $B$, i.e.,
 the ordered set where the  ordering
of $A$  supersedes that of $B$.
 We will at times write $AB := A\oplus B$.   $A \prec B$ will denote 
ordered inclusion, i.e. $A\subset B$ and the ordering coincides.


\subsection{The Duhamel Formula} \label{ss:duh}
For any fixed $n_0\geq 1$, the Duhamel formula states:
\begin{align}
        e^{-it H } =& \sum_{m=0}^{n_0 -1} (-i)^m \int_0^{t*} \[ ds_j\]_0^m
                 e^{-is_0 H_0}V_{\omega} e^{-i s_1 H_0} V_{\omega}
                \cdots V_\omega e^{-is_m H_0} \notag  \\
        &+ (-i)^{n_0} \int_0^{t*} \[ds_j\]_0^{n_0} e^{-is_0 H} V_{\omega}
 e^{-is_1H_0}
                V_{\omega} \cdots V_\omega e^{-is_{n_0} H_0} \;,
        \label{E:duham}
\end{align}
 where $H$ is the (full) Hamiltonian given in (\ref{E:hamiltonian}) and:

\begin{align}
        \int_0^{t*} \[ds_j\]_m^n := \int_0^t \ldots 
\int_0^t \Big(\prod_{j=m}^{n} ds_j \Big)\, 
                \delta\Big(t- \sum_{j=m}^n s_j\Big)\;,
                \label{E:simplex}
\end{align} where $m\leq n$, and the star refers to the constraint 
$t = \sum s_j$.  $ V_\omega $ is the  potential given in (\ref{E:hamiltonian})
 and $H_0=-\frac{1}{2}\Delta$.
Expanding the potential $V_\omega = \sum_{\a=1}^N V_\a$ in the Duhamel
 formula, we generate many terms.  We can label these terms by a sequence 
of obstacles, say,  
$\balpha = (\a_1, \a_2, \cdots, \a_n)$. The terms without $e^{-itH}$ 
(in the first line of (\ref{E:duham})) will be called {\it fully expanded},
the others will be called {\it truncated}.

We write the Duhamel formula in momentum space. 
The kernel of the typical fully expanded
term is of the form:
\begin{equation} \label{E:duhamtypical}
        \int_0^{t*} \[ds_j\]_0^n  e^{-i s_0 p_0^2/2} 
        \hat V_{\a_1}(p_0-p_1) e^{-i s_1 p_1^2/2}  \hat V_{\a_2}(p_1-p_2)
        \cdots e^{-i s_n p_n^2/2}
\end{equation}
with the intermediate momenta $p_1, p_2, \ldots , p_{n-1}$ integrated out.
The truncated terms are of the same form with $ e^{-i s_0 p_0^2/2}$ replaced
with $e^{-is_0H}$:
\begin{equation} \label{E:duhamtypicaltrun}
        \int_0^{t*} \[ds_j\]_0^n  e^{-i s_0 H}
        \hat V_{\a_1}(\cdot -p_1) e^{-i s_1 p_1^2/2}  \hat V_{\a_2}(p_1-p_2)
        \cdots e^{-i s_n p_n^2/2}.
\end{equation}

The obstacles in $\balpha = (\a_1, \a_2, \ldots, \a_n)$ are allowed to repeat. 
 We can relabel them by a sequence of {\it centers} 
\begin{equation*}
        A:= (\a_1, \a_2, \ldots, \a_m) , \quad  x_{\a_j} 
\in \omega \hbox{ for all } j 
\end{equation*} 
and a sequence of non-negative numbers
\begin{equation*}
        (k_1, k_2, \ldots, k_m)\; ,
\end{equation*} 
where $k_j +1$ denotes the number of times $\a_j$ repeats itself 
consecutively (we say that $k_j$ is the number of internal recollisions).
 The sequence $A$ has the property that $\a_j \neq \a_{j+1}$.
In order words, the original collision sequence is given by
\begin{equation}
        (\underbrace{\a_1, \ldots, \a_1}_{k_1+1},       
        \underbrace{\a_2, \ldots, \a_2}_{k_2+1}, \ldots,
        \underbrace{\a_m, \ldots, \a_m}_{k_m +1} ) \; .
\label{center}
\end{equation}
 We shall divide the set of momenta into {\it internal} ones and
 {\it external} ones.  The internal momenta are running between the 
same obstacles; the external ones are the rest.  The internal momenta
 will be integrated out first (resummation of loop diagrams). Hence
 repeated consecutive collisions with the same obstacle, 
{\it internal collisions},  will be considered as a single 
(physical) collision and will be referred to as a collision
 with a {\it center}.  When we speak of 
the number of collisions, we will actually be referring to the 
number of collisions with centers. For example,  in the sequence 
(\ref{center}) there are $\sum_{j=1}^m(k_j+1)$ collisions,
 $m$ centers and there are $k_j$ internal momenta running
among the $k_j+1$ collisions with the same center $\alpha_j$.

Collision histories will be recorded with the ordered set $A$.  Typically
 we will use the variable $m$ to denote the cardinality of $A$, $m = |A|$. 
 Next, let $J$ be a set of lexicographically ordered double indices for 
the internal momenta 
\begin{equation}\label{Jdef1}
        J=J_{m,\bk} : = \Bigl( 11, 12, \ldots, 1k_1, 21, 22, \ldots 2k_2,
        \ldots mk_m\Bigr)\; ,
\end{equation} 
where $\bk = (k_1, k_2, \ldots, k_m)$ denotes the number 
of the internal 
momenta for the obstacle $\a_j$.  We shall denote the internal momenta 
by $\fq_J:=(q_{j \ell})_{j\ell \in  J_{m,\bk}}$.     

Since $ \widehat V_{\a} := \mathcal{F} V_{\a} = e^{-ip x_\a} \widehat V(p) $, 
we are able to separate the random part of the potential, in the form 
of a random phase, from the deterministic part.  Consequently, denote
 the random phase corresponding to collision history $A$ by
\begin{equation} \label{E:defchi}
        \chi(A;\fp_{0,m}) := \prod_{j=1}^{m} e^{-ix_{\alpha_j}(p_{j-1}-p_j)}.
\end{equation}Note that it is independent of the internal momenta.  
Then the  deterministic part of the potentials is given by
\begin{equation} \label{E:defL}
        L(\fp_{0,m}, \fq_{J_{m,\bk}}) : = \prod_{j=1}^{m}
                        \hV(p_{j-1}-q_{j1})\hV(q_{j1}-q_{j2})
                \ldots \hV(q_{jk_j}-p_j).
\end{equation}In the case where $k_j=0$, we only have the term: 
 $\hV(p_{j-1} -p_j)$.  Notice that this expression is independent of
 the location of the obstacles; that information is contained in 
the random phase, $\chi$.  

Given a set of momenta, $\fr_{0,m}$, define the free evolution kernel as: 
\begin{equation} \label{E:Kdef}
        K(t;r_0, \fr_m): = (-i)^m \int_0^{t*}  \[d s_j\]_{j=0}^m 
\prod_{j = 0}^m e^{-is_j r_j^2/2}.
\end{equation}Notice that this expression is independent of the 
order of the momenta.  
Considering  (\ref{E:duhamtypical}) and using the previously 
established notation for internal
 and external momenta, the free evolution kernel associated with
 the collision sequence $A$
 and internal momenta $\bk$ is:
\begin{align*}
        K(t; \fp_{0,m}, \fq_{J_{m,\bk}}) :=&
        K(t; p_0, q_{11},\ldots, q_{1k_1}, p_1, 
      q_{21}, \ldots, p_{m-1}, q_{m1},\ldots, q_{mk_m},p_m).
\end{align*}
Define the fully summed (for internal collisions) free evolution kernel as:
\begin{equation} \label{E:defscriptK}
        {\mathcal  K}(t; \fp_{0,m}):=\sum_{k_1,\ldots, k_m=0}^\infty
         \int d\fq_{J_{m,\bk}}
        K(t; p_0, \fp_{m}, \fq_{J_{m,\bk}}) 
        L(p_0, \fp_{m}, \fq_{J_{m,\bk}}).
\end{equation}
With these notations,
 we can express the fully expanded wave function with a
 collision sequence $A$ (and resummation of loop diagrams) and its 
associated propagator by 
\begin{equation} \label{E:defpsiA}
        \psi_A(t, p_0) := \scriptU^{\circ}_{A}(t)\psi_0 (p_0) :=
                        \int d\fp_m \,{\mathcal K}(t; p_0, \fp_{m}) 
\chi(A, p_0,\fp_{m})
                        \psi_0(p_{m})\;,
\end{equation}
where $\psi_0$ is the initial wave function in momentum space.  
 It is important to note that the first momentum, $p_0$, is not summed
 for internal momenta.  The circle in the notation $\scriptU^{\circ}_{A}(t)$
refers to the fact that it is a fully expanded propagator.

Define the fully expanded wave function with  $m$ collisions (this is 
always counted according to the collision centers) without recollision, 
and its associated propagator by: 
\begin{equation}
        \psi_{m}^{\mathrm{no\,  rec}} := \scriptU^{\circ}_m(t) \psi_0 (p_0):= 
                \sum_{A: |A|= m}^{\mathrm{no \, rec }}  \psi_A 
                \label{E:defpsim} \, .
\end{equation}  The "no rec'' in $\sum_{ |A|= m}^{\mathrm{no \, rec}}$ reminds us that
we sum over sets $A$ without repetition (recollision), i.e. 
$\alpha_i\neq \alpha_j$, $i\neq j$.

%
%

\subsection{Error Terms and Time Division}

Let $m_0=m_0(\ep)$ be an $\ep$-dependent parameter to be chosen later.  
The Duhamel formula consists of sum of terms of the form (\ref{E:duhamtypical})
and (\ref{E:duhamtypicaltrun}). It allows the flexibility to expand certain
truncated terms (\ref{E:duhamtypicaltrun}) further and stop the expansion for other
terms.
In the truncated terms
 we will continue the expansion only for
 terms whose number of centers is less
 than $m_0$ and that are non-repeating.  In other words, we stop the Duhamel expansion
 whenever the number of external collisions reaches $m_0$ or if there is a genuine, 
non-internal recollision.  The result is the decomposition:   
\begin{equation} \label{E:1stdecomp}
        e^{-i t H} \psi_0  = \sum_{m=0}^{m_0-1}\psi^{\mathrm{no\,rec}}_m(t) + 
                        \Psi^{\mathrm{error}}_{m_0}(t) :=
                        \psi^{\mathrm{no\,rec}}_{< m_0}(t)
+\Psi^{\mathrm{error}}_{m_0}(t) \, .
\end{equation}

The truncated terms will be estimated by using the unitarity of 
the full Hamiltonian evolution:
$$
       \Big\| \int_0^t e^{-isH}\psi_s \; ds
 \Big\|^2\leq t \int_0^t \|\psi_s\|^2 ds \; ,
$$
where the additional $t$ factor is the price for using this crude bound.
We are able to reduce this price by dividing the total time interval $[0,t]$
into $n$ pieces.  We will eventually choose $n=n(\ep)$ in a precise way.
We refer to this method as the {\it time division argument}.

As a first step, observe that for $n\ge 1$,
\begin{align}
        e^{-it H} \psi_0 = \Big(\prod_{k=1}^n e^{-i \frac{t}{n} H}\Big) \psi_0
       = e^{-i \frac{(n-k)t}{n}H} \(e^{-i\frac{kt}{n}H}\psi_0 \)
                \label{E:timedivmain} 
\end{align}
for any $0\leq k\leq n$.
Before each new time evolution of length $\frac{t}{n}$ we successively
separate the main term and the error term. 
For a fixed $n\ge 1$, we define the main term
up to time $\frac{kt}{n}$  as the
sum of the non-recollision terms with less than $m_0$ collisions:
\begin{align}\label{phimain}
        \varphi^{\mathrm{main}}_k(t) :=& \sum_{m=0}^{m_0-1}     
                \psi_m^{\mathrm{no\,rec}}\(\frac{kt}{n}\) = \sum_{m=0}^{m_0-1} 
                \scriptU^{\circ}_{m}\(\frac{kt}{n}\)\psi_0 \; .
\end{align}
We follow the time evolution of the main term only, i.e., we define
\begin{align}
        \varphi_k (t) : =& e^{-i\frac{t}{n} H}
 \varphi_{k-1}^{\mathrm{main}}(t) \notag \\ 
                \varphi_k^{\mathrm{error}}(t) :=& 
\varphi_k (t) - \varphi_k^{\mathrm{main}}(t).
                \label{E:timedivdecomp}
\end{align}

By (\ref{E:timedivmain}),
\begin{align}
        e^{-i t H}\psi_0
        =& \sum_{k=1}^n e^{-i \frac{(n-k)t}{n}H}\varphi_k^{\mathrm{error}}(t) + 
                \varphi_n^{\mathrm{main}}(t). 
                \label{E:timedivtotal}  
\end{align} Since 
$\varphi_n^{\mathrm{main}}(t)=\psi_{< m_0}^{\mathrm{no\,rec}}(t)$, 
comparison with (\ref{E:1stdecomp}) allows us to write
\begin{align}
        \Psi_{m_0}^{\mathrm{error}} =  
                \sum_{k=1}^n e^{-i \frac{(n-k)t}{n}H}\varphi_k^{\mathrm{error}}.
                \label{E:deferror}
\end{align}Our estimates of the error term will
 initiate from this expression.  The idea is that versing our wave functions as
 $e^{-it_1 H} e^{-it_2 H} \psi_0$, we can expand each full 
evolution propagator independently using the Duhamel formula and we can
gain control of how closely in time the collisions occur.  Classical intuition 
tells us that it is improbable for a path to have many collisions in a very 
short  time. The time division technique employed here will exploit this. 
 With this agenda in mind, we will now
define the time-divided propagators.

\medskip

Let $A$ be an ordered set of size $m$ and $t>0$.  Suppose $m_1, m_2\ge 0$
and  $t_1,t_2\ge 0$ satisfy $m=m_1+m_2$, $t=t_1+t_2$.
  Define the fully expanded,
time-divided      propagator associated to $A$ as:
\begin{align}
        \scriptU^{\circ}_{m_1,m_2;A}&(t_1, t_2)\psi_0(p_0)  \notag \\ 
        :=&  \int d\fp_m \, \mathcal{K}_{m_1}(t_1;\fp_{0,m_1})
                \mathcal{K}_{m_2} (t_2; \fp_{m_1,m})
                \chi(A; \fp_{0,m}) \psi_0 (p_m)
                        \label{E:freenorectdivdef} .
\end{align}
We will also write 
\begin{align}
        \mathcal{K}_{m_1,m_2}(t_1, t_2; \fp_{0,m}) := 
 \mathcal{K}_{m_1}(t_1;\fp_{0,m_1})
                \mathcal{K}_{m_2} (t_2; \fp_{m_1,m}) \label{E:deftdivK}.
\end{align}
With summation over $A$ with non-repeating indices and recalling that $m=m_1+m_2$
we define
\begin{align}
        \scriptU^{\circ}_{m_1,m_2}(t_1,t_2)\psi_0(p_0) :=&
                \sum_{A: |A|=m}^{\mathrm{no\,rec}}
                \scriptU^{\circ}_{m_1,m_2;A}(t_1,t_2)\psi_0(p_0) .
 \label{E:Umm0}
\end{align}
If $A_1$ is the ordered set of the first $m_1$ elements of $A$ and $A_2$ 
the last $m_2$ elements of $A$, notice that
\begin{align*}
        \scriptU^{\circ}_{m_1,m_2;A}(t_1, t_2)\psi_0(p_0) =& \,
       \scriptU^{\circ}_{A_1}(t_1) \scriptU^{\circ}_{A_2}(t_2)\psi_0 (p_0) .
\end{align*}

%

Next, we define wave functions starting from a potential (according to 
the field theory jargon, we call them "amputated'').  They will be denoted by 
tildes. The amputated wave function with collision sequence $A$
 and its associated propagator is
\begin{equation}
        {\widetilde\psi}_A(t, p_0) : = \scriptUt_A(t)\psi_0(p_0) :=
                        \int dp \,  \hat 
                        V_{\a_1}(p_0-p) \psi_{A \setminus \{\a_1\}}(t,p)
\end{equation} if $m=|A|\ge 1$.  Note that we do not 
sum the last potential for internal collisions.  
The time-divided amputated propagator associated with $A$ can 
be defined for $m_1\geq 1$:
\begin{align*}
        \scriptUt_{m_1,m_2;A}(t_1,t_2)\psi_0(p_0) := 
                \int dp \;  \hat V_{\a_1}(p_0-p) \,
                \scriptU^{\circ}_{m_1-1,m_2; A \setminus \{\a_1\}}(t_1,t_2)\psi_0(p)\;.
\end{align*}
This allows us to define the time-divided full propagator that will
be denoted by $\scriptU$ without cirle or tilde:
\begin{align*}
        \scriptU_{m_1,m_2;A}(t_1,t_2)\psi_0(p_0) := \int_0^{t_1} ds \, e^{-i(t_1-s)H} 
                \scriptUt_{m_1,m_2;A}(s,t_2)\psi_0(p_0) .
\end{align*}
Similarly to (\ref{E:Umm0}), we define the propagators ${\scriptU}_{m_1,m_2}$
and $\widetilde{\scriptU}_{m_1,m_2}$ as
\begin{align}
        \overset{(\sim)}{\scriptU}_{m_1,m_2}(t_1,t_2) \psi_0(p_0) :=& 
\sum_{A:  |A|=m}^{\mathrm{no \,rec}}
               \overset{(\sim)}{ \scriptU}_{m_1,m_2;A}(t_1,t_2)\psi_0(p_0)  \; .
\label{E:ufull}
\end{align}

%
%

\subsection{Properties of the Kernel}

By (\ref{E:defscriptK}), any analysis of $\mathcal{K}$ will involve the free 
evolution kernel $K$, which was defined in (\ref{E:Kdef}).  We now give several 
ways to do so.  Define for $t>0$:
\begin{equation}  \label{E:defeta}
        \eta(t):= \begin{cases}
                        1   & \text{for $t \leq 1$} \\
                        t^{-1} &\text{for $t > 1$}
                        \end{cases}
\end{equation}We will typically write $\eta_j := \eta(t_j)$, $\eta'_j := \eta(t'_j)$
 and $\eta :=\eta(t)$.

\begin{lemma} [$\a$-Representation] \label{L:Kidentity} We have the following 
identity for $\eta >0$
\begin{equation} \label{E:Kidentity}
        K(t; \fr_{0,m}) =\frac{i}{2\pi } e^{ \eta t} \int_{\mathbb{R}} d\a \,
 e^{-i \a t} \prod_{j =0}^m
                 \frac{1}{\alpha - p_j^2/2 +i\eta}  \, .
\end{equation}Consequently, for $\eta(t)$ defined in (\ref{E:defeta}):
\begin{equation*}
        |K(t; \fr_{0,m})| \leq C \int_{\mathbb{R}} 
                d\a \prod_{j= 0}^m \frac{1}{| \alpha - p_j^2/2 + i \eta(t) |} \, .
\end{equation*}
\end{lemma}
\noindent The proof is given in \cite{EY1}.  The second statement is a 
consequence of $e^{\eta(t) t} \leq C$.  The variable $\a$ 
(and $\widetilde\a, \beta, \widetilde\beta$)
will typically be used for one dimensional integration on $\mathbb{R}$. 
 In the future we will not explicitly denote the integration domain
 for these variables
with the convention that it is always over the real line. 

With this in mind, we can write:          
\begin{align}
        \mathcal{K}(t; \fp_{0,m}) =& \frac{i}{2\pi} e^{\eta t} 
                \int d\a \;\frac{ e^{-i\a t} }{\a -p_0^2/2+ i\eta}
                \prod_{j=1}^m \frac{B(\a, p_{j-1},p_j)}{\a - p_j^2/2 + i \eta}
\label{E:KF}  \\
        B_\eta(\a, p_{j-1}, p_j) :=&\sum_{k_j=0}^\infty \int
 \[dq_{j k}\]_{k=1}^{k_j}  \hV(p_{j-1}-q_{j1})
                \notag  \\
        &\quad \times   \frac{\hV(q_{j1} -q_{j2})}{\a - q_{j1}^2/2 + i\eta }
                \cdots  \frac{ \hV(q_{jk_j} - p_j)}{\a - q_{jk_j}^2/2 + i\eta }
            \label{E:defB},
\end{align}where the $k_j=0$ term in the sum is $\hV(p_{j-1} - p_j)$. 
The dependence of $B_\eta$ on the regularization parameter will often be
suppressed in the notation, unless it becomes crucial,
 and we use $B(\a, p_{j-1}, p_j) :=B_\eta(\a, p_{j-1}, p_j)$.

 The formula (\ref{E:defB}) has the
interpretation that summing over internal collisions, in effect, 
changes our potential from $\hV(p_{j-1} - p_j)$ to $B_\eta(\a, p_{j-1}, p_j)$. 
 Moreover, the smoothness and decay properties of $\hV$
will be passed onto $B$.  This will be made precise in Lemma \ref{L:Blemma},
 which implies, in particular,  that 
\begin{align*}
      \sup_{\eta, p,r} 
 | \< p-r\>^{30} \<\grad_p\>^2\<\grad_r\>^2 B_\eta(\a,p,r) | \leq M\lambda_0 ,
\end{align*}where $\lambda_0 $ is defined in (\ref{E:deflambda}) and $M$ is 
independent of $\a$ and $\eta \leq 1$.

We also remark that with $\alpha = p_j^2/2$ we have
\begin{equation}
   \lim_{\eta\to 0+0} B_\eta\Big( \frac{p_j^2}{2}, p_{j-1}, p_j\Big)
   = T_{\mathrm{scatt}}(p_{j-1}, p_j) \; .
\label{tscat}
\end{equation}
The existence of this limit follows from Lemma \ref{L:Blemma}.
 The identification
with the scattering $T$-matrix follows from the standard 
Born series expansion
(see Theorem XI.43 of \cite{RS}).

As Lemma \ref{L:Kidentity} will be a fundamental tool in our estimates,
 we will collect some facts which will assist in the estimate of the terms 
on the right-hand side of (\ref{E:Kidentity}). 
 They follow from simple calculus and we leave their proofs to the reader.
\begin{proposition} \label{P:crossingtools}
Recall (\ref{E:deflog}) and $\eta:= \eta(t)$ in (\ref{E:defeta}).  
Then the following estimates hold:
\begin{align}
        \sup_{\a, p} \int_{\mathbb{R}} 
                \frac{ d\a}{\< \a\> | \a- p^2/2 + i \eta |} \leq&
 C \log t  \label{E:alphaint} \\
        \sup_{\a} \int \frac{dp \, \<\a\>}{\< p \>^4 | \a - p^2/2 + i \eta  |} \leq&
 C \log t
                \label{E:intadecay} \\
        \sup_{p,\a} \frac{\<\a\>}{\<p\>^{4} | \a- p^2/2 + i \eta  |} \leq& Ct
                \label{E:inftydecay} \\
        \sup_{r,\a}\int \frac{dp}{\<p-r\>^4 | \a - p^2/2 + i \eta  |} \leq& C \log t
                 \label{E:intnodecay} \\
        \sup_{p,\a} \frac{1}{| \a - p^2/2 + i \eta |} \leq  & C  t
                \label{E:inftynodecay} \, ,
\end{align} where $C$ is independent of $t$.  $\;\;\Box$
\end{proposition}
The next result will be the key estimate
to control  the so-called {\it crossing terms}.  
\begin{proposition} \label{P:crossing}
Under the assumptions of Proposition \ref{P:crossingtools}, we have:
\begin{align*}
        \sup_{\a,\a'} \int \frac{ dp}{\< p\>^4 \<p+q\>^4} 
\frac{\<\a\>}{| \a- p^2/2 + i\eta|} 
                \frac{\<\a'\>}{|\a' -(p+q)^2/2 + i\eta|} 
\leq \frac{C (\log t)^2}{|q| + \eta} ,
\end{align*}where $C$ is independent of $t$.
\end{proposition}
\begin{proof}
We change to spherical coordinates and measure the angular component of $p$ against
 the fixed vector $q$.
If $|q|>0$ and $r:=|p|$, we have:  
\begin{align*}
        \int \frac{ dp}{\< p\>^4 \<p+q\>^4}&\frac{ \<\a\>}
    {| \a- p^2/2 + i\eta|} 
                \frac{\<\a'\>}{|\a' -(p+q)^2/2 + i\eta|} \\
        =&\int_0^\infty \frac{ \langle \alpha \rangle r^2 \; d r}
        {\<r\>^4| \a- r^2/2 + i\eta|  } 
         \int_{-1}^1 \frac{dz}{\< (r^2 + q^2 + 2r|q| z)^{1/2} \>^4}  \\
     &\times \frac{\<\a' \>}{| \a' - (r^2 + q^2 + 2r|q|z )/2 + i\eta |} \\
        \leq & \; \frac{C}{|q|} \int_0^{\infty} \frac{ r\; dr}{\<r\>^4} \frac{\< \a\>}
     {| \a - r^2/2 +i\eta|} 
                \int_0^{\infty} \frac{dz}{\<z\>^2} 
\frac{\<\a' \>}{ |\a'- z + i\eta|} \\
        \leq &\; \frac{C \log t}{|q|} \int_0^{\infty} 
\frac{ r\; dr}{\<r\>^4} \frac{\< \a\>}{| \a - r^2/2 +i\eta|} \\
        \leq & \;\frac{C (\log t)^2}{|q|}.
\end{align*}
Combining this with the trivial estimate:
\begin{align*}
        \int \frac{ dp}{\< p\>^4 \<p+q\>^4}& \frac{ \<\a\>}{| \a- p^2/2 + i\eta|} 
                \frac{\<\a'\>}{|\a' -(p+q)^2/2 + i\eta|} \\
        \leq& \; \frac{1}{\eta}  \int \frac{dp}{\<p\>^4} 
\frac{\<\a\>}{ |\a - p^2/2 + i \eta|}
\end{align*}
which holds for all $q$, we prove the lemma using (\ref{E:intadecay}).
\end{proof}
                
The next result shows that the free kernel enjoys a "semi-group" property. 
 It will be crucial in giving us flexibility to estimate the kernel
 in different ways.  
\begin{proposition} \label{P:semigroup}
Let $m\geq1$ and $I_1, I_2 \subset \{ 0, \ldots, m\}$ such that
 $I_1 \cap I_2 = \emptyset$ and $I_1 \cup I_2 = \{0, \ldots, m\}$. 
 That is, $I_1$ and $I_2$ partition $\{0, \ldots , m\}$.  Recalling
 the notation (\ref{E:simplex}), one has the identity
\begin{align}
        K(t; \fr_{0,m}) = -i  \int^{t*}_0 ds_1 ds_2 K(s_1; \fr_{I_1})
 K(s_2; \fr_{I_2})\;,
        \label{E:Ksemigroup}
\end{align}where $\fr_{I_k} := (r_j)_{j\in I_k}$.  If one of the sets, 
say $I_2$, is empty, we will define $K(s_2; \fr_{I_2}):= i \delta(s_2)$. 
 In this case, the decomposition is trivial.
\end{proposition}

Proposition \ref{P:semigroup} follows directly from the definition, 
(\ref{E:Kdef}).  An immediate consequence is
\begin{align}
        \mathcal{K}(t; \fp_{0,m}) =& \int_0^{t*} ds d\tau \,K(s;p_0, \fp_m) 
                F(\tau; p_0, \fp_m) \notag \\
        F(\tau;  \fp_{0,m}) :=&  
                -i\sum_{k_1, \ldots, k_m=0}^{\infty} \int d\fq_{J_{m,\bk}}
 K(\tau, \fq_{J_{m,\bk}}) 
                L(p_0, \fp_m, \fq_{J_{m,\bk}}) 
                \label{E:defF} ,
\end{align}
with $ L(p_0, \fp_m, \fq_{J_{m,\bk}}) $ defined in (\ref{E:defL}) and 
where the term corresponding to $k_1 = \cdots = k_m =0$ is
 $\delta(\tau) L(\fp_{0,m})$.
This decomposition isolates the external momenta in the complete free 
kernel
from the {\it effective potential},  $F(\tau)$,  that is obtained
 after integrating out
the internal momenta. This term
 will be estimated in Lemma \ref{L:tdivfdecay}.

Using (\ref{E:Ksemigroup}) we can combine the decompositions given in 
(\ref{E:defF}) and (\ref{E:defB}).  

\begin{proposition} \label{P:mixedK} Let $0\leq \mu_1< \mu_2 \leq m$.  We have:
\begin{align*}
        \mathcal{K}(t; \fp_{0,m}) =& \int_0^{t*} dt_1 dt_2 \,
                \mathcal{K}(t_2; \fp_{\mu_1+1,\mu_2})\frac{e^{\eta(t_1) t_1} }{2\pi}   \\
        &\times \int d\a  \, e^{-i\a t_1} 
                \prod_{j=0}^{\mu_1} \frac{B(\a,p_{j},p_{j+1})}{\a- p_{j}^2/2 + i \eta_1}
                \prod_{j=\mu_2+1}^m \frac{B(\a, p_{j-1},p_j)}{\a - p_j^2/2 + i \eta_1}  \, .
\end{align*}
\end{proposition}

%
%

\section{Error Estimate}

The goal of this section is to prove:
\begin{lemma} \label{L:errorest} 
Let $m_0 = m_0(\ep)$ be chosen by (\ref{E:m0choice}).  For 
$\Psi_{m_0}^{\mathrm{error}}(t)$ defined in (\ref{E:1stdecomp}), we have:
\begin{equation*}
        \lim_{\ep\to 0}\lim_{L\to \infty} 
\bE \| \Psi_{m_0}^{\mathrm{error}}(T\ep^{-1}) \|^2 =0\;.
\end{equation*}
\end{lemma}

Since our main term is comprised of only terms with collision histories
 which contain no recollisions, terms resulting from the Duhamel 
expansion which have collision histories with recollisions are included 
in the error term.  It is the estimate of the error term where we will 
need to analyze the size of recollision terms.  Recall that we already 
sum our wave functions in the main term for immediate recollisions 
(internal collisions), thereby eliminating them from subsequent analysis.  

Given $m_1, m_2 \geq 0$ and $A$ of size $m:=m_1+m_2$ with no repeating
 indices, again denote by $A_1$ the ordered set containing the first 
$m_1$ elements of $A$ and $A_2$ containing the remaining $m_2$ elements
 of $A$.  Write $\a_{k}$ for the $k$-th element of $A$.  

For $2\leq \k \leq m$, we define the amputated propagator with collision
 history of $A_2$ in $(0,t_2]$ and $A_1$ from $(t_2, t_1+t_2]$ with 
recollision $\a_\k$ to be 
\begin{align*}
        \scriptUt^{\mathrm{rec},\k}_{m_1,m_2;A}(t_1, t_2)\psi_0(p_0)  
                :=&
                 \int dr_0 \,  
                 \hat{V}_{\a_\k} (p_0 -r_0)\,  
\scriptU_{m_1,m_2;A}^{\circ}(t_1, t_2)\psi_0(r_0) \; ,
\end{align*}
recalling the definition of the time-divided
propagator $\scriptU_{m_1,m_2;A}^{\circ}$
from (\ref{E:freenorectdivdef}).
The superscript $\kappa$
decodes the location of the
recollision.

The corresponding full propagator is then:  
\begin{align}
        \scriptU^{\mathrm{rec},\k}_{m_1,m_2;A}(t_1,t_2) :=& \int_0^{t_1} ds \, 
e^{-i(t_1-s)H}
                \scriptUt^{\mathrm{rec},\k}_{m_1,m_2;A} (s, t_2) \;. 
        \label{E:defonerec}
\end{align}
Summing over $A$ and $2\leq \k \leq m$ removes their
 respective indices in the above propagators:
\begin{align}
        \scriptU^{\mathrm{rec}}_{m_1,m_2}(t_1,t_2) :=\sum_{\k=2}^m 
\sum_{A: |A|=m}^{\mathrm{no\,rec}} 
                \scriptU^{\mathrm{rec},\k}_{m_1,m_2;A} (t_1,t_2).
\label{U1mm}
\end{align}

Using this definition for the recollision term, together
with definition of the fully expanded non-recollision term
$\mathcal{U}_{m_1,m_2}^{\circ}$
from (\ref{E:Umm0}) and the truncated  term with a full
propagator $ \mathcal{U}_{m_0,m}$ from
(\ref{E:ufull}), we have the following decomposition of
 the $k$-th error term:

\begin{lemma} \label{L:timediverrorexpsn} Given (\ref{E:timedivdecomp}),
 and $1 \leq k \leq n$, define $t_1:=t/n$ and $t_2:= (k-1)t/n$.  We have:
\begin{align}
        \varphi_k^{\mathrm{error}}(t)   =&      
                \sum_{m_1=1}^{m_0-1} \sum_{m_2=m_0-m_1}^{m_0-1}
                \mathcal{U}_{m_1,m_2}^{\circ}(t_1, t_2) \psi_0
                +\sum_{m=0}^{m_0-1} \mathcal{U}_{m_0,m}(t_1, t_2)\psi_0 
                \notag \\
        &\quad +  \sum_{m_1=0}^{m_0-1}\sum_{m_2= (2-m_1)_+}^{m_0-1}
                \mathcal{U}^{\mathrm{rec}}_{m_1,m_2}(t_1, t_2)\psi_0 \notag \\
        =:&  \varphi_k^{\mathrm{error,1}}(t)+
                \varphi_k^{\mathrm{error,2}}(t)+ 
\varphi_k^{\mathrm{error,3}}(t)
                \label{E:timediverrordecomp} \, ,
\end{align}where $(a)_+ := \max(a,0)$. 
\end{lemma}Notice that our definitions imply that $\scriptU^{\circ}_m(0) = 0$ 
for $m>0$ and $\scriptU^0_0 (0) = \operatorname{Id}$.  Consequently, any
 time-divided propagator of the form $\scriptU_{m_1,m_2}(t_1, 0)$ will be 
zero unless $m_2=0$.  In this case, we have $\scriptU_{m_1,0}(t_1,0) = 
\scriptU_{m_1}(t_1)$.  
 
\begin{proof}
The proof is just a careful Duhamel expansion.  Recall
from (\ref{phimain}) and (\ref{E:timedivdecomp}) that
\begin{align*}
        \varphi_k(t) = e^{-i t_1 H} \varphi^{\mathrm{main}}_{k-1}(t) 
        =\sum_{m=0}^{m_0 -1} e^{-i t_1 H} 
\mathcal{U}_m^{\circ}( t_2)\psi_0 \, .
\end{align*} We now use the Duhamel formula to expand the full propagator. 
 We will stop the expansion when the new potential term represents a
 recollision or after $m_0$ new external collisions.  As before, internal
 collisions do not count when we speak of total collisions and we compensate
 by summing over them at each step of the expansion.  Performing this, we have:
\begin{align*}
        \sum_{m=0}^{m_0 -1}e^{-i t_1 H} \mathcal{U}_m^{\circ}( t_2) 
        =& \sum_{m_1,m_2 = 0 }^{m_0-1} \mathcal{U}^{\circ}_{m_1,m_2}(t_1, t_2) 
                + \sum_{m_2=0}^{m_0 -1} 
\mathcal{U}_{m_0, m_2}(t_1, t_2)        \\
         &+\sum_{m_1=0}^{m_0-1}\sum_{m_2=(2-m_1)_+}^{m_0-1} 
                \mathcal{U}^{\mathrm{rec}}_{m_1,m_2}(t_1, t_2).
\end{align*}
Finally one can verify 
\begin{align*}
        \sum_{\stackrel{m_1, m_2 \geq 0}{m_1 + m_2 = m}}
         \mathcal{U}^{\circ}_{m_1, m_2}(t_1,t_2) = 
\mathcal{U}^{\circ}_{m}(t_1+t_2),
\end{align*}
which collects the main terms and completes the proof of 
the lemma.  Thus, at each step $k$, where $1\leq k \leq n$, we use the
 Duhamel formula to expand the additional factor $e^{-it_1H}$.  We keep the wave functions 
which have total collisions $m$ where $m\leq m_0-1$ and there are no 
recollisions.  Any other cases are collected in the  error terms.       
\end{proof} 

We will now systematically estimate each of the three terms in
 (\ref{E:timediverrordecomp}).  For the orientation of the reader
we note that 
the first error term, $\varphi_k^{\mathrm{error,1}}(t)$, is a fully
expanded term with at least $m_0$ total number of collisions.
The second term, $\varphi_k^{\mathrm{error,2}}(t)$, contains
a full propagator after $m_0$ collisions in the short
time interval $[t_2, t_2+t_1]$. Finally the last error
term, $\varphi_k^{\mathrm{error,3}}(t)$, contains the recollisions.

%
%

\subsection{Preliminary estimates}

Recalling (\ref{E:freenorectdivdef}), we see that all of the randomness 
present in the wave function $\scriptU^{\circ}_{m_1,m_2;A}(t_1,t_2)\psi_0$
 is contained in the random phase, 
$\chi(A;\fp_{0,m})$. We start with discussing the expectation
value of these random phases.
Notice the randomness is
 unaffected by the time division - the time division is fully
 recorded in the kernel, 
$\mathcal{K}_{m_1,m_2}(t_1, t_2)$.

\subsubsection{Expectation of the random phases}

The net effect of 
expectation of our random phases will be to  induce various linear
relations (so called {\it pairing relations}) among our external momenta. 
For the precise formulation, we introduce the notation
 \begin{equation}
\varrho_{\Lambda}^{(n)} := \frac{N(N-1)\ldots (N-n+1)}{|\Lambda|^{n}} 
= \prod_{j=0}^{n-1} \varrho \Big(1- \frac{j}{N}\Big)\; 
\label{def:rhon}
\end{equation}
for the density of $n$-particle clusters, where we recall
the single-obstacle density $\varrho = N/|\Lambda|$
and its scaling $\varrho=\varrho_0\ep$.
Note that for a fixed $n$ and $\ep$:
\begin{align}
        \lim_{L\to \infty} \varrho^{(n)}_{\Lambda} = \lim_{L\to \infty} 
        \prod_{j=0}^{n-1} \varrho\Big( 1- \frac{j}{N} \Big) = \varrho^n \; .
\label{rhorho}
\end{align}
We also denote by $S(b)$ the permutation group on $b$ elements.

\begin{lemma}[Simple Set Expectation] \label{L:bcd}
Recall the notation introduced in (\ref{E:defptoell}).  
Suppose $G \in  L^2(\mathbb{R}^{3(m+1)};\mathbb{C})$ and
 the random phase $\chi$ is given in  (\ref{E:defchi}). Then
\begin{align}
        \bE \Big\| &\sum_{A: |A|=m}^{\mathrm{no\,rec}} \int d\fp_m \chi(A; p_0, \fp_m) 
                G(p_0,\fp_m) \Big\|_{L^2(dp_0)}^2 \notag \\   
        =& \sum_{b=0}^m\sum_{\sigma\in\operatorname{S}(b)}
                \sum_{\bell,\bell'}  
                \varrho^{(2m-b)}_{\Lambda} 
                \int  dp_0 d\fp_b d\fp'_b  G(p_0^{\ell_0},\fp_b^{\bell_{b}})
 \overline{G(p_0^{\ell'_0}, {\fp'}_b^{ \bell'_{b}})}
                \Delta_{\sigma}(p_0,\fp_b,\fp'_b)  \label{E:bcd}
\end{align}where $\bell_{0,b} := (\ell_0, \ldots, \ell_b)$, $\Sigma_{\bell,\bell'}$ 
is the sum over such vectors with components in the non-negative integers such that
 $\Sigma_{j=0}^b \ell_j = \Sigma_{j=0}^b \ell'_j =m-b$, and 
\begin{align}
        \Delta_{\sigma}(p_0, \fp_b, \fp'_b) :=& \prod_{j=1}^b 
                \delta[( p_{j-1}-p_j) - (\pp_{\sigma(j)-1}-\pp_{\sigma(j)})]
        \label{E:defpairing}.
\end{align}
\end{lemma} In the future, we will refer to $\Delta_{\sigma}$ 
as the {\it pairing function} and to its constituent delta functions as the 
{\it pairing relations}.  

\begin{proof}
In what follows all summations on ordered sets (such as $A$ or $A'$)
 will be understood to be summed over sets with non-repeating indices. 
 That is, we will drop the "no rec" from our summations.  We begin by 
expanding the squared sum:
\begin{align*}
        \Big|& \sum_{A:|A|=m} \int d\fp_m \chi(A; p_0,\fp_m)
 G(p_0, \fp_m) \Big|^2 \\
        &= \sum_{A:|A|=m} \sum_{A':  |A'|=m} \int d\fp_m d\fp'_m 
                \chi(A;p_0, \fp_m) \overline{ \chi(A'; p_0, \fp'_m)} 
                G(p_0,\fp_m) \overline{ G(p_0, \fp'_m)}\; .
\end{align*}The key to this Lemma is writing the sum over possible $A$ and
 $A'$, ordered sets of size $m$ with no repetition, as a sum over their
 possible intersections and then over their disjoint complements.  
Explicitly,
\begin{align*}
        \sum_{A:|A|=m} \sum_{A':  |A'|=m} = \quad \sum_{b=0}^m
 \sum_{B: |B| =b} 
                \sum_{\sigma\in \operatorname{S}(b) } \sum_{ (A, A')} ,
\end{align*}
where the last sum is over $A, A'$ of size $m$ such that 
$A\cap A' = B$, $B\prec A$ and $\sigma(B) \prec A$.  

We introduce the vector
$\bell_{0,b} := (\ell_0, \ell_1, \ldots, \ell_b)\; $ as follows: let 
 $\ell_0 = j$ if $\alpha_{j+1} \in B$ and $\a_k \notin B$ for $k \leq j$. 
 Similarly, denote by $\ell_k$ the number of $\a_j$ between the $k$-th and
 $(k+1)$-th members of $A$ which are in not $B$. 
 By this definition, if $\a_k$ is the $j$-th member of $B$ and $\a_{k+1} $ 
is the $(j+1)$-th member of $B$, then $\ell_j =0$.   In other words, the
 vector $\bell_{0,b}$ counts the number of $\a_j$'s in between members of
 $B$. Thus  $\bell_{0,b}$  describes precisely how $B$ is embedded in $A$.
Consequently, we have the relation:
\begin{align}
        \sum_{j=0}^{b} \ell_j = m - b \label{E:summingells}.
\end{align}
See Figure 1
 for the corresponding Feynman diagram.  The bullets refer to centers,
the lines between them are free propagators carrying a momentum.
The filled bullets are single centers that do not appear anywhere
else in the expansion, therefore the incoming and outgoing momenta
are the same. The elements of the set $B$ (unfilled bullets)
involve momentum transfer.
 Define $\bell'_{0,b}$ in a similar way for $A'$.

\begin{figure} \label{fig:basic}
        \includegraphics[scale=0.7]{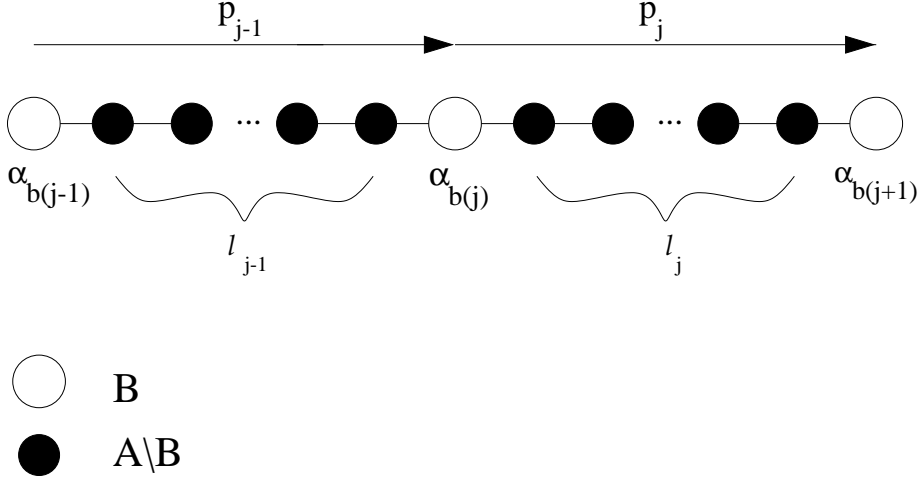}
        \caption{Basic Feynman Diagram}
\end{figure}

We next take the expectation of the $L^2$ norm.  Using
the independence of the variables $x_\alpha$, the expectation
\begin{align*}
        \bE_{x_{\a_j}}  e^{i x_{\a_j} p } :=& 
\frac{1}{|\Lambda|} \int_{\Lambda} 
                dx_{\a_j} e^{i x_{\a_j} p } 
                = \frac{\delta( p) }{|\Lambda|},
\end{align*}
 and (\ref{E:defchi}), we have
\begin{align}
        \bE \chi(A; \fp_{0,m}) &\overline{ \chi(A';\fp'_{0,m})} \notag \\
         =|\Lambda|^{-(2m-b)} &\prod_{j=1}^b \delta[( p_{b(j)-1} - p_{b(j)}) 
                - (\pp_{b'\circ\sigma(j)-1}-\pp_{b'\circ\sigma(j)})] \notag \\
         \times &\prod_{j=0}^b \Big[
                \prod_{k= b(j)+1}^{b(j+1) -1}\delta(p_{k-1}-p_k)
                \prod_{k= b'(j)+1}^{b'(j+1)-1}\delta(\pp_{k-1}-\pp_k) \Big]  
        \label{E:pairingfunctions} , 
\end{align} where we defined
\begin{align}
        b(j) :=\sum_{\iota=0}^{j-1} \ell_{\iota} + j  &&
        b'(j) := \sum_{\iota=0}^{j-1} \ell'_{\iota} + j \label{E:defb} \; ,
\end{align}
and we set our convention as $\Sigma_m^{m-1} =0 $ 
and $\Pi_m^{m-1} =1$.

We now integrate over the variables not involved in the pairing
 relations; specifically we integrate over $\fp_m\setminus \{ p_{b(j)} \}_{j=0}^{b} $
 and their prime counterparts.  Of the variables left, we relabel 
$p_{b(j)} \to p_j$ and $\pp_{b'(j)} \to \pp_j$, for $j=0,1, \ldots, b$. 
 Consequently
\begin{align*}
        \bE \Big\| \sum_{A:|A|=m}& \int d\fp_m \chi(A; p_0,\fp_m)
 G(p_0, \fp_m) \Big\|^2 \\
        =& \sum_{b=0}^m \sum_{B:|B|=b} \sum_{\sigma\in \operatorname{S}(b) }
                \sum_{(A,A')} |\Lambda|^{-(2m-b)} \\
        &\qquad 
                \times\int dp_0 d\fp_b d\fp'_b
 G(p_0^{\ell_0},\fp_b^{\bell_b}) 
                \overline{ G(p_0^{\ell'_0}, {\fp'_b}^{\bell'_b})} 
                \Delta_{\sigma}(p_0,\fp_b, \fp'_b)   .
\end{align*}
Lemma \ref{L:bcd} then follows since the total number  of  obstacles 
is $N$, hence the ways of choosing $B$, $A$ and $A'$ such that 
$B\prec A$, $\sigma(B) \prec A'$ and $A\cap A' = B$, for a fixed 
$\sigma$, $\bell_{0,b}$, and $\bell'_{0,b}$ is $\frac{N!}{(N-(2m-b) )!}$.  

\end{proof}

 In typical applications of this lemma, with $m_1+m_2=m$, we will set
\begin{align*}
        G(p_0, \fp_m) =\mathcal{K}_{m_1,m_2}(t_1, t_2;  \fp_{0,m}) 
 \psi_0(p_m).
\end{align*}
where we recall the definition (\ref{E:deftdivK}).
For a fixed $\bell_{0,b}$, the integrand in (\ref{E:bcd}) implies
 that we will have to make estimates on 
$\mathcal{K}_{m_1,m_2}(t_1, t_2;  \fp_{0,b}^{\bell_{0,b}})$. 
 To do this, we will introduce more notation.

Let $\beta = \beta(m_1,\bell)$ be such that $0 \leq \beta \leq m_1$ 
and satisfy
\begin{align}
        b(\b) \leq m_1 \leq b(\b+1)-1 \; ,
        \label{E:defbeta}
\end{align}
and we define
\begin{align}
        \ell_{\b1} := m_1 - b(\b)  && 
        \ell_{\b2} := b(\b+1) -1 -m_1
        \label{E:defellbeta12}.
\end{align}
In particular $\ell_\beta= \ell_{\b1} + \ell_{\b2}$.
In other words, $\beta$ is the number of $B$-elements before
the time division, and $\ell_{\b1}, \ell_{\b2}$ describe
how the time division line divides the $(A\setminus B)$-elements
between the $\beta$-th and $(\beta+1)$-th $B$-elements 
(see Figure 2; the dashed vertical line indicates the time division).
We define the {\it primed} versions analogously. 

Recalling (\ref{E:defptoell}) and (\ref{E:deftdivK}), we have the expression:
\begin{align*}
        \mathcal{K}_{m_1,m_2}&(t_1, t_2; \fp_{0,b}^{\bell_{0,b}})  \\ 
        =&\, \mathcal{K}(t_1; \underbrace{p_0,\ldots, p_0}_{\ell_0+1},
                \ldots, \underbrace{ p_\b, \ldots, p_\b}_{\ell_{\b1}+1}) 
                \mathcal{K}(t_2; \underbrace{p_\b,\ldots, p_\b}_{\ell_{\b2}+1},
                \ldots, \underbrace{ p_b, \ldots, p_b}_{\ell_{b}+1}) \,         .
\end{align*}

\begin{figure}\label{fig:tdiv}
        \includegraphics[scale=0.55]{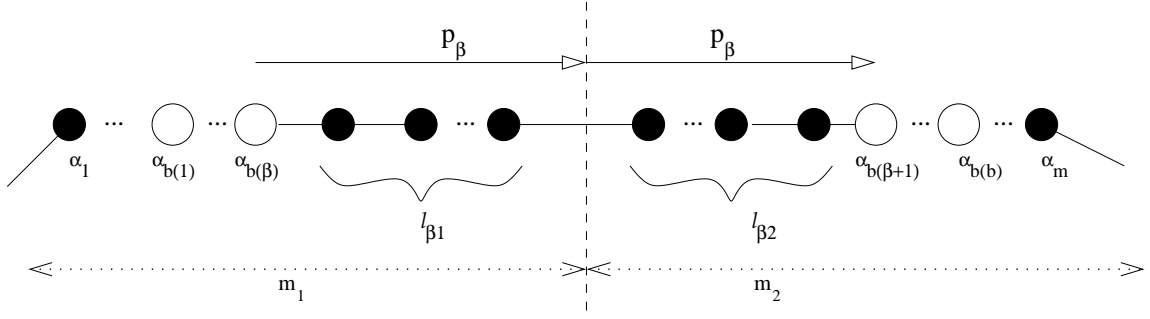}
        \caption{Time Division}
\end{figure}

In accordance with (\ref{E:defF}), one can check that:   
\begin{align*}
        \mathcal{K}_{m_1,m_2}&(t_1, t_2; \fp_{0,b}^{\bell_{0,b}}) \\ 
        =&
                \int_0^{t_1*} ds_1 d\tau_1 \int_0^{t_2*}ds_2 d\tau_2 \, 
                K_{m_1,m_2}(s_1,s_2; \fp_{0,b}^{\bell_{0,b}}) 
                F_{m_1,m_2}(\tau_1, \tau_2;\fp_{0,b}^{\bell_{0,b}})  
\end{align*}for 
\begin{align}
         K_{m_1, m_2}(s_1,s_2; \fp_{0,b}^{\bell_{0,b}}) :=& 
K(s_1; \fp_{0,\b}^{ \bell_{0,\b1}})
                K(s_2; \fp_{\b,b}^{ \bell_{\b2,b}}) \notag \\
         F_{m_1,m_2}(\tau_1,\tau_2; \fp_{0,b}^{\bell_{0,b}}) :=& 
 F(\tau_1; \fp_{0,\b}^{\bell_{0,\b1}})
                F(\tau_2; \fp_{\b,b}^{\bell_{\b2,b}}) \;. 
                 \label{E:deftdivF}
\end{align} 
In the future, we will omit the subscripts on $\mathcal{K}_{m_1, m_2}$, $K_{m_1,m_2}$ 
and $F_{m_1,m_2}$ when they are obvious from the context.

In what follows, we will adopt the following convention.  We will use upper case index 
variables  when summing over index sets of the form 
$$
 ( 0, \ldots, \b-1,\b1, \b2,\b+1 \ldots, b).$$  
Moreover, define the upper case momenta in the following way.  If $\fp_{0,b}$ 
is a set of momenta, the corresponding upper case momenta are defined by:
\begin{align}
        P_J :=&  p_J \qquad \mbox{for $J \neq \b1, \b2$} \notag \\
        P_{\b1} =& P_{\b2} := p_{\b} \label{E:upperlower} .
\end{align}Using this convention, we can write:        
\begin{align*}
        K_{m_1,m_2}(s_1, s_2; \fp_{0,b}^{\bell_{0,b}}) 
                = (-i)^{b} \int_0^{s_1*} \[ d\sigma_{J} \]_0^{\b1} 
                \int_0^{s_2*} \[d\sigma_{J} \]_{\b2}^b \prod_{J=0}^b e^{-i \sigma_J P_J^2/2} 
                \frac{(i\sigma_J)^{\ell_J}}{\ell_J!} \;,
\end{align*}where the notation implies that the product is
\begin{align*}
         \prod_{J=0}^b e^{-i \sigma_J P_J^2/2} \frac{(i\sigma_J)^{\ell_J}}{\ell_J!} 
         =& e^{-i (\sigma_{\b1}+ \sigma_{\b2}) p_\b^2/2} \frac{ (i\sigma_{\b1})^{\ell_{\b1}} 
                (i\sigma_{\b2})^{\ell_{\b2}}}{\ell_{\b1}! \ell_{\b2}!}\\ 
        &\times
                \prod_{j=0; j\neq \b}^b e^{-i \sigma_j p_j^2/2} 
\frac{(i\sigma_j)^{\ell_j}}{\ell_j!} \;.
\end{align*}

\subsubsection{Estimates on the effective potential}

The next result estimates the size of the effective potential
$ F(\tau)$ obtained after integrating out the internal momenta
(see (\ref{E:defF})).
\begin{lemma} \label{L:tdivfdecay}Let $0\leq b \leq m$ and $I\subset \{ 0, \ldots, b\}$
 with $|I |=n \leq b+1$ and $\xi := (\xi_1, \ldots, \xi_n) \in 
\{ 0, 1, 2\}^n$ be a multi-index.  Let $\bell_{0,b}$ 
be as in the statement of Lemma \ref{L:bcd}.  
If $G$ is twice differentiable then there is a universal constant
 $M$ such that:
\begin{align*}
        \Big| D^{\xi}_{\fp_I}\Big(  
                F(\tau ; \fp_{0,b}^{\bell_{0,b}}) G(\fp_{0,b} ) \Big) \Big| 
        \leq \frac{(M \lambda_0)^m} {\< \tau \>^{3/2} } 
                  \sup_{\xi' \in \{ 0, 1, 2\}^n} | D^{\xi'}_{\fp_I} 
                 G(\fp_{0,b} )  | 
                \prod_{j=1}^b \frac{1}{\<p_{j-1}-p_j \>^{26}} \;.
\end{align*}
\end{lemma}This lemma is a consequence of 
the dispersive estimates on the free propagator.  In particular, 
\begin{align*}
        \Big| \int dp \, e^{-is p^2/2} f(p) 
\Big| \leq \|e^{is \Delta/2} \check{f} \|_{L^{\infty}} \leq 
                Cs^{-3/2} \|\check{f}\|_{L^1}   \leq C  s^{-3/2} \| f \|_{H^2} .
\end{align*}
We can combine this with the trivial bound
 $ | \int dp \, e^{-isp^2/2} f(p) | \leq \| f\|_{L^1}$  to get:
\begin{align*}
        \Big| \int dp \, e^{-is p^2/2} f(p)
 \Big| \leq C\<s\>^{-3/2} ( \| f \|_{L^1} +\|f\|_{H^2} ).
\end{align*}We will frequently need to apply this estimate iteratively.  
To do this precisely, we make some definitions.  Suppose
 $I \subset \{0, \ldots, b\}$ of length $n$ and write
 $I = (i_1, \ldots, i_n)$.  Denote by $\xi$ a multi-index of length
 $n$ where $\xi_j \in \{0,2\}$.  Define the following operations
 on functions:  
\begin{align}\label{Ndef}
        N^\xi_{d\fp_I} := N^{\xi_{i_1}}_{dp_{i_1}} \circ 
\cdots \circ N^{\xi_{n}}_{dp_{i_n}}
                &&
        N^{0}_{dp_j} := \int dp_j |  \, \cdot \ | \notag \\
        N^{2}_{dp_j} := \Big(\int dp_j | \, \cdot \ |^2 \Big)^{1/2} &&
        \< D_{\fp_{I}}^{\xi} \> := \prod_{j=1}^n 
\<\grad_{p_{i_j}} ^{\xi_{i_j}}\> 
\end{align} 
Now let $f \in  \mathcal{S}(\mathbb{R}^{3(b+1)}; \mathbb{C})$ 
and define:
\begin{align}
        |\!|\!| f |\!|\!|_{d\fp_{I}} :=& \sum_{\xi\in \{0,2\}^{n}} 
                N^\xi_{d\fp_{I}}  \< D_{\fp_I }^{\xi}\> f   \; . 
                \label{E:deftripnorm}
\end{align}
With this language, 
\begin{align}
        \Big| \int dp e^{-is p^2/2} f(p) \Big| \leq
 C\<s\>^{-3/2} |\!|\!| f|\!|\!|_{dp}.
        \label{E:basicdisp}
\end{align}
We now move on to prove Lemma \ref{L:tdivfdecay}.

\begin{proof}Write $\bk:= \bk_m$ and $J:= J_{m,\bk_m}$
(recall the definition from (\ref{Jdef1})). 
 {F}rom (\ref{E:defF}), we have:
\begin{align*}
        D^{\xi}_{\fp_I} \Big( F&(\tau; \fp_{0,b}^{\bell_{0,b}})
G(\fp_{0,b}) \Big)
        = -i \sum_{k_1,\ldots, k_m =0}^\infty \int d\fq_{J}
                K(\tau; \fq_{J})
                D^{\xi}_{\fp_I}   
\Big( L(\fp_{0,b}^{\bell_{0,b}}, \fq_{J}) G(\fp_{0,b})   \Big)  .
\end{align*}
For a fixed $\bk$, we have from (\ref{E:Kdef}):
\begin{align*}
        \int d\fq_{J}&
                K(\tau; \fq_{J})
                D^{\xi}_{\fp_I} \Big(  
 L(\fp_{0,b}^{\bell_{0,b}}, \fq_{J}) G(\fp_{0,b})        \Big)   \\      
        =&\int_0^{\tau*} \[ d\sigma_{jk}\]_{jk \in J}
                \times (-i)^{\| \bk\|}\int d\fq_{J} \prod_{jk \in J}
 e^{-i \sigma_{jk} q_{jk}^2/2} 
                D^{\xi}_{\fp_I}         
\Big( L(\fp_{0,b}^{\bell_{0,b}}, \fq_{J}) G(\fp_{0,b}) \Big),
\end{align*}where $\| \bk \| := \sum_{j=1}^m k_j $.  
Applying (\ref{E:basicdisp}) iteratively, we have:
\begin{align*}
        \Big| \int d\fq_{J}
                K(\tau; &\fq_{J})
                D^{\xi}_{\fp_I}\Big( L(\fp_{0,b}^{\bell_{0,b}}, \fq_{J})
 G(\fp_{0,b}) \Big)      \Big| \\
        \leq& C^{\| \bk \|} \Big|\!\Big|\!\Big|
 D^{\xi}_{\fp_I}\Big( L(\fp_{0,b}^{\bell_{0,b}}, \fq_{J}) 
                G(\fp_{0,b}) \Big) 
                \Big|\!\Big|\!\Big|_{d\fq_J}\int_0^{\tau*} 
            \[ d\sigma_{jk}\]_{ J}
                \prod_{jk \in J} 
                \frac{1}{\< \sigma_{jk}\>^{3/2}} 
                 \\
        \leq&  \frac{ C^{ \| \bk\| } }{\< \tau\>^{3/2}} 
               \Big|\!\Big|\!\Big|   D^{\xi}_{\fp_I} 
  \Big( L(\fp_{0,b}^{\bell_{0,b}}, \fq_{J}) G(\fp_{0,b}) 
\Big)\Big|\!\Big|\!\Big|_{d\fq_J}\;,
\end{align*}
where the estimate is due to the multiple time integration
defined in (\ref{E:simplex}).  
Using the Leibniz rule and the triangle inequality,
\begin{align*}
         \Big|\!\Big|\!\Big|
         D^{\xi}_{\fp_I}   \Big(L(\fp_{0,b}^{\bell_{0,b}}, \fq_{J}) 
G(\fp_{0,b}) \Big) \Big|\!\Big|\!\Big|_{d\fq_J} \leq&
                \sum_{\xi'\leq\xi} \binom{\xi}{\xi'}|\!|\!|
 D_{\fp_I}^{\xi-\xi'} L(\fp_{0,b}^{\bell_{0,b}}, \fq_{J}) 
                D^{\xi'}_{\fp_I}G(\fp_{0,b})|\!|\!|_{d\fq_J}\;,
\end{align*}
where the notation $\xi'\leq \xi$ indicates componentwise ordering and 
$\binom{\xi}{\xi'} := \binom{\xi_1}{\xi'_1}\cdots\binom{\xi_n}{\xi'_n}$. 
 The form of $L(\fp_{0,b}^{\bell_{0,b}},\fq_J)$ in (\ref{E:defL})
allows us to write:
\begin{align*}
        |\!|\!|  &D^{\xi-\xi'}_{\fp_I} L(\fp_{0,b}^{\bell_{0,b}}, \fq_{J}) 
 |\!|\!|_{d\fq_J} \\  
        \leq& (M\lambda_0)^{m +\| \bk \| } \int d\fq_{J}
                \prod_{j=1}^m \frac{1}{\< r_{j-1} -q_{j1}\>^{30}}
                \frac{1}{\< q_{j1}-q_{j2}\>^{30}}\cdots 
                \frac{1}{\<  q_{j k_j}-r_{j} \>^{30}}\bigg|_{\fr_{0,m} = 
     \fp_{0,b}^{\bell_{0,b}}} \\
        \leq& (M\lambda_0)^{m +\| \bk \| }
                \prod_{j=1}^b \frac{1}{\<p_{j-1}-p_j\>^{26}}\int d\fq_J
                \prod_{j=1}^m \frac{1}{\< r_{j-1} -q_{j1}\>^{4}}
                \cdots 
                \frac{1}{\<  q_{j k_j}-r_{j} \>^{4}}
\bigg|_{\fr_{0,m} = \fp_{0,b}^{\bell_{0,b}}}.
\end{align*}
Summing over $k_j$ and using that $\lambda_0 \ll1$, 
we obtain the lemma.        
\end{proof}
Define:
\begin{align}
        J_1 :=& (11, \ldots, 1k_1, \ldots, m_1 1, \ldots, m_1k_{m_1})
 \notag \\
        J_2 :=& (m_1+1\, 1, \ldots, m_1+1 \, k_{m_1+1} , \ldots, m1, 
\ldots, mk_m),
\label{Jdef}
\end{align}where the first double index in $J_2$
has $m_1+1$ as the first element and $1$ as the second, etc. 
 This implies $J_1 \oplus J_2 = J_{m,\bk_m}$.  Expanding 
(\ref{E:deftdivF}) and using (\ref{E:defF}) yield:
\begin{align*}
         F_{m_1,m_2}(\tau_1,\tau_2; \fp_{0,b}^{\bell_{0,b}}) =&
         -\sum_{k_1, \ldots, k_m=0}^{\infty} 
                \int d\fq_{J_{m,\bk_m} }
 L(\fp_{0,b}^{ \bell_{0,b}}, \fq_{J_{m,\bk_m}})
                 K(\tau_1;\fq_{J_1}) K(\tau_2; \fq_{J_2})  .
\end{align*}
Again, the degenerate term of $k_1=\cdots = k_m =0$ of
 the last sum is defined as 
$\delta(\tau_1) \,\delta(\tau_2) L(\fp_{0,b}^{\bell_{0,b}})$. 
 A simple corollary to Lemma \ref{L:tdivfdecay} is the estimate:
\begin{align}
        \Big| D^{\xi}_{\fp_I}\Big(  
                F_{m_1,m_2} (&\tau_1,\tau_2  ; \, 
                \fp_{0,b}^{\bell_{0,b}}) G(\fp_{0,b} ) \Big) \Big|
                \notag  \\
        &\leq \frac{(M \lambda_0)^{m_1+m_2} } {\< \tau_1 \>^{3/2}
 \<\tau_2\>^{3/2} } 
                  \sup_{\xi' \in \{ 0, 1, 2\}^n} | D^{\xi'}_{\fp_I} 
                 G(\fp_{0,b} )  | 
                \prod_{j=1}^b \frac{1}{\<p_{j-1}-p_j \>^{26}} 
\label{E:tdivfdecay}. 
\end{align}

\subsection{Estimate of $\varphi^{\mathrm{error, 1}}(t)$ }

We now estimate the first error term, 
$\varphi^{\mathrm{error, 1}}(t)$, in Lemma \ref{L:timediverrorexpsn}.  
We will omit the $L\to\infty$ limit from the rest of this section
with the understanding, that this limit is taken in every estimate
before any other limits.

\begin{lemma} \label{L:0,0} Recall that $t = T\ep^{-1}$ 
and $\varrho= \ep \varrho_0$.  Let $m_0 = m_0(\ep) \gg 1$, $n=n(\ep)\gg 1$
 (we will make precise choices later in (\ref{E:m0choice}) and
(\ref{E:nchoice})) and suppose  
$1\leq m_1, m_2 \leq m_0-1$  such that $m=m_1+m_2 \geq m_0$.  
Then for $t_1+t_2=t$ with $t_2 \geq t_1$, we have the bound:
\begin{align}
        \bE \|  \mathcal{U}_{m_1,m_2}^{\circ}&(t_1,t_2)\psi_0\|^2 
                \notag \\
        \leq& C (M\lambda_0)^m \<T\>^m \Big[
                \frac{(\varrho t_1)^{m_1}
 (\varrho t_2)^{m_2}  }{m_1 ! m_2 ! }  
                +  m! \varrho  (\log t)^{m+ \O(1)}
 (\varrho t_1)^{m_1-1} (\varrho t_2)^{m_2}
          \Big] \label{E:0,0}.
\end{align}Consequently, for $k\geq1$:
\begin{align}
        \bE \| \varphi^{\mathrm{error,1}}_k(t) \|^2  \leq
                C^{m_0} \<T \>^{2m_0} 
                \Big[ \frac{1}{n}\frac{1}{m_0!} +  (2m_0)! \varrho
 (\log t)^{2m_0 +\O(1)}\Big] \; .
        \label{E:error1}        
\end{align}
\end{lemma}

\begin{proof}
By definition (\ref{E:freenorectdivdef}), we have:
\begin{align*}
        \scriptU^{\circ}_{m_1, m_2;A}&(t_1, t_2)\psi_0 (p_0) \\
        =& \int d\fp_{m} \,  \mathcal{K}(t_1; \fp_{0,m_1}) 
                \mathcal{K}(t_2;\fp_{m_1,m})\chi(A; \fp_{0,m}) 
\psi_0(p_m) .
\end{align*}We apply Lemma \ref{L:bcd} to get:
\begin{align}
        \bE \| \mathcal{U}_{m_1, m_2}^{\circ}& (t_1,t_2) \psi_0 \|^2 
\notag \\
                \leq& \sum_{b=0}^m \sum_{\sigma\in \operatorname{S}(b)}
 \sum_{\bell, \bell'}
                \varrho^{2m-b} \Big| \int dp_0 d\fp_b d\fp'_b  \,
 \Delta_{\sigma}(p_0, \fp_b,\fp'_b) 
                \psi_0(p_b) \overline{\psi_0(p'_b) } \notag  \\
        &\times \mathcal{K}_{m_1,m_2}(t_1,t_2; p_0^{\ell_0},\fp_b^{\bell_b}) 
                \overline{\mathcal{K}_{m_1,m_2}(t_1,t_2; p_0^{\ell'_0}, 
                {\fp'_b}^{\bell'_b}) } \Big|    \notag \\
        :=&  \, \mbox{(Direct)} + \mbox{(Crossing)}   \; ,  
 \label{E:norectimedivdecomp}
\end{align}where 
\begin{align}
        \mbox{(Direct)}  :=&    
                \sum_{b=0}^m \sum_{\bell,\bell'} \varrho^{2m-b} 
 \Big| \int dp_0 d\fp_b \,      
                |\psi_0(p_b) |^2 \notag \\
         &\times\mathcal{K}(t_1,t_2; p_0^{\ell_0},\fp_b^{\bell_b}) 
                \overline{\mathcal{K}(t_1,t_2; p_0^{\ell'_0}, 
\fp_b^{\bell'_b}) } \Big| \notag \\
        \mbox{(Crossing)} :=& \sum_{b=2}^m \sum_{\sigma\in 
\operatorname{S}(b)\setminus
                \operatorname{Id}} \sum_{\bell, \bell'}
                \varrho^{2m-b} \Big| \int dp_0 d\fp_b d\fp'_b \,
 \Delta_{\sigma}(p_0, \fp_b,\fp'_b) 
                \psi_0(p_b) \overline{\psi_0(p'_b) } \notag  \\
        &\times \mathcal{K}(t_1,t_2; p_0^{\ell_0},\fp_b^{\bell_{b}}) 
                \overline{\mathcal{K}(t_1,t_2; p_0^{\ell'_0}, 
{\fp'_b}^{\bell'_b}) } \Big|
        \label{E:defdircross}.
\end{align} 
The decomposition depends on whether $\sigma$ is trivial (identity) or not.
When $\sigma$ is trivial, then the pairing functions 
(\ref{E:defpairing}) reduce to the relations 
$\prod_{j=1}^b\delta(p_j-\pp_j)$.   
This decomposition will correspond to the two terms on the right hand 
side of estimate (\ref{E:0,0}).  

We will treat the Direct term first.  
Applying the Schwarz inequality and using that 
$\Sigma_{\bell'}1 = \Sigma_{\bell} 1= \binom{m}{b} \leq 2^m$,
\begin{align*}
        \mbox{(Direct)}\leq& \sum_{b=0}^m \sum_{\bell} 2^m 
\varrho^{2m-b} \int d\fp_{0,b}
                |\psi_0(p_b) |^2  |\mathcal{K}(t_1, t_2; 
\fp_{0,b}^{\bell_{0,b}})|^2.
\end{align*}Using (\ref{E:defF}), we can write
\begin{align*}
        \mbox{(Direct)} 
                \leq&2^m  \sum_{b=0}^m  \sum_{\bell} 
                \varrho^{2m-b} \int_0^{t_1*} ds_1 d\tau_1 
\int_0^{t_1*} ds'_1 d\tau'_1
                \int_0^{t_2*} ds_2 d\tau_2  \int_0^{t_2*} ds'_2 d\tau'_2  \\
        &\times \int d\fp_{0, b} 
|\psi_0(p_b)|^2 K(s_1,s_2; \fp_{0,b}^{\bell_{0,b}})
        \overline{K(s'_1,s'_2; \fp_{0,b}^{\bell_{0,b}})} \\
        &\times F(\tau_1,\tau_2; \fp_{0,b}^{\bell_{0,b}}) 
                \overline{F(\tau'_1,\tau'_2; \fp_{0, b}^{\bell_{0,b}})} .
\end{align*}We now estimate the free kernel. With the index
 convention introduced in (\ref{E:upperlower}), we use 
(\ref{E:Kdef}) and (\ref{E:Ksemigroup}) to write:  
\begin{align}
        \int d\fp_{0,b} \,& |\psi_0(p_b)|^2 K(s_1,s_2;
 \fp_{0,b}^{\bell_{0,b}})
                \overline{K(s'_1,s'_2; \fp_{0,b}^{\bell_{0,b}})}
                F(\tau_1,\tau_2; \fp_{0,b}^{\bell_{0,b}}) 
                \overline{F(\tau'_1,\tau'_2; \fp_{0,b}^{\bell_{0,b}})} 
 \notag \\
        =& \int d\fp_{0,b}  \,  F(\tau_1,\tau_2; \fp_{0,b}^{\bell_{0,b}}) 
                \overline{F(\tau'_1,\tau'_2;\fp_{0,b}^{\bell_{0,b}})} 
 |\psi_0(p_b)|^2 \notag \\
        &\times
                \int_0^{s_1*} \[d\sigma_J\]_0^{\b1}    \int_0^{s'_1*}    
                \[ d\sigma'_J \]_{0}^{\b1} 
                \int_0^{s_2*} \[d\sigma_J\]_{\b2}^b \int_0^{s'_2*} 
 \[ d\sigma'_J \]_{\beta2}^b         
                \notag\\
        &\times  \prod_{J=0}^b e^{-i (\sigma_J - \sigma'_J) P_J^2/2} 
                \frac{ (\sigma_J \sigma'_J)^{\ell_J}}{ (\ell_J!)^2}    \;.   
        \label{E:timedivfreesplit}   
\end{align}
For notational convenience, assume $\b\neq b$.  The case $b=\beta$ is 
estimated in the same way.  By the decay of the initial wave 
function in momentum space and the triangle inequality, we have:
\begin{align*}
        \Big| \int d\fp_{0,b}  &|\psi_0(p_b)|^2 F(\tau_1,\tau_2;
 \fp_{0,b}^{\bell_{0,b}}) 
                \overline{F(\tau'_1,\tau'_2; \fp_{0,b}^{\bell_{0,b}})} 
                \prod_{J=0}^b e^{-i (\sigma_J - \sigma'_J) P_J^2/2}\Big| \\
        \leq& \|  \psi_0 \|_{30,0}^2   \sup_{p_b} \Big| \int d\fp_{0,b-1} 
                F(\tau_1,\tau_2;\fp_{0,b}^{\bell_{0,b}}) 
                \overline{F(\tau'_1,\tau'_2; 
\fp_{0,b}^{\bell_{0,b}})}\<p_b\>^{-60} \\
        &\times
                e^{-i [(\sigma_{\beta1}-\sigma'_{\beta1})  
                + (\sigma_{\beta2} - \sigma'_{\beta2})]p_\beta^2/2}
                 \prod_{j=0; j\neq \beta}^{b-1} e^{-i (\sigma_j -\sigma'_j)
 p_j^2/2} \Big| \\
        \leq& \|  \psi_0 \|_{30,0}^2  C^b    \sup_{p_b} |\!|\!|  
F(\tau_1,\tau_2; \fp_{0,b}^{\bell_{0,b}}) 
                \overline{F(\tau'_1,\tau'_2;\fp_{0,b},\ell)} \<p_b\>^{-60} 
|\!|\!|_{d\fp_{0,b-1}} \\
        &\times \< (\sigma_{\beta1}-\sigma'_{\beta1})  
                + ( \sigma_{\beta2} - \sigma'_{\beta2}) \>^{-3/2}
                \prod_{j=0;j\neq \b}^{b-1} \< \sigma_j - \sigma'_j
 \>^{-3/2} ,
\end{align*}
where the last estimate used (\ref{E:basicdisp}) iteratively. 
 Applying this to (\ref{E:timedivfreesplit}), using ${\sigma'}_J  \leq s'_1$
 and $\sigma'_J\leq s'_2$ for $J\leq \b1$ and $J>\b2$, respectively,
 and performing the integration over $\[d\sigma'_J\]_0^{\b1}$ and
 $\[d\sigma'_J\]_{\b2}^b$, we have:
\begin{align*}
        \mbox{(\ref{E:timedivfreesplit})}
        \leq&  C^{b} \sup_{p_b} |\!|\!|  F(\tau_1,\tau_2; 
\fp_{0,b}^{\bell_{0,b}}) 
                \overline{F(\tau'_1,\tau'_2;\fp_{0,b}^{\bell_{0,b}})} 
\<p_b\>^{-60} |\!|\!|_{d\fp_{0,b-1}} \\
        &\times 
                 \frac{{s'}_1^{\ell_0 + \cdots + \ell_{\b1}}} 
                {\ell_0! \cdots \ell_{\b1}!} \frac{{s'}_2^{\ell_{\b2}
 +\cdots +\ell_b}}
                {  \ell_{\b2}! \cdots \ell_b! } 
                 \int_0^{s_1*}  \[d\sigma_J\]_0^{\beta1}     
\int_0^{s_2*} \[ d\sigma_J\]_{\beta2}^b
                 \prod_{J=0}^b \sigma_J^{\ell_J} ,
\end{align*}
where we have also used the trivial estimate $1/(\ell_j)! \leq 1$. 
 Using the identity 
$$\int_0^s (s-\sigma_j)^m \sigma_j^{\ell_j} d\sigma_j = 
\frac{\ell_j ! m!}{(m+\ell_j+1)!} s^{m+\ell_j+1} ,
$$ 
we have:
\begin{align*}
        \mbox{(\ref{E:timedivfreesplit})}
        \leq&  \frac{s_1^{m_1} s_2^{m_2} t^{m-b} C^{b}}{m_1! m_2!} 
                \sup_{p_b} |\!|\!|  F(\tau_1,\tau_2; \fp_{0,b}^{\bell_{0,b}}) 
                \overline{F(\tau'_1,\tau'_2;\fp_{0,b}^{\bell_{0,b}})}
 \<p_b\>^{-60} |\!|\!|_{d\fp_{0,b-1}} 
                 \, .
\end{align*}
Using the definition of $|\!|\!| \cdot |\!|\!|$ and  (\ref{E:tdivfdecay}) we conclude
\begin{align*} 
        \sup_{p_b} |\!|\!|  F(\tau_1,&\tau_2; \fp_{0,b}^{\bell_{0,b}}) 
                \overline{F(\tau'_1,\tau'_2;\fp_{0,b}^{\bell_{0,b}})} 
\<p_b\>^{-60} |\!|\!|_{d\fp_{0,b-1}} 
                \leq \frac{C(M\lambda_0)^m}{\langle \tau_1 \rangle^{3/2} 
                \langle \tau_2 \rangle^{3/2} \<\tau'_1\>^{3/2} \<\tau'_2\>^{3/2}} \; ,
\end{align*} which implies
\begin{align}
        \mbox{(Direct)} \leq& (M\lambda_0)^m  
                \sum_{b=0}^m \sum_{\bell} \varrho^{2m-b} 
          \int_0^{t_1*} ds_1 d\tau_1 \int_0^{t_1*} ds'_1 d\tau'_1 
          \int_0^{t_2*} ds_2 d\tau_2 \int_0^{t_2*} ds'_2 d\tau'_2  \notag  \\
        &\times \frac{ t^{m-b}} {\<\tau_1\>^{3/2} \langle \tau'_1 \rangle^{3/2}
                \<\tau_2\>^{3/2}\langle \tau'_2 \rangle^{3/2}}
                \frac{ s_1^{m_1}}{ m_1 !}
                \frac{s_2^{m_2}}{m_2 ! } \notag \\
        \leq&C (M\lambda_0)^m \< T\>^m 
                \frac{(\varrho t_1)^{m_1} (\varrho t_2)^{m_2} }
                {m_1! m_2!}   \; .  \label{E:timedivest1.2} 
\end{align}
The last inequality uses $\Sigma_{b=0}^m \Sigma_{\bell} 1= 2^m$.  
This proves the  estimate on the Direct term.

\medskip

It remains to estimate the Crossing term in (\ref{E:defdircross}). 
 We proceed in the spirit of the "indirect" term estimates in \cite{EY1} 
which are based on the $\a$-representation of the free kernel (Lemma \ref{L:Kidentity}).  
In particular, from (\ref{E:defB}), we have the representation
\begin{align*}
        \mathcal{K}_{m_1,m_2}(t_1, t_2; \fp_{0,m}) 
                 =& -\frac{e^{\eta_1 t_1 + \eta_2 t_2} }{4 \pi^2} 
                \int \frac{d\a_1 d\a_2 \quad 
e^{-i(\a_1 t_1-\a_2 t_2)}}{(\a_1 - p_0^2/2 + i \eta_1)
                (\a_2 - p_{m_1}^2/2 + i\eta_2)} \\
        &\prod_{j=1}^{m_1} \frac{B(\a_1, p_{j-1},p_j)}{\a_1 - p_j^2/2 + i \eta_1} 
                \prod_{j=m_1+1}^{m}  \frac{B(\a_2, p_{j-1},p_j)}{\a_2 - p_j^2/2 + i \eta_2}.
\end{align*}where $\eta_j:= \eta(t_j)$. 
 To shorten our expressions, define for $k=\{1,2\}$:
\begin{align}
        [\a_k, p] := 
                         \a_k - p_k^2/2 + i \eta_k   \, ,  
                \label{E:greenshortdef}
\end{align}
and its absolute value is denoted by
$$
|\a_k,p| := | [\a_k,p] | .
$$
 Analogous 
definitions are introduced for the primed versions with the same
$\eta_k$ regularizations:
\begin{align}
       [\a_k', p] := 
                         \a_k' - p_k^2/2 + i \eta_k, \qquad |\a_k',p| := | [\a_k',p] |  \, .
\end{align}
  Note that the regularization
$\eta_k$ is not explicitly accounted for in the notation. However,
 the short notations $[\alpha, p]$, $|\a, p|$
will always be used
in a context when $\alpha$ equals to one of the variables
$\alpha_1, \alpha_2, \alpha_1', \alpha_2'$ and  the index of $\alpha$
indicates the index of the regularizing $\eta$.

Consequently, it remains to bound:
\begin{align}
         \sum_{b=2}^m& \sum_{\sigma\neq \operatorname{Id}} 
\sum_{\bell,\bell'}
                \varrho^{2m-b} \Big| \int  d\fp_{0,b} d\fp'_b \,
 \psi_0(p_b) \overline{\psi_0(p'_b)} 
                \Delta_\sigma(p_0, \fp_b, \fp'_b) \notag\\
        & \times \int d\balpha  d\balpha'  \,
                e^{-i (\balpha -\balpha')\cdot(t_1,t_2)}  B(\a_1, \fp_{0,\b})
 B(\a_2, \fp_{\b,b}) 
                \overline{ B(\a'_1, p_0,\fp'_{\b'})} 
                \overline{ B(\a'_2, \fp'_{\b',b})} 
                \notag \\
        &\times \prod_{J=0}^{\b1} 
                \frac{ B(\a_1,P_J,P_J)^{\ell_J} }{[\a_1, P_J]^{\ell_J+1}}  
                \prod_{J =\b2}^b  \frac{B(\a_2,P_J,P_J)^{\ell_J}}
                {[ \a_2 , P_J ]^{\ell_J+1}}  \notag \\
        &\times  \prod_{J=0}^{\b'1}
                \frac{  \overline{B(\a'_1,\PP_J,\PP_J)}^{\ell'_J}  }
                {[ \a'_1, \PP_J ]^{\ell'_J+1}}   
                \prod_{J= \b'2}^b 
  \frac{  \overline{B(\a'_2,\PP_J,\PP_J)}^{\ell'_J}}
                {[\a'_2 , \PP_J ]^{\ell'_J+1}}   \Big|
        \label{E:expandedcrossing}
\end{align}where $\balpha:=(\a_1,\a_2)$, $\balpha':=(\a_1', \a_2')$ and
\begin{align}
        B(\a, \fp_{n,m} ) := \prod_{j=n+1}^m B(\a, p_{j-1}, p_j) &&\mbox{for $n\leq m$}.
        \label{E:defBvector}
\end{align}
We will now proceed as in  Lemma 3.5 of \cite{EY1} by exploiting the pairing 
relations and estimating each almost singular integral in a particular way. 
 The technical Lemma \ref{L:Blemma} (to be proven later in Section \ref{sec:est})
and the triangle inequality imply
\begin{align}
        \sup_{\a_1, \a_j}\Big|  B(\a_1,\fp_{0,\b} ) B(\a_2, \fp_{\b, b} ) \Big| 
                \leq \prod_{j=1}^b \frac{ M\lambda_0}{\< p_{j-1} -p_j \>^{30}} 
                \leq \prod_{i=1}^{6} \frac{1}{\<p_{k_i}\>^{4}} 
                \prod_{j=1}^b \frac{ M\lambda_0}{\< p_{j-1} -p_j \>^{4}} , \label{E:getdecay}
\end{align}where $k_i$ are between $0$ and $b$ and can be chosen at will.  
The same statements hold for the primed momenta.

In general the pairing structure can be quite complicated.  However, we know from
 Lemmas 2.4 and 2.8 of \cite{EY1} that we can express the primed momenta as 
linear combinations of the non-primed ones, in particular:
$$\Delta_{\sigma}(p_0, \fp_b, \fp'_b) =\prod_{j=1}^b \delta ( \pp_j - l_j(\fp_{0,b})) , $$ 
for some linear functions $l_j$.  Moreover, we always have the condition $\pp_b = p_b$. 
  The assumption that $\sigma \neq \operatorname{Id}$ implies that there is a 
$0<\kappa <b$ such that $l_{\k}(\fp_{0,b})$ is nontrivial.  That is, there are distinct 
indices $\kappa_1, \ldots, \kappa_{\iota}$ such that $p'_{\kappa} = \pm p_{\kappa_1} 
\pm \cdots \pm p_{\kappa_\iota}$ where the right hand side contains at least 3 terms. 
 Hence we can always choose $\kappa_1, \kappa_2$ such that
$\k_1, \k_2 \neq b$, and $\k_1 \neq\b$.  

Suppose first that
$\kappa\neq \beta'$.  Let
\begin{align*}
        \a(j) := \begin{cases}
                        \a_1 &\text{for $ 0\leq j <\b$} \\
                        \a_2 &\text{for $ \b\leq j \leq b$}
                 \end{cases} 
\end{align*}and define $\a'(j)$ analogously, with $\b'$ in place of $\b$.  
Define $\a(\k_1,\k_2)^c$ so that 
$$
\{ \a(\k_1), \a(\k_2), \a(\k_1,\k_2)^c \} = \{\a_1, \a_2\}.
$$
  In the case where $\{ \a(\k_1) , \a(\k_2) \} = \{ \a_1, \a_2\}$ choose 
$\a(\k_1,\k_2)^c = \a_1$.  Similarly define ${\a'(\k)}^c$ so that $\{\a'(\k),
 {\a'(\k)}^c \} = \{ \a'_1, \a'_2\}$.  {F}rom (\ref{E:expandedcrossing}), we need to bound:
\begin{align}
        \int & d\fp_{0,b} d\fp'_{b-1}\, \<p_b\>^{60}| \psi_0(p_b)|^2 
                \prod_{j=1}^{b-1} \delta( p'_j = l_j(\fp_{0,b})) 
                \int d\balpha d\balpha' \notag \\
        &\times 
                \prod_{J=0}^{\b1} \frac{1}{|\a_1, P_J|^{\ell_J+1}} 
                \prod_{J=\b2}^{b} \frac{1}{|\a_2, P_J|^{\ell_J+1}}
                \prod_{J=0}^{\b'1} \frac{1}{|\a'_1, P'_J|^{\ell'_J+1}} 
                \prod_{J=\b'2}^{b} \frac{1}{|\a'_2, P'_J|^{\ell'_J+1}} \notag \\
        &\times
                \prod_{i=1}^6 \frac{1}{\<p_{k_i}\>^{4}  \<p'_{k'_i}\>^4 }
                \prod_{j=1}^b \frac{M\lambda_0}{\<p_{j-1}-p_j\>^{4} \<\pp_{j-1}-\pp_j\>^4}
        \label{E:crossingstart}   .
\end{align}First choose $k'_1 = \k$, $k_1 = \k_1$ and $k_2 =\k_2$.  
We begin by using (\ref{E:summingells}) and making the bound:
\begin{align*}
        \sup_{\stackrel{\a'_1,\a_2}{p_0,p_b,\fp'_{b-1} }} &\Big[  
                \frac{\< {\a'(\k)}^c \>}{\<p'_{k'_2}\>^{4}}
                \frac{1}{|\a'_1,\pp_{\b'}|^{\ell'_{\b'1}} |\a'_2, \pp_{\b'}|^{\ell'_{\b'2}}} 
                \frac{1}{|\a'_2,p_b|^{\ell'_b}} \frac{1}{|\a'(\k),\pp_{\k}|^{\ell'_\k}} \\
        &\times \frac{1}{|\a'_2, \pp_{\b'}|} 
                \prod_{\stackrel{0\leq j\leq b-1}{j\neq \k,\b'}}
 \frac{1}{|\a'(j), \pp_j|^{\ell'_j+1}}
                 \Big] \leq C t_1^{m_1-1}t_2^{m_2}  \,. 
\end{align*}
(In case of $\kappa=\beta'$  the first term on the second line is omitted.)
 Indeed, this estimate follows as we can pick $k'_2$ such that $ |{\a'(\k)}^c , 
\pp_{k'_2}|^{-1}$ is a factor in the above product.  
We then apply (\ref{E:inftydecay}) to obtain
$$ 
        \sup_{{\a'(\k)}^c, \pp_{k'_2}}
        \frac{\< {\a'(\k)}^c\> }{\< \pp_{k'_2}\>^4 |{\a'(\k)}^c , \pp_{k'_2}|} 
        \leq \begin{cases} C t_1 &\text{if $\k \geq \b'$} \\    
                                     C t_2 &\text{if $\k < \b'$}
                \end{cases}     
$$
while using (\ref{E:inftynodecay}) on the remaining factors and applying $t_1\leq t_2$.
After estimating the initial wave function, $\| \psi_0\|^2_{30,0} \leq C $, we 
obtain
\begin{align*}
        \mbox{(\ref{E:crossingstart})} \leq& C t_1^{m_1 -1} t_2^{m_2}
                \sup_{p_b} \int d\fp_{0,b-1} d\fp'_{b-1} \prod_{j=1; j\neq \k}^{b-1} 
                \delta( \pp_j = l_j(\fp_{0,b}) ) \\
        &\times \int \frac{d\a_1}{\<\a_1\>  |\a_1,p_\b| }  
                \int \frac{d\a_2}{\<\a_2\>  |\a_2,p_b| }
                \int \frac{d\a'_1}{\<\a'_1\> |\a'_1,\pp_{\b'}|} 
                \int \frac{d\a'_2}{\<\a'_2\> |\a'_2,p_b| }  \\
        &\times
                \frac{ \delta(\pp_\k =\pm p_{\k_1}\pm p_{\k_2} 
 \pm \cdots)}{ | \a'(\k),\pp_\k| \, 
                |\a(\k_1),p_{\k_1}| \, |\a(\k_2), p_{\k_2}|} 
                \frac{\<\a(\k_1)\> \<\a(\k_2)\>\<\a'(\k)\>}{\< \pp_\k\>^4 
                \<p_{\k_1}\>^4 \<p_{\k_2}\>^4} \, 
       \frac{\<\a(\k_1,\k_2)^c \> }{\<p_{k_3}\>^4} \\
        &\times 
                \underbrace{\frac{1}{|\a_1,p_\b|^{\ell_{\b1}} 
                |\a_2, p_\b|^{\ell_{\b2}}}
                \prod_{j=0;j\neq\b}^b \frac{1}{|\a(j), p_j|^{\ell_j}} }_{=:(i)} \\
        &\times \underbrace{
                \prod_{\stackrel{j=0}{j\neq\k_1,\k_2}}^{b-1} \frac{1}{|\a(j),p_j|} 
                \prod_{i=4}^6 \frac{1}{\< p_{k_i}\>^4}
                \prod_{j=1}^{b} \frac{M\lambda_0}{\< p_{j-1} -p_j\>^4}}_{=:(ii)} .
\end{align*}By our assumptions that $m_1, m_2 \geq 1$ and $m_1+m_2\geq m_0$, we can
 choose $k_3$ so that the factor $|\a(\k_1,\k_2)^c, p_{k_3}|^{-1}$ appears in either
 $(i)$, $(ii)$ or both.  We now use (\ref{E:inftynodecay}) to estimate the factors in
 $(i)$.  If $|\a(\k_1,\k_2)^c, p_{k_3}|^{-1}$ appears in (i) (for some choice of $k_3$), 
we estimate this term by (\ref{E:inftydecay}):
$$  
        \sup_{\a(\k_1,\k_2)^c, p_{k_3}}  \frac{\<\a(\k_1,\k_2)^c \>}{ \< p_{k_3}\>^4 
                |\a(\k_1,\k_2)^c, p_{k_3}|} 
        \leq \begin{cases} C t_1 &\text{if $\k_1 \geq \b$} \\   
                                     C t_2 &\text{if $\k_1 < \b$} \; .
                \end{cases}      
$$
Either way, we apply (\ref{E:summingells}) to produce the bound $(i) \leq t^{m-b}$. 
 Next, we integrate $\pp_j$ for $1\leq j \leq   b-1$ except for $\k$, thus removing
 their corresponding delta functions.  We then bound:
\begin{align}
        \mbox{(\ref{E:crossingstart})} \leq& C t_1^{m_1-1} t_2^{m_2} t^{m-b}
        \sup_{p_b} \Big[ \int   \frac{d\a_2}{\<\a_2\> \, | \a_2,p_b| }
                \frac{d\a'_2}{\<\a'_2\> \, |\a'_2,p_b| } 
                \int \frac{d\a_1}{\<\a_1\>}  \int \frac{d\a'_1}{\<\a'_1\>} 
\int d\fp_{0,b-1}  \notag \\
        &\times
                \frac{ \delta(\pp_\k = p_{\k_1}\pm p_{\k_2} \pm \cdots)}{|\a'(\k),\pp_\k| \, 
                |\a(\k_1),p_{\k_1}| \, |\a(\k_2), p_{\k_2}|} 
                \frac{\<\a(\k_1)\> \<\a'(\k)\> \<\a(\k_2)\>}{\< \pp_\k\>^4 
                \<p_{\k_1}\>^4 \<p_{\k_2}\>^4}\notag  \\
        &\times \frac{1}{|\a'_1,l_{\b'}(\fp_{0,b}) |} \frac{1}{|\a_1,p_{\b}|} \times 
 \, (ii') \Big]  
        \label{E:beforetripleint}
\end{align}where $(ii')$ is $(ii)$ multiplied by $\< \a(\k_1,\k_2)^c\> /\<p_{k_3}\>^4$
 if that factor was not used in the estimate of $(i)$.

In this case we choose $k_3$ so that $| \a(\k_1,\k_2)^c, p_{k_3}|$
appears  in (ii).  We next integrate $\pp_{\k}$ which identifies 
$\pp_{\k} = p_{\k_1} \pm p_{\k_2}\pm \cdots$.  If $g = g( |p_{\k_1}|)$ 
is a non-negative function of $|p_{\k_1}|$, we have the estimate with $r=|p|$:
\begin{align}
        \int \frac{dp_{\k_1}}{\< p_{\k_1} \>^4} & 
\frac{ \< \a'(\k) \>  g( |p_{\k_1}| )}{ | \a'(\k), p_{\k_1} \pm p_{\k_2} \pm \cdots| \;
 |\alpha(\kappa_1), p_{\kappa_1}|
\, \< p_{\k_1} \pm p_{\k_2} \pm \cdots \>^4 } \notag \\
  &\leq   \int_0^\infty  \frac{ dr \, r }{ \< r\>^4 }
 \frac{1}{ |\a(\k_1), r |} \frac{g(r)}{| p_{\k_2} \pm \cdots|}  ,
\label{this}
\end{align}
where we have abused the notation and wrote $| \a, |p| | = | \a, p|$. 
 Indeed this follows from parametrizing the angular component of $p_{\k_1}$ 
relative to that of $\pm p_{\k_2} \pm\cdots$ 
and performing the angular integration exactly as in the proof of
 Proposition \ref{P:crossing}.

To apply (\ref{this}), we choose $k_4= \k_1-1$  in (\ref{E:getdecay}), and
we can make 
the last line in (\ref{E:beforetripleint}) independent of the angular variable of $p_{\k_1}$ 
by estimating $\<p_{\k_1-1}-p_{\k_1}\>^{-1}\leq 1$. 
The decay in  the variable $p_{k_4}=p_{\kappa_1-1}$ is lost, but it is restored by
the additional factor $\< p_{k_4}\>^{-4}$.
 Our choice of $k_4$ will assure that we have enough decay factors to perform the 
necessary integrations.  We obtain
\begin{align*}
        \mbox{(\ref{E:crossingstart})} \leq& C t_1^{m_1-1} t_2^{m_2} t^{m-b}
        \log t  \times  \sup_{p_b} \Big[ \int   \frac{d\a_2}{\<\a_2\>  |\a_2,p_b| }
                \frac{d\a'_2}{\<\a'_2\> |\a'_2,p_b| } \int \frac{d\a_1}{\<\a_1\>}  
                \int \frac{d\a'_1}{\<\a'_1\>}  \\
        & \int d\fp_{0,\k_1-1} d\fp_{\k_1+1,b-1}
                \frac{\<\a(\k_2)\>}{| \a(\k_2) , p_{\k_2}|}
                \int \frac{d |p_{\k_1}| \, \, | p_{\k_1}|}{\< |p_{\k_1}| \>^4 } 
                \frac{\<\a(\k_1)\>}{|\a(\k_1), |p_{\k_1}| |}
                \frac{1}{| p_{\k_2} \pm \cdots|} \\
        & \frac{1}{|\a'_1,l_{\b'}|} \frac{1}{|\a_1,p_{\b}|} \times  \, (ii'') \Big] ,
\end{align*}where $(ii'')$ is the same as $(ii')$ with $\<p_{\k_1-1} - p_{\k_1}\>^{-4}$
 majorized by $1$ and $k_4 = \k_1-1$.  

We then apply (\ref{E:alphaint}) twice to make the bound:
$$
        \sup_{\fp_{0,b} }\Big(\int \frac{d\a'_1}{\<\a'_1\> |\a'_1, l_{\b'}|} 
                \int \frac{d\a'_2}{\<\a'_2\> |\a'_2,p_b|}\Big) \leq C (\log t)^2 .
$$We can now integrate $|p_{\k_1}|$ and then choose coordinates for $p_{\k_2}$
 so that its angular component is parametrized relative to that of
 $p_{\k_3} \pm \cdots \pm p_{\k_\iota}$. 
 Choosing $k_5= \k_2-1$ and using $\<p_{\k_2-1}-p_{\k_2} \>^{-4} \leq 1$ makes the
 remaining terms independent of this angle.  We then integrate the angle, as done before,
 allowing us to integrate the remaining $p_j$ except for $p_\b$.  The integration is
 handled using (\ref{E:intnodecay}) in all instances except possibly one:  in the
 case where $(ii')$ contains $\< \a(\k_1,\k_2)^c\> /\<p_{k_3}\>^4$, we use 
(\ref{E:intadecay}) to handle this term.  Since $|\a_1,p_\b| \neq |\a(\k_2), |p_{\k_2}| \;|$ 
we can use (\ref{E:alphaint}) to bound
$$
\int \frac{d\a_1}{\<\a_1\> |\a_1,p_\b|} \leq C \log t , 
$$ 
while integrating 
$|\a_2, p_{\b}|^{-1}$ and completing the integration of $|p_{\k_2}|$ produces $\log$
 factors.  The order in which this is done will depend on whether or not $\b = \k_2$. 
 Finally we use (\ref{E:alphaint}) to estimate 
$$
        \sup_{p_b} \int \frac{d\a_2}{\<\a_2\> |\a_2, p_b|} 
                \leq C \log t   .
$$
Collecting these estimates completes the proof in the case where $\k \neq \b'$.  
The case $\k = \b'$  is easier to handle and can done as above.  Consequently,
\begin{align*}
        \mbox{(\ref{E:crossingstart})}\leq C(M\lambda_0)^{2b} t^{m-b} t_1^{m_1-1} t_2^{m_2}
                (\log t)^{m+\O(1)} ,
\end{align*}which applied in (\ref{E:expandedcrossing}), proves the first statement 
of Lemma \ref{L:0,0}. The second statement can be easily deduced from the first.
\end{proof}

\subsection{Estimate of $\varphi^{\mathrm{error,2}}_k$}

We next prove the amputated version of the preceding lemma which will be used, 
by setting $m_1 = m_0$, to estimate $\varphi^{\mathrm{error, 2}}_k$ in 
(\ref{E:timediverrordecomp}).

\begin{lemma} \label{L:0,1} Suppose, $m_1 >2$, $ 0 \leq m_2 < m_0$ and define $m=m_1+m_2$. 
 Let $t_1\leq t_2 $ and $t_1 +t_2= t = T\ep^{-1}$ for  $k\geq 1$.  We then have the bound:
\begin{align*}
        \sup_{0\leq s \leq t_1} &\bE \|  \scriptUt_{m_1,m_2}(s, t_2) \psi_0 \|^2 \\ 
        \leq& C(M\lambda_0)^{m} \<T\>^m    
                \Big[ \frac{\varrho (\varrho t_1)^{m_1-1}
 (\varrho t_2)^{m_2}}{(m_1-1) ! m_2!} + m! \varrho^2 (\varrho t_1)^{m_1-2}
 (\varrho t_2)^{m_2} (\log t)^{m+\O(1)}\Big]
\end{align*}It then follows that:
\begin{align*}
        \bE \| \varphi_{k}^{\mathrm{error,2}}(t) \|^2 \leq  C^{m_0} \<T\>^{2m_0} 
\Big[  \frac{t}{n^{m_0} m_0!} 
                + \frac{ (2m_0)! (\log t)^{2m_0+\O(1)} }{ n^{m_0}} \Big]        
\end{align*}
\end{lemma}
\begin{proof}
The proof of the first statement is almost identical to the proof of Lemma
 \ref{L:0,0} only that we replace
\begin{align*}
        \mathcal{K}_{m_1,m_2}(t_1, t_2; \fp_{0,m}) \qquad\mbox{with}\qquad \hV(p_0-p_1) 
                \mathcal{K}_{m_1-1,m_2}(t_1, t_2; \fp_{1,m}) .
\end{align*}The missing $p_0$ in the latter free kernel effectively eliminates a power 
of $t_1$ from the estimate of Lemma \ref{L:0,0} and also reduces the effect of
$m_1$ by one in the estimate.

As a technical note, the crossing estimates which are done with the aid of the 
$\a$-representation (Lemma \ref{L:Kidentity}) require the kernel to have at least 
two momentum variables.  Usually this amounts to requiring $m_j > 0$ for $j=1,2$.  
In the previous lemma, this was avoided by assumption.  However, in this case, it is 
possible that $m_2 =0$.  Accordingly, we do not expand the kernel $\mathcal{K}_{m_2}(t_2)$ 
with Lemma \ref{L:Kidentity} but use the trivial estimate
 $|\mathcal{K}_{0}(t_2; r_b) |  \leq 1 $ thus reducing our estimates to 
those without time division.  Otherwise, the proof of the first statement 
follows in the exact same way as the previous lemma.

To prove the second statement, we recall from the defintion (\ref{E:ufull})
that
\begin{align*}
        \scriptU_{m_1,m_2}(t_1,t_2) \psi_0 = \int_0^{t_1} ds \; e^{-i(t_1 -s) H} 
                \scriptUt_{m_1,m_2}(s,t_2)\psi_0 \; .
\end{align*} A simple consequence of the unitarity of $e^{-i(t_1 -s) H} $ implies
\begin{align}
        \bE \|  \scriptU_{m_1,m_2}(t_1,t_2) \psi_0 \|^2 \leq
                t_1^2 \sup_{0\leq s \leq t_1} \bE \| \scriptUt_{m_1,m_2}(s,t_2)\psi_0\|^2.
                \label{E:crudeunitary}
\end{align}The first part of the lemma with $t_1 = t/n$, $t_2 = (k-1)t/n$ and 
$m_1 = m_0$ yields:
\begin{align*}
        \bE \| \varphi_{k}^{\mathrm{error,2}}(t) \|^2 \leq& 
                m_0^2 t_1^2  \sum_{m_2 =0}^{m_0-1} 
\bE \|\scriptUt_{m_0,m_2}(t_1,t_2)    \psi_0\|^2\\
        \leq& C^{m_0} \<T\>^{2m_0} \Big[  \frac{t}{n^{m_0} m_0!} 
                + \frac{ (2m_0)! (\log t)^{2m_0+5} }{ n^{m_0}} \Big]  \; .   
\end{align*}
\end{proof}

%
%
\subsection{Estimate of $\varphi^{\mathrm{error,3}}_k$} \label{ss:error3}
We now move on to estimate the third error term in 
(\ref{E:timediverrordecomp}).  As a rule of thumb, a genuine
 recollision will allow us to argue as in the estimates of the 
crossing term in Lemma \ref{L:0,0} to eliminate a power of $t_1$. 
 However we will obtain a factor of $t_1^2$ when we apply crude 
estimates such as (\ref{E:crudeunitary}).  Since the amputation
 effectively eliminates one power of $t_1$ (as in Lemma \ref{L:0,1}), 
this term will be $\O(n^{-2})$ when $m_1$ is small.  After summing on
 $k$ in (\ref{E:deferror}), our error term will be $\O(1)$ at best, 
which is not sufficient.  
Consequently, we are forced to continue the Duhamel
 expansion.  

The idea is that we will keep expanding until we either 
obtain another genuine recollision or we get a new collision center. 
 The latter will produce another factor of $n^{-1}$ so that after 
summation on $k$ in (\ref{E:deferror}) our term will be $\O(n^{-1})$ 
and by choosing $n$ to be sufficiently large, this term will vanish
 in the limit.  The case of a second recollision should be smaller 
by a power of time, which guarantees that this term vanishes in the limit.
  Intuitively, in order to have recollisions, obstacles need to be within
 a close vicinity of one another.  Hence terms with these collision 
histories should be small since the probability of such configurations
 is higher order.  If the obstacles were not within a close proximity
 with one another, then the wave function would need to travel very 
far to recollide and again, classically, we should be able to argue 
that the respective term is higher order.    

However, there is a technical difficulty which presents itself here. 
 Viewing things classically, it is possible that two obstacles are 
$\O(1)$ distance apart.  When this happens, our wave can collide with 
these obstacles one after another in succession (of two or more times) 
and give the appearance of undergoing only one recollision.  Though the
 probability that the obstacles have this configuration is 
$\varrho = \O(\ep)$, this factor is not sufficient to compensate
the loss in our unitary estimate. 
 Consequently, we need effectively sum up the two-obstacle Born series
 to account for this.  Not all pairs of recollisions need to be treated
 in this manner.  If the original collision sequence is given by 
$(\a_1, \ldots, \a_m)$ and we obtain a new collision center which is a 
recollision at $\a_{\k_1}$ followed by another new collision center which 
is a recollision at $\a_{\k_2}$, we will immediately be able to argue that
 the terms corresponding to the case where $\k_2> \k_1$ are small on the 
basis that this collision pattern is higher order.  Indeed, in order to
 have a genuine recollision and not an internal collision, we need
 $\k_1 \geq 2$.  This implies that there is at least one more obstacle 
in the vicinity of $\a_{\k_1}$ and $\a_{\k_2}$.  The probability of this 
configuration occurring is higher order.  Hence we will only sum the 
two-obstacle Born series in the case where $\k_2< \k_1$ (they can never 
be equal since we already summed over internal collisions).   

Before we precisely describe the final stopping rule for our Duhamel 
expansion, we need to define propagators associated to more complicated 
collision patterns.  Given $A$ of size $m$, let $m_1$ and $m_2$ be
 non-negative integers with $m_1+m_2=m$.  For $n_1\geq 2$ define:
\begin{align*}
        \mathcal{A}( n_1) := \underbrace{( \ldots ,\a_{\k_1}, \a_{\k_2}, \a_{\k_1}) }_{n_1 } 
\end{align*}This will be the sequence of centers associated to the pair collision 
mentioned above.  The propagator associated with the pair recollision is defined as 
\begin{align}
        &\scriptU^{\circ;\k_1,\k_2}_{[n_1], m_1,m_2;A}(t_1,t_2)\psi_0 (p_0)  \notag \\
        &\qquad := \int  d\fu_{0 ,n_1-2} d\fr_{0,m} 
                \mathcal{K}_{n_1+m_1,m_2}(t_1,t_2; p_0,\fu_{0,n_1-2}, \fr_{0,m})  \notag  \\
        &\qquad \qquad \times
                \chi(\mathcal{A}(n_1)\oplus A; p_0, \fu_{0,n_1-2}, \fr_{0,m}) \psi_0(r_m)\, ,
        \label{E:pairrecdef}
\end{align}
The number $n_1$ in brackets indicates the number of pair recollisions.
For the free propagator kernel we applied
he definition  (\ref{E:deftdivK}) in the following form 
$$
  \mathcal{K}_{n_1+m_1,m_2}(t_1,t_2; p_0,\fu_{0,n_1-2}, \fr_{0,m})
 = \mathcal{K}_{n_1+m_1}(t_1; p_0,\fu_{0,n_1-2}, \fr_{0,m_1})
  \mathcal{K}_{m_2}(t_2; \fr_{m_1,m}) \;
$$ 
(see Figure 3 for the order of momentum variables).

\begin{figure}\label{fig:2ob}
        \includegraphics[scale=0.6]{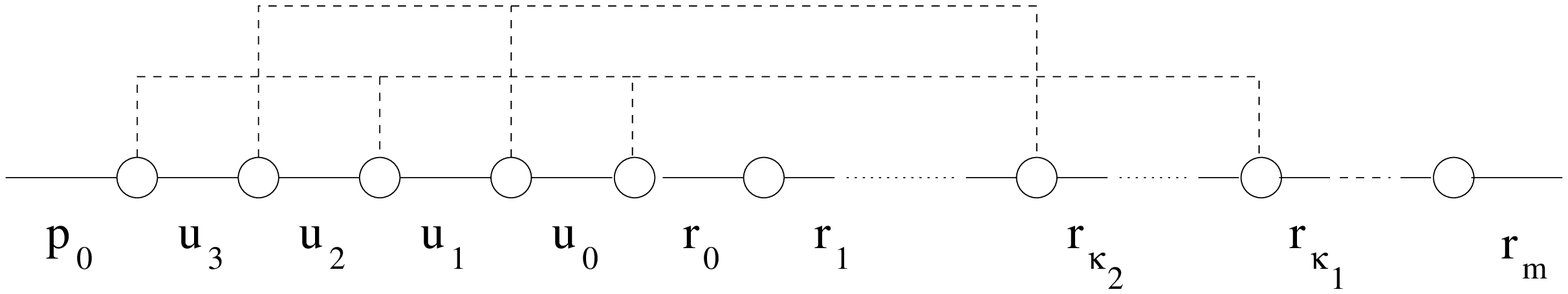}
        \caption{Resummation of two obstacles}
\end{figure}

Summing over $1\leq \k_2 <\k_1\leq m$ and $A$ gives:
\begin{align*}
        \scriptU^{\circ}_{[n_1],m_1,m_2} (t_1,t_2) := \sum_{\stackrel{\k_1,\k_2
                }{1\leq \k_2<\k_1\leq m}}\sum_{A:  |A|=m} 
                \scriptU^{\circ;\k_1,\k_2}_{ [n_1] ,m_1,m_2;A}(t_1,t_2)\; .
\end{align*}
The propagators for $n_1 = 1$ are defined as:
\begin{align*}
        \scriptU^{\circ;\k}_{[1],m_1,m_2;A}(t_1,t_2)\psi_0 :=& \int \!\! d\fr_{0,m} \,
                \mathcal{K}_{m_1+1,m_2}(t_1,t_2;p_0, \fr_{0,m}) 
                \chi(\a_{\k}\oplus A; p_0,\fr_{0,m}) \psi_0(r_m) \, , \\
        \scriptU^{\circ}_{[1],m_1,m_2}(t_1,t_2) :=& \sum_{\k=2}^m \sum_{A: |A|=m} 
                \scriptU^{\circ;\k}_{[1],m_1,m_2;A}(t_1,t_2) .
\end{align*}
Note that these are the fully expanded versions of the truncated
one recollision terms (\ref{E:defonerec})
and (\ref{U1mm}).

We will also need to define the amputated version of the two recollision propagator:
\begin{align}
        &\scriptUt^{\k_1,\k_2}_{[n_1],m_1,m_2;A}(t_1,t_2)\psi_0 \notag \\
        &\qquad :=    \int  d\fu_{0 ,n_1-2} d\fr_{0,m} \hV(p_0-u_0)
                \mathcal{K}_{n_1+m_1-1,m_2}(t_1,t_2;\fu_{0,n_1-2}, \fr_{0,m})  \notag  \\
        &\qquad \qquad \times
                \chi(\mathcal{A}(n_1)\oplus A; p_0, \fu_{0,n_1-2}, \fr_{0,m}) \psi_0(r_m) 
\label{tworec}
\end{align}
and 
$$
\scriptUt^{\k_1,\k_2}_{[n_1],m_1,m_2}(t_1,t_2) =\sum_{A: |A|=m} 
\scriptUt^{\k_1,\k_2}_{[n_1],m_1,m_2;A}(t_1,t_2) \; .
$$

The next propagators are associated with the pair recollision pattern
 followed by a new collision with $\alpha_0$. For $n_1\ge 2$ we define:
\begin{align}
        &\scriptUt^{\k_1,\k_2}_{[n_1], m_1,m_2;\a_0, A} (t_1,t_2)\psi_0 (p_0) \notag \\
        &\qquad \quad := \int dp_1 d\fu_{0,n_1-2} d\fr_{ 0,m} \hV(p_0-p_1) 
                \mathcal{K}_{n_1+m_1,m_2}(t_1,t_2;p_1, \fu_{n_1-2}, \fr_{0,m})  \notag \\
        &\qquad \qquad \times \chi( \a_0\oplus \mathcal{A}(n_1)\oplus A; 
                \fp_{0,1}, \fr_{0,m}) \psi_0(r_m)
        \label{E:pairrecplusonedef}
\end{align}and the summed up version:
\begin{align*}
        \scriptUt_{1, [n_1],m_1,m_2}(t_1,t_2):=\sum_{\stackrel{\k_1,\k_2
                }{1\leq \k_2<\k_1\leq m}}\sum_{\hat{A}:  |\hat{A}|=m+1}
        \scriptUt^{\k_1,\k_2}_{[n_1], m_1,m_2;\a_0,A} (t_1,t_2)   
\end{align*}where the sum is over sets $\hat{A} := \a_0 \oplus A$ 
with non-repeating indices.
Note that the order of subscripts, $1, [n_1],m_1,m_2$ indicate the chronological
order of the collision types, the bracket indicates the number of pair recollisions.
  
For the special case $n_1 =1 $ the  propagators are defined as:
\begin{align*}
        \scriptUt^{\k}_{[1], m_1,m_2;\a_0, A}(t_1,t_2)\psi_0 :=& \int d\fp_{0,1} d\fr_{0,m} 
                \mathcal{K}_{m_1+1,m_2}(t_1,t_2;p_1, \fr_{0,m})  \notag \\
        & \times \chi( \a_0 \oplus \a_{\k} \oplus A; 
                \fp_{0,1}, \fr_{0,m}) \psi_0(r_m)\,.
\end{align*}and
\begin{align*}
        \scriptUt_{1,[1],m_1,m_2}(t_1,t_2) := \sum_{\k=2}^m \sum_{\hat{A}:  |\hat{A}|=m+1}
                \scriptUt^{\k}_{ [1], m_1,m_2;\a_0,A}(t_1,t_2).
\end{align*}
Finally we have the propagators corresponding to the pair 
recollisions followed by a genuine recollision with $\alpha_{\k_3}$:
\begin{align}
        &\scriptUt^{\k_1,\k_2,\k_3}_{ * , [n_1], m_1,m_2;A}(t_1,t_2)\psi_0 (p_0) \notag \\
        &\qquad \quad := \int dp_1 d\fu_{0,n_1-2} d\fr_{0,m} \hV(p_0-p_1)
                \mathcal{K}_{n_1+m_1,m_2}(t_1,t_2;p_1, \fu_{n_1-2}, \fr_{0,m}) 
                \notag \\
        &\qquad\qquad \times
     \chi(  \a_{\k_3}\oplus \mathcal{A}(n_1)\oplus A; \fp_{0,1},\fu_{0,n_1-2}, \fr_{0,m}) 
                \psi_0(r_m)
        \label{E:pingpongdef},
\end{align}
and
\begin{align*}
        &\scriptUt_{*, [n_1],m_1,m_2}(t_1,t_2) :=
                \sum_{\k_1=2}^m \sum_{\k_2=1}^{\k_1-1} 
                \sum_{\stackrel{\k_3=1}{\k_3\neq\k_1,\k_2}}^m 
                \sum_{A: |A|=m}
                \scriptUt^{\k_1,\k_2,\k_3}_{ *,[n_1] ,m_1,m_2;A}(t_1,t_2) .
\end{align*}
The condition on $\k_3$ assures us that the new recollision is unrelated to the 
pair recollision. The star indicates the new recollision that is
independent of the pair recollisions.

If $\scriptUt(t_1,t_2)$ is any one of the amputated propagators defined above, 
its corresponding full propagator is defined as:
\begin{align}
        \scriptU(t_1,t_2) := \int_0^{t_1} ds \, e^{-i(t_1-s)H} \scriptUt(s,t_2)\,.
\label{UU}
\end{align}
In our notation, summation over appropriate ranges of a particular index removes 
that index.  For example, when the pair recollision indices $\k_1, \k_2$ do not 
appear explicitly, then the summation over $1\leq \k_2<\k_1 \leq m$ has been performed.
  If we sum over a different set of $\k_1, \k_2$ (as we will below), the summation will
 appear explicitly.  

We now give a precise stopping rule for the expansion of the recollision
term  $\scriptU^{\mathrm{rec}}_{m_1,m_2}(t_1,t_2)$ defined in (\ref{U1mm}). 
 Dropping the explicit dependence 
on $(t_1,t_2)$ in our propagators, we expand beyond the first recollision center and
we obtain
\begin{align*}
        \scriptU^{\mathrm{rec}}_{m_1,m_2} =& \,
                \scriptU^{\circ}_{[1],m_1,m_2}        + \sum_{2\leq \k_1, \k_2\leq m }  
                 \scriptU^{\k_1,\k_2}_{[2],m_1,m_2}
                + \scriptU_{1,[1],m_1,m_2} \; .
\end{align*}
The first term corresponds to the fully expanded term after the first
recollision. The second term is the pair recollision.
The third term is a single recollision ($n_1=1$) followed by
a fresh collision. 

The second term will be split according to $\kappa_1<\kappa_2$ or
$\kappa_2 < \kappa_1$. In the easier case, when $\kappa_1<\kappa_2$,
one can  use the unitarity on the full evolution 
already after the second recollision ($n_1=2$). When $\kappa_2<\kappa_1$
we have to continue the expansion of this term. We stop
when we obtain a brand new
collision center or if we have a recollision at a center 
$\a_{\k_3} \neq \a_{\k_1} , \a_{\k_2}$.  Internal recollisions are not counted
 (they are summed as before) and we only expand according to centers. 
 Formally, this gives the identity:
\begin{align}
          \scriptU^{\mathrm{rec}}_{m_1,m_2} =& \sum_{n_1=1}^{\infty} 
                \scriptU^{\circ}_{[n_1], m_1,m_2} + \sum_{2\leq \k_1<\k_2\leq m}
                \scriptU^{\k_1,\k_2}_{[2],m_1,m_2}  \notag \\
        &+\sum_{n_1=1}^\infty \scriptU_{1,[n_1],m_1,m_2} 
                + \sum_{n_1=2}^{\infty} \scriptU_{*,[n_1],m_1,m_2} \; .
        \label{E:error3decomp}
\end{align}
From the estimates below it will follow  that these 
series converge to $\scriptU^{\mathrm{rec}}_{m_1,m_2}$.


\begin{lemma} \label{L:error3bound}
If $1\leq k \leq n$, we have:
\begin{align*}
        \bE \| \varphi_{k}^{\mathrm{error, 3}}(t) \|^2 \leq C \<T\>^{4m_0} (\log t)^{\O(1)} 
                \Big[ \frac{1}{n^{3}} + (2m_0)! \varrho (\log t)^{2m_0} \Big] .
\end{align*}
\end{lemma}
\begin{proof}
Applying the Schwarz inequality to (\ref{E:error3decomp}), we have the bound:
\begin{align*}
        \bE \|  & \scriptU^{\mathrm{rec}}_{m_1,m_2}(t_1,t_2) \psi_0 \|^2 \\
        \leq& 
                \sum_{n_1=1}^{\infty} n_1^2 \, 
                \bE \| \scriptU^{\circ}_{[n_1], m_1,m_2}(t_1,t_2)\psi_0\|^2
                + \sum_{2\leq \k_1<\k_2\leq m} m^2 \bE\| \scriptU^{\k_1,\k_2}_{[2],m_1,m_2}
                (t_1,t_2)\psi_0\|^2 \\
        &+\sum_{n_1=1}^\infty  n_1^2 \, \bE\| \scriptU_{1,[n_1],m_1,m_2} (t_1,t_2)\psi_0 \|^2
                + \sum_{n_1=2}^{\infty}n_1^2\, 
  \bE \| \scriptU_{*,[n_1],m_1,m_2}(t_1,t_2)\psi_0 \|^2 \,.
\end{align*}

These four terms are estimated in the following technical
lemmas, whose proofs are given in the next section.
Here we present only a short explanation after each lemma.

\begin{lemma}\label{L:pairrec}                  
Suppose $m_1 + m_2 \geq 2$ and $n_1\geq 1$.  Then
\begin{align*}
        \bE\| \scriptU^{\circ}_{[n_1],m_1,m_2}(t_1,t_2)\psi_0 \|^2 \leq  
        (M\lambda_0)^{n_1+m} \<T\>^{2m-2} 
                \varrho (\log t)^{\O(1)}\Big[ \frac{\<T\>}{n} + m! \varrho (\log t)^m \Big]  
\end{align*}
\end{lemma}
This term is small by a factor $\varrho$ that comes from the recollision.
The first term in the square bracket corresponds to the direct term. 
Since there is a new collision in the short time interval $[t_2, t_2+t_1)$,
this will provide an extra factor $\varrho t_1$ and hence the factor $1/n$.
All the other crossing terms carry an extra $\varrho$.

\begin{lemma}\label{L:pairrecplus1}
Suppose $m_1 + m_2 \geq 2$ and $n_1\geq 1$.  We have the bound:
\begin{align}
        \sup_{0\leq s \leq t_1}& 
    \bE \| \scriptUt_{1,[n_1],m_1,m_2}(s,t_2)\psi_0\|^2 \notag \\
        \leq& (M\lambda_0)^{n_1+m} \varrho^2 \<T\>^{2m-2} (\log t)^{\O(1)} \Big[
               \frac{\<T\>}{n}   + m! \varrho (\log t)^m\Big] 
        \label{E:pairrecplus1}
\end{align}
\end{lemma} 
This estimate is similar to the one in Lemma \ref{L:pairrec};
 the additional $\varrho$ factor comes from the amputation.

\begin{lemma} \label{L:nested2rec} Let $m_1+m_2\geq2$.   Then:
\begin{align*}
        \sup_{0\leq s \leq t_1}
                \bE \Big\| \sum_{2\leq \k_1<\k_2\leq m} 
                \scriptUt^{\k_1,\k_2}_{[ 2], m_1,m_2}(s,t_2)\psi_0 \Big\|^2 
        \leq&  (M\lambda_0)^m m! \varrho^3 \<T\>^{2m-3} (\log t)^{m+ \O(1)}.
\end{align*}
\end{lemma}

In this estimate we gain $\varrho^2$ from the two recollisions
and an additional $\varrho$ from the amputation. We will not have
to distinguish direct and crossing terms.

\begin{lemma} \label{L:pingpong2rec} Given $n_1 \geq2$ and $m_1+m_2 \geq 2$ 
we have the bound 
\begin{align*}
        \sup_{0\leq s \leq t_1}
                \bE& \|  \scriptUt_{ *, [n_1],m_1,m_2}(s,t_2)\psi_0 \|^2            
                \leq     m! (M\lambda_0)^{n_1+m} \varrho^3 \<T\>^{2m-3} \log(t)^{ m + \O(1)}
\end{align*}
for the propagator defined in (\ref{E:pingpongdef}).
\end{lemma}  
The amputated propagator we estimate here 
corresponds to the case of having two genuine recollisions.  
Each recollision will yield a factor of $\varrho$ by utilizing a 
nontrivial pairing relation.  Recalling the discussion at the beginning 
of Section \ref{ss:error3}, we will treat the pair recollision as 
one genuine recollision.  Hence we gain a factor of $\varrho^2$ from 
the recollisions and an extra $\varrho$ from the amputation.

\bigskip

Applying  these Lemmas, recalling the relation between $\scriptU$ and
 $\scriptUt$ from (\ref{UU}), using the unitarity estimate and
 performing the sums over $n_1$ and using $M\lambda_0 <1$, we get:
\begin{align*}  
        \bE \|   \scriptU^{\mathrm{rec}}_{m_1,m_2}(t_1,t_2) \psi_0 \|^2 
                \leq& (M\lambda_0)^m \<T\>^{2m-2} (\log t)^{\O(1)} \Big[ \frac{\<T\>^3}{n^3} 
                + m! \varrho \<T\>^2 (\log t)^m \Big] .
\end{align*}Consequently,
\begin{align*}
        \bE \| \varphi_{k}^{\mathrm{error,3}}(t) \|^2 =& 
                \bE \Big\|\sum_{m_1 = 0}^{m_0-1} \sum_{m_2 = (2-m_1)_+}^{m_0-1} 
                \scriptU^{\mathrm{rec}}_{m_1,m_2}(t_1,t_2) \psi_0 \Big\|^2  \\
        \leq& C \<T\>^{4m_0} (\log t)^{\O(1)} 
                \Big[ \frac{1}{n^{3}} + (2m_0)! \varrho (\log t)^{2m_0} \Big] \, ,
\end{align*}which proves  Lemma \ref{L:error3bound}.           
\end{proof}

\begin{proof} [{\bf Proof of Lemma \ref{L:errorest}}]
Recall that $\varrho= \varrho_0 \ep$ and $t= T\ep^{-1}$.  We have from (\ref{E:deferror}),
\begin{align*}
        \Psi_{m_0}^{\mathrm{error}}(t) = 
        \sum_{k=1}^{n} e^{-i \frac{(n-k)t}{n}}\varphi_{k}^{\mathrm{error}}(t)\; .
\end{align*}Using the unitarity of $H$, and the Schwarz inequality, we have:
\begin{align*}
        \bE \| \Psi_{m_0}^{\mathrm{error}}(t) \|^2 \leq C \sum_{k=1}^n k^{3/2} \, \Big(
                \bE \| \varphi_k^{\mathrm{error\,1}}(t) \|^2 
                + \bE \| \varphi_k^{\mathrm{error\,2}}(t) \|^2
                +\bE \| \varphi_k^{\mathrm{error\,3}}(t) \|^2 \Big).
\end{align*}Lemmas \ref{L:0,0}, \ref{L:0,1} and \ref{L:error3bound} imply:
\begin{align*}
        \bE \| \Psi_{m_0}^{\mathrm{error}}(t) \|^2 
                \leq& C^{m_0} \<T\>^{4m_0} (\log t)^{\O(1)}\Big[n^{-1/2}+
                \frac{n^{3/2}}{m_0 !}  \\
        &+ (2m_0)! n^{5/2}\ep (\log t)^{2m_0} 
                + \frac{1}{n^{m_0-5/2} \, m_0! \ep}+ \frac{(2m_0)! 
(\log t)^{2m_0}}{n^{m_0-5/2}}
                \Big] \; .
\end{align*}It is easy to see that setting our parameters to
\begin{align}
        m_0 :=& \frac{ | \log \ep |}{10 \log |\log \ep|}  \label{E:m0choice} \\ 
        n:=& \ep^{-\frac{1}{100}}  \label{E:nchoice}
\end{align}guarantees that $\bE \| \Psi_{m_0}^{\mathrm{error}}(T/\ep)\|^2 \to 0$ as $\ep \to 0$.  
\end{proof}


\section{Proof of the technical error estimates}

In this section we prove the four technical Lemmas that
were needed to complete the argument in the  preceding section.
We will discuss Lemma \ref{L:pairrec} in details, then 
we explain the necessary modifications to prove the
other three Lemmas. Since several arguments are very similar,
we will not repeat them in each case.

\subsection{Proof of Lemma \ref{L:pairrec} for $n_1\ge 2$}

We start by computing the expectation value
as in Lemma \ref{L:bcd}:
\begin{align*}
        \bE \Big\|&\sum_{1\leq \k_2<\k_1\leq m} 
                \sum_{A:|A|=m}^{\mathrm{no\, rec}} 
                \scriptU^{\circ;\k_1,\k_2}_{[n_1], m_1,m_2;A}(t_1,t_2)\psi_0 \Big\|^2 \\
        =&\sum_{\stackrel{1\leq \k_2<\k_1\leq m}{1\leq \k'_2< \k'_1\leq m}} 
                \sum_{\stackrel{B: |B|=b}{0\leq b\leq m}}
                \sum_{\sigma\in\operatorname{S}(b)}     
                \sum_{(A, A')}  \\
        &\qquad \bE \int dp_0 \, 
    \scriptU^{\circ;\k_1,\k_2}_{n, m_1,m_2;A}(t_1,t_2)\psi_0(p_0)\,
                \overline{\scriptU^{\circ;\k'_1,\k'_2}_{[n_1], m_1,m_2;A'}(t_1,t_2)
    \psi_0(p_0)}\; ,
\end{align*}
where the sum on $(A,A')$ is short for summing over ordered sets $A, A'$ of size $m$ 
with non-repeating elements, such that $A   \cap  A'= B$, $B\prec A$ and 
$\sigma(B) \prec  A'$.   Using independence of the obstacles and the 
Schwarz inequality, we have:
\begin{align*}
        \bE\| &\scriptU^{\circ}_{[n_1],m_1,m_2}(t_1,t_2)\psi_0 \|^2 \\
        \leq& \sum_{\stackrel{1\leq \k_2<\k_1\leq m}{1\leq \k'_2< \k'_1\leq m}} 
                \sum_{\stackrel{B: |B|=b}{0\leq b\leq m}}
                \sum_{\sigma\in\operatorname{S}(b)}     
                \sum_{(A, A')}  \\
        &\quad \bE_{B} \Big\{ \| \bE_{A\setminus B} \, \scriptU^{\circ;\k_1,\k_2}_{[n_1], 
                m_1,m_2;A}(t_1,t_2)\psi_0 \|^2 + \|\bE_{A'\setminus B} \,
            \overline{\scriptU^{\circ;\k'_1,\k'_2}_{[n_1], m_1,m_2;A'}(t_1,t_2)
  \psi_0}\|^2 \Big\} \\
        \leq& \sum_{\stackrel{B: |B|=b}{0\leq b\leq m}} 
                \sum_{\stackrel{A: |A|=m}{B\prec A}}     
                \sum_{1\leq\k_2<\k_1<m}  C(N,m,b)  \,
                \bE_B \| \bE_{A\setminus B} \,
                \scriptU^{\circ;\k_1,\k_2}_{[n_1], m_1,m_2;A}(t_1,t_2)\psi_0 \|^2  \\
        :=& \, \mbox{(I)} + \mbox{(II)}         .
\end{align*}for
\begin{align}
        \mbox{(I)} :=& \sum_{\stackrel{B: |B|=b}{0\leq b\leq m}}                
                \sum_{\stackrel{A: |A|=m}{B\prec A}}  \sum_{\stackrel{ 1\leq \k_2<\k_1\leq m}
                {\{\a_{\k_1},\a_{\k_2} \}  \subseteq B} }        
                C(N,m,b)        \,
                \bE_B \| \bE_{A\setminus B} \,
                \scriptU^{\circ;\k_1,\k_2}_{[n_1], m_1,m_2;A}(t_1,t_2)\psi_0\|^2  \notag  \\
        \mbox{(II)} :=& \sum_{\stackrel{B: |B|=b}{0\leq b\leq m}}               
                \sum_{\stackrel{A: |A|=m}{B\prec A}}  
                \sum_{\stackrel{ 1\leq \k_2 <\k_1 \leq m}{  \{\a_{\k_1},\a_{\k_2} \} 
     \nsubseteq B} }C(N,m,b) \,
                \bE_B \| \bE_{A\setminus B} \,
                \scriptU^{\circ;\k_1,\k_2}_{[n_1], m_1,m_2;A}(t_1,t_2)\psi_0 \|^2
        \label{E:pairrec12}
\end{align}and $C(N,m,b) := \binom{N-m}{m-b} \frac{m! (m-1) (m-2)}{2} $. 

\bigskip

We will first treat term (I).  Recalling (\ref{E:defb}), define $(\g_1,\g_2)$ such that
 $\k_1 = b(\g_1)$ and $\k_2 = b(\g_2)$.  This means that $\alpha_{\kappa_1}$ falls in between
the $\gamma_1$-th and $(\gamma_1+1)$-th element of $B$ in
the sequence of centers, and similar statement holds for $\alpha_{\kappa_2}$.

Taking expectation in the proof of Lemma \ref{L:bcd}, we
  obtain the bound:
\begin{align}
        \mbox{(I)} \leq& C^m \sum_{b=2}^m \sum_{\bell_{0,b}} 
                \sum_{1\leq \gamma_2<\gamma_1\leq b} 
                b! \, \mathcal{W}_{n_1}(t_1,t_2; \gamma_1,\gamma_2, b, \bell_{0,b}) ,
\label{summ}
\end{align}
where
\begin{align} \label{E:pairrecI}
        \mathcal{W}_{n_1}&(t_1,t_2; \gamma_1,\gamma_2, b, \bell_{0,b})   \\
        :=&     \varrho^{2m-b}\Big| \int dp_0 d\fu_{0,n_1-2} d\fu'_{0,n_1-2} 
                d\fr_{0,b} d\fr'_{0,b} 
                \Delta(\fu_{0,n_1-2},\fu'_{0,n_1-2}, \fr_{0,b}, \fr'_{0,b}) \notag \\
        &\times \psi_0(r_b) \overline{\psi_0(\rr_b) }
                \mathcal{K}(t_1, t_2; p_0, \fu_{0,n_1-2}, \fr_{0,b}^{\bell_{0,b}})
                \overline{ \mathcal{K}(t_1,t_2; p_0, \fu'_{0,n_1-2},
    {\fr_{0,b}'}^{\bell_{0,b}})} \Big|\; , \notag
\end{align}
where the vector $\bell_{0,b}$ is defined in the proof of Lemma \ref{L:bcd} and 
\begin{align}
        \Delta(&\fu_{0,n_1-2},\fu'_{0,n_1-2}, \fr_{0,b}, \fr'_{0,b}) \notag \\
        &=\delta\Big[ -\Sigma_{j=0}^{n_1-2} (-1)^{n_1-j} (u_j -\uu_j) 
                -(r_0-\rr_0) +(r_{\g_2} - \rr_{\g_2}) \Big] \notag \\
        &\times \prod_{j=1}^{\g_2-1} \delta[ \rr_{j} = r_j -(r_{0} - \rr_{0}) ] 
                \prod_{j=\g_2+1}^{\g_1-1} \delta[ \rr_{j} = r_j -(r_{\g_2} - \rr_{\g_2}) ] 
                \prod_{j=\g_1}^b \delta(\rr_j = r_j) \notag \\
        &:=\Delta_1(\fu_{0,n_1-2}, \fu'_{0,n_1-2}, r_0,\rr_0, r_{\g_2},\rr_{\g_2}) 
                \times \Delta_{2,b}(\fr_{0,b},\fr'_{0,b}) \; .      
        \label{E:pairrecdeltaI} 
\end{align}
The decomposition above separates the first pairing relation, 
$\Delta_1$, which is the only relation containing the variables
 $\fu_{0,n_1-2}$ and $\fu'_{0,n_1-2}$, from the remaining $b-1$ relations, $\Delta_{2,b}$.
Note that $\mathcal{W}_{n_1}$ also depends on $m_1, m_2$, but 
these parameters are determined by the variables $b$ and $\bell_{0,b}$
so they will be omitted from the notation.

 Figure 4
shows the Feynman diagram when $(\g_1, \g_2, n_1) = (3, 1, 4)$. The dashed
lines on the picture indicate identical centers. The time division
line is not shown; it can cut the sequence of filled obstacles 
($r$-momenta lines) anywhere as in Figure 2.
\begin{figure}\label{fig:pairrec}
        \includegraphics[scale=0.48]{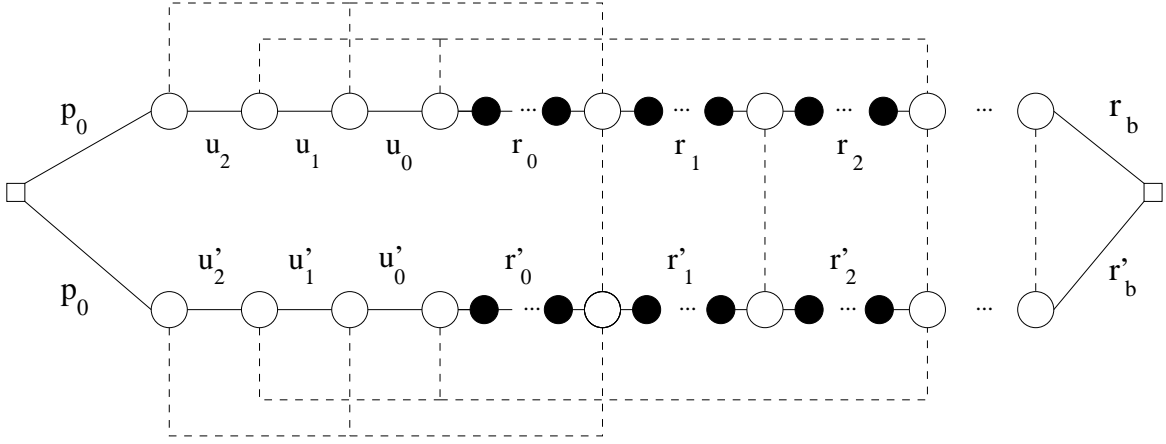}
        \caption{Feynman diagram example for Term (I)}
\end{figure}

We will now bound $\mathcal{W}_{n_1}(t_1,t_2; \g_1,\g_2, b,\bell_{0,b})$ 
by considering several cases in the following subsections.

\subsubsection{Term (I), case  $(\g_1,\g_2) = (2,1)$, $b\leq4$, $2\leq n_1 \leq 8$}\label{sss:2,1smalln1}

Recall the notation introduced in (\ref{E:upperlower}).  We apply
 (\ref{E:KF}) to expand our time-divided
 kernels.\footnote{In the special case of $m_2=0$, we use the trivial
 estimate $| \mathcal{K}(t_2; r_b) | \leq 1$ and our subsequent
 estimates will be similar but easier as we do not have divided time.} 
 Using the notation defined in (\ref{E:greenshortdef}), we write:
\begin{align}
        \mathcal{W}_{n_1}&(t_1,t_2;2,1, b,\bell_{0,b}) \notag \\
        =&\varrho^{2m-b} \int d\balpha  d\balpha'  \Big|\int dp_0
                 d\fr_{0,b} d\fr'_{0,b} \,
                 \Delta_{2,b}(\fr_{0,b}, \fr'_{0,b}) \psi_0(r_b) \overline{ \psi_0(\rr_b)} 
                 \frac{1}{[\a_1, p_0]}\frac{1}{ \overline{[\a'_1, p_0]}}        \notag \\
        &\times 
                \prod_{J=0}^{\b1} 
 \frac{B(\a_1,R_J,R_J)^{\ell_J} }{ [\a_1, R_J]^{\ell_J +1} }
                \frac{\overline{B(\a'_1,R'_J,R'_J)}^{\ell_J} }{ \overline{[\a'_1, R'_J]}^{\ell_J+1 } } 
                \prod_{J=\b2}^b  \frac{B(\a_2,R_J,R_J)^{\ell_J} }{[\a_2, R_J]^{\ell_J +1} }
                \frac{\overline{B(\a'_2,R'_J,R'_J)}^{\ell_J}}{\overline{[\a'_2, R'_J]}^{\ell_J+1 }}
                \notag \\  
        &\times   B(\a_1, \fr_{0,\b}) \overline{ B(\a'_1, \fr'_{0,\b})}          
                B(\a_2, \fr_{\b,b}) \overline{ B(\a'_2, \fr'_{\b,b})} \notag \\
        &\times\Big[\int d\fu_{0,n_1-2} d\fu'_{0,n_1-2} 
                \Delta_1(\fu_{0,n_1-2}, \fu'_{0,n_1-2}, r_0,\rr_0, r_{\g_2},\rr_{\g_2})         
                \notag  \\
        &\quad \times B(\a_1; p_0, \fu_{0,n_1-2}, r_0)
                \overline{ B(\a'_1; p_0, \fu'_{0,n_1-2} , \rr_0) }
                \prod_{j=0}^{n_1-2} \frac{1}{[\a_1,u_j]}
 \frac{1}{  \overline{[\a'_1, \uu_j]}}  \Big]
                \Big|. \label{E:(2,1)}   
\end{align}
Notice that by (\ref{E:defbeta}) and (\ref{E:defb}), $\beta$ 
is determined by $m_1$, $m_2$ and $\bell_{0,b}$.  
We also defined $\balpha := (\a_1, \a_2)$ with an
 analogous definition for $\balpha'$.  

Taking absolute values into all of the integrals, we 
use $\| \psi_0\|_{30,0}^2 \leq C$ and Lemma \ref{L:Blemma} to get the bound: 
\begin{align}
        \mathcal{W}_{n_1}&(t_1,t_2; 2,1, b,\bell_{0,b}) \notag \\
        \leq&C (M\lambda_0)^{2m-2b} \varrho^{2m-b} \sup_{r_b} \int dp_0  d\fu_{0,n_1-2}  
                d\fu'_{0,n_1-2} d\fr_{0,b-1} d\fr'_{0,b} \, \Delta_2(\fr_{0,b},\fr'_{0,b})
                  \notag \\
        &\times\int d\balpha d\balpha'  
                \frac{1}{|\a_1, p_0| \, |\a'_1,p_0| }
     \Delta_1(\fu_{0,n_1-2}, \fu'_{0,n_1-2}, r_0, \rr_0, r_{1}, \rr_{1})
                \prod_{j=0}^{n_1-2} \frac{1}{|\a_1, u_j| \, | \a'_1, u'_j|}\notag \\
        &\times 
                \prod_{J=0}^{\b1} \frac{1}{| \a_1, R_J|^{\ell_J +1} }
   \frac{1}{ | \a'_1, R'_J|^{\ell_J+1 }}
                \prod_{J=\b2}^{b} \frac{1}{| \a_2, R_J|^{\ell_J +1} }
    \frac{1}{ | \a'_2, R'_J|^{\ell_J+1 }}
                \notag \\
        &\times \frac{M\lambda_0}{\<p_0-u_0\>^4} \frac{1}{\<u_0\>^4} 
                \prod_{j=1}^{n_1-2} 
                \frac{M\lambda_0}{\<u_{j-1}- u_j\>^4} \prod_{j=1}^b 
                \frac{M\lambda_0}{\<r_{j-1} -r_j\>^4} \prod_{i=1}^5\frac{1}{\<r_{k_i}\>^4}
                \notag \\
        &\times \frac{M\lambda_0}{\<p_0-u'_0\>^4} \frac{1}{\<u'_0\>^4}
                \prod_{j=1}^{n_1-2} \frac{M\lambda_0}{\<u'_{j-1}- u'_j\>^4} \prod_{j=1}^b 
                \frac{M\lambda_0}{\<r'_{j-1} -r'_j\>^4} 
                \prod_{i=1}^5\frac{1}{\<r'_{k'_i}\>^4} \; ,
        \label{E:pairrecsmalln1}
\end{align}
where the last two lines were obtained by estimating the 
potential terms using Lemma \ref{L:Blemma} and (\ref{E:getdecay}). 
 As in the crossing estimate of Lemma \ref{L:0,0}, the indices 
$k_i, k'_i$, for $1\leq i \leq 5$ will be chosen in the following 
estimates.  Let $k_1= k'_1 = 0$, $k_2=k'_2 = \g_2$ and $k_3 = k'_3 = \b$,
the other $k$-values are arbitrary.

We begin by using (\ref{E:inftynodecay}) and (\ref{E:summingells}) 
to estimate 
\begin{align}
        &\prod_{J=0}^{\b1} \frac{1}{|\a_1, R_J|^{\ell_J } 
                |\a'_1, R'_J|^{\ell_J }} \prod_{J=\b2}^b  
\frac{1}{|\a_2, R_J|^{\ell_J } 
                |\a'_2, R'_J|^{\ell_J }} \leq  t^{2m-2b}  ,
        \label{E:inftyell}
\end{align}
where we also include the factor 
$\frac{\<\a_2\> \<\a'_2\>}{\<r_\b\>^4 \<\rr_\b\>^4}$ 
and apply (\ref{E:inftydecay}) twice if $\ell_{\b2} >0$.  

Let $\sigma \in \{0,1\}$ with $\sigma \neq \b$. 
Note that for an appropriately chosen $v$, we have
$$
\Delta_1(\fu_{0,n_1-2}, \fu'_{0,n_1-2},
 r_0,\rr_0, r_{\g_2},\rr_{\g_2}) = \delta(u_0 - u'_0 \pm \rr_{\sigma} + v) \; .
$$
We use the bound:  
\begin{align}
        \sup_{v, \a_1, \a'_1} \int& 
\frac{du_0 du'_0 dr'_\sigma}{ \<u_0\>^4 \<\uu_0\>^4 
                \<r'_{\sigma}\>^4}             
 \frac{\<\a_1\> \<\a'_1\> \<\a'(\sigma)\>
                \delta(u_0 - u'_0 \pm \rr_{\sigma} + v)  }{ |\a'_1, u'_0| \, 
                | \a'(\sigma), \rr_{\sigma}| \, |\a_1, u_0|  } \notag  \\ 
        &\leq C (\log t)^2 \sup_{v, \a_1, \a'_1}
\int \frac{d \rr_\sigma}{ \< \rr_\sigma \>^4} 
                \frac{ \<\a'(\sigma)\>}{ |\a'(\sigma),\rr_{\sigma}| } 
\frac{1}{| \rr_\sigma \pm v| + \eta}
                \notag \\
        &\leq C (\log t)^3 \; ,
        \label{E:crossingDelta1}
\end{align}
where $\a'(\sigma) = \a'_1$ for $\b\neq0$ and 
$\a'(\sigma) =\a'_2$ for $\b=0$.  In the second line we use  
Proposition \ref{P:crossing} to perform the $du_0$ and
$du_0'$ integrals.  
This allows us to estimate the remaining integrals of 
$\fu_{n_1-2}, \fu'_{n_1-2}$ using (\ref{E:intnodecay}).   
Using (\ref{E:inftynodecay}) we make the estimate 
\begin{align}
        \frac{1}{|\a'_1,p_0| } \prod_{j=2}^{\b-1} \frac{1}{|\a'_1, r'_j| }
                \prod_{j=\b+1;\neq 1}^{b-1} \frac{1}{| \a'_2, r'_j|}   \leq 
                \begin{cases}
                        t_1 t^{b-3} &\text{$1<\b <b $} \\
                        t_1 t^{b-2} &\text{otherwise ,} 
                \end{cases} 
\end{align}and then integrate over the variables $\fr'_{2,b}$ which gets rid
 of the (trivial) pairings in $\Delta_2$.  We now apply (\ref{E:intnodecay}) 
to handle the integration in $p_0$ and $r_j$, for all $j\neq \b, b$. The
 remaining estimates depend on $\b$.  

If $\b\leq1$, we apply (\ref{E:alphaint}) to get
\begin{align*}
        \sup_{r_\b, \rr_\b}\int \frac{d\a_1 d\a'_1}{\<\a_1\>\<\a'_1\>} \frac{1}{|\a_1,r_\b|} 
                \frac{1}{|\a'_1,r'_\b|} \leq C (\log t)^2 \, ,
\end{align*}allowing us to use (\ref{E:intadecay}) on the remaining 
factors of $r_\b$ and $\rr_\b$.  This leaves us to apply (\ref{E:alphaint}) 
twice more to handle the factors in $r_b$ and the integration in $\a_2$ and $\a'_2$.   

When $1<\b<b$, we apply (\ref{E:intnodecay}) to integrate the remaining factors 
in $\fr'_{0,1}$. The four $r_\b$ factors are handled by applying (\ref{E:inftydecay}) 
on $|\a'_2, r_\b|^{-1}$, (\ref{E:alphaint}) on $|\a_1,r_\b|^{-1}$ and 
$|\a'_1,r_\b |^{-1}$, followed by applying (\ref{E:intadecay}) on $|\a_2,r_\b|^{-1}$.  
The remaining factors are handled as before.

Finally, if $\b=b$, we integrate the remaining factors in $\fr'_{0,1}$ then treat
 the last four factors in $r_b$ by  a combination of (\ref{E:alphaint}), 
(\ref{E:intadecay}) and (\ref{E:inftydecay}) as we did before.  In all cases for
 $\b$, we get the estimate:
\begin{align}
        \mathcal{W}_{n_1}(t_1,t_2; 2,1, b,\bell_{0,b}) \leq&  (M\lambda_0)^{n_1+m}
                 \varrho (\varrho t_1) (\varrho t)^{2m-b-2} (\log t)^{ b+ 2 n_1 
                +\O(1)} .
\label{West}
\end{align}
This estimate is sufficient
 when $ 2\leq n_1 \leq 8$ since $n_1 = \O(1)$ in our power of log.  However for 
$n_1>8$, we will need to introduce the two-obstacle Born series term to assure
 that our power of $\log t$ does not grow too much.  This is treated in the next case.

\subsubsection{Term (I), case  $(\g_1,\g_2) = (2,1)$, $b\leq4$, $n_1 > 8$}\label{sss:2,1bign1}
We begin by expressing our first pairing relation in (\ref{E:pairrecdeltaI}) as
\begin{align*}
        \Delta_1 (\fu_{0,n_1-2}, \fu'_{0,n_1-2},\ldots) = 
                \int d\nu \,  e^{i \nu [ -\Sigma_{j=0}^{n_1-2} (-1)^{n_1-j} (u_j -\uu_j) 
                        -(r_0-\rr_0) +(r_{1} - \rr_{1}) ]} .
\end{align*}Defining:
\begin{align}
        \scriptB_{n_1,\nu }(\a_1,p_0,r_0) :=& 
                \int d \fu_{0,n_1-2}  B(\a_1, p_0,u_0)
                \frac{B(\a_1,u_0,u_1) e^{i\nu u_0}}{\a_1-u_0^2/2 + i\eta_1}
                \frac{B(\a_1, u_1, u_2) e^{-i\nu u_1}}{\a_1 - u_1^2/2 + i\eta_1}  \notag \\
        &\times \cdots\times
                \frac{B(\a_1,u_{n_1-2},r_0) \exp{\big[
   (-1)^{n_1-2}i \nu u_{n_1-2} \big]} }{\a_1 - u_{n_1-2}^2/2+i\eta_1} , 
                \label{E:defR}
\end{align}the bracketed integral in (\ref{E:(2,1)}) can be expressed as:
\begin{align*}
        \int& d\fu_{0,n_1-2} d\fu'_{0,n_1-2}\, \Delta_1(\fu_{0,n_1-2}, \ldots)   
                \prod_{j=0}^{n_1-2}\frac{1}{[\a_1,u_j]}\frac{1}{ \overline{[\a'_1, \uu_j]}}\\
        &\qquad B(\a_1; p_0, \fu_{0,n_1-2}, r_0) 
\overline{ B(\a'_1; p_0, \fu'_{0,n_1-2} , \rr_0)}
                  \\
        =& \int d\nu \, e^{-i\nu[(r_0-\rr_0) -(r_1-\rr_1)]} 
                \scriptB_{n_1,\nu}( \a_1, p_0, r_0) 
\overline{ \scriptB_{n_1, \nu} (\a'_1,p_0,\rr_0)}.
\end{align*}By Lemma \ref{L:Rest}, we have the bound:
\begin{align*}
        \sup_{r_1,\rr_1}\Big|  \int d\nu \,& e^{-i\nu[(r_0-\rr_0) -(r_1-\rr_1)]} 
                \scriptB_{n_1,\nu}( \a_1, p_0, r_0) 
\overline{ \scriptB_{n_1, \nu} (\a'_1,p_0,\rr_0)}   
                \Big|  \\
        &\leq \frac{(M\lambda_0)^{n_1}}{\< p_0 - r_0 \>^{30} \<p_0-\rr_0\>^{30}}
\end{align*}which, in light of (\ref{E:getdecay}), allows us to follow 
the technique presented the case $2\leq n_1 \leq 8$ to complete the estimate.
  Here, we no longer have the variables $\fu_{0,n_1-2}, \fu'_{0,n_1-2}$ nor 
the pairing relation $\Delta_1$ which relates them.  In place of (\ref{E:crossingDelta1})
 we simply use (\ref{E:inftydecay}) to estimate the factors containing $\rr_{\sigma}$. 
 We mention that we will need to apply (\ref{E:intadecay}) to bound $|\a_1,p_0|^{-1}$ 
and (\ref{E:inftydecay}) to bound $|\a'_1, p_0|^{-1}$.  The rest of the details differ 
trivially from the previous case and are left as an exercise.  The result is, for 
$ n_1\geq 2 $ and $b\leq 4$, the bound:
\begin{align}
        \mathcal{W}_{n_1}(t_1,t_2; 2,1, b,\bell_{0,b}) \leq& (M\lambda_0)^{n_1+m} 
                \varrho (\varrho t_1)(\varrho t_2)^{2m-b-2} (\log t)^{b  
                +\O(1)}\nonumber \\ 
        \leq&  (M\lambda_0)^{n_1+m}\frac{\varrho}{n} (\varrho t)^{2m-b-1} (\log t)^{ \O(1)} . 
\label{West1}
\end{align}The first inequality actually holds for all $b$.  However, immediate
 application in (\ref{E:pairrecI}), after summation with the $b!$ prefactor, 
creates a factor of $m!$ in our estimate which is too large.  This will be 
avoided by observing that for large $b$, most of our pairings are direct pairings
 ($ \rr_j = r_j$) which can be estimated with time division, as in Lemma \ref{L:0,0}. 
 These estimates should produce a $(m!)^{-1}$ which would be more than enough in our case. 
 However, we only capture $(b!)^{-1}$, which is adequate for our estimates.  We will
 treat this in the next case. 

\subsubsection{Term (I), case $(\g_1,\g_2) = (2,1)$, $b > 4$} \label{sss:2,1b>4}

Returning to (\ref{E:pairrecI}), we consider first the case of $0\leq \b \leq 2$. 
 Proposition \ref{P:mixedK} and (\ref{E:defF}) imply:
\begin{align*}
        \mathcal{K}(t_2&; \fr_{\b,b}^{\bell_{\b2,b}})\\
        =& \frac{1}{2\pi} \int_0^{t_2*} \[dt_{2j}\]_1^3 \,  
                K(t_{22}; \fr_{3,b-1}^{\bell_{3,b-1}}) 
                F(t_{23}, \fr_{3,b-1}^{\bell_{3,b-1}}) e^{\eta_{21} t_{21}}
                \int d\a_2 \, e^{-i\a_2 t_{21}} \\
        &\times 
                \prod_{J=\b2}^2 \frac{ B(\a_2, R_J, R_J)^{\ell_J}}{[\a_2, R_J]^{\ell_J+1}}
                \frac{B(\a_2, r_b, r_b)^{\ell_b}}{[\a_2, r_b]^{\ell_b+1}} 
                B(\a_2, \fr_{\b, 3}) B(\a_2, r_{b-1},r_b)  .
\end{align*}
We recall the convention mentioned after (\ref{E:defeta}), i.e., $\eta_{21}=\eta(t_{21})$.
Subsequently, applying (\ref{E:KF}) to 
$\mathcal{K}(t_1; p_0, \fu_{0,n_1-2}, \fr_{0,\b}^{\bell_{0,\b1}})$, we get:
\begin{align}
        \mathcal{W}_{n_1}&(t_1,t_2; 2,1, b,\bell_{0,b}) \notag \\
        \leq &\varrho^{2m-b} \int_0^{t_2*} \[d t_{2j}\]_1^3   
      \int_0^{t_2*} \[dt'_{2j}\]_1^3 
                \int d\fr_{3,b-1} K(t_{22}; \fr_{3,b-1}^{\bell_{3,b-1}}) 
                \overline{K(t'_{22}; \fr_{3,b-1}^{\bell_{3,b-1}}) } \notag \\
        &\qquad F(t_{23}, \fr_{3,b-1}^{\bell_{3,b-1}}) 
                \overline{ F(t'_{23}, \fr_{3,b-1}^{\bell_{3,b-1}}) }
                A_{n_1}(t_1, t_{21}, t'_{21}; r_3, r_{b-1}) \; ,\label{E:(2,1)bigb}
\end{align}
where
\begin{align*}
        A_{n_1}(&t_1, t_{21}, t'_{21}; r_3, r_{b-1}) \\
        :=&\int dp_0 d\fr_{0,2} d\fr'_{0,1} dr_b\, | \psi_0 (r_b)|^2   
                \int d\balpha  d\balpha' 
         e^{-i [\balpha\cdot (t_1, t_{21}) - \balpha'\cdot(t_1,t'_{21}) ]}
                e^{ \boldsymbol{\eta} \cdot (t_1,t_{21}) +
 \boldsymbol{\eta}' \cdot (t_1, t'_{21}) }    
                 \\
        &\times  \frac{1}{[\a_1, p_0]}\frac{1}{ \overline{[\a'_1, p_0]}} 
                \frac{ B(\a_2, r_b, r_b)^{\ell_b} } {[\a_2,r_b]^{\ell_b+1} } 
                \frac{\overline{B(\a'_2,r_b,r_b)}^{\ell_b}} {
                \overline{ [\a'_2, r_b]}^{\ell_b+1}}  \\
        &\times \prod_{J=0}^{\b1}\frac{B(\a_1,R_J,R_J)^{\ell_J}}{ [\a_1, R_J]^{\ell_J +1} }
                \frac{ \overline{B(\a'_1,R'_J,R'_J)}^{\ell_J} }{
                \overline{ [\a'_1, R'_J]}^{\ell_J+1 } } 
                \prod_{J=\b2}^2  \frac{B(\a_2,R_J,R_J)^{\ell_J} }{[\a_2, R_J]^{\ell_J +1} }
                \frac{\overline{B(\a'_2,R'_J,R'_J)}^{\ell_J}}{
                \overline{[\a'_2, R'_J]}^{\ell_J+1 }}   \\
        &\times 
                B(\a_1,r_{0,\b}) B(\a_2, r_{\b,3}) \overline{B(\a'_1,\rr_{0,\b})
                B(\a'_2, \rr_{\b,1},r_{2,3})} B(\a_2,r_{b-1},r_b)
 \overline{B(\a'_2, r_{b-1},r_b)} \\
        &  \times \Big[  \int d\fu_{0,n_1-2} d\fu'_{0,n_1-2} \,
                \Delta_1(\fu_{0,n_1-2}, \fu'_{0,n_1-2}, \fr_{0,1},\fr'_{0,1} ) \\
        &\qquad B(\a_1; p_0, \fu_{0,n_1-2}, r_0)
                \overline{ B(\a'_1; p_0, \fu'_{0,n_1-2} , \rr_0)}   
                \prod_{j=0}^{n_1-2} \frac{1}{[\a_1,u_j]}
\frac{1}{  \overline{[\a'_1, \uu_j]}}  \Big]
\end{align*}and for $\balpha := (\a_1, \a_2)$, $\balpha' := (\a'_1, \a'_2)$,
 $\boldsymbol{\eta} :=
(\eta_{1}, \eta_{21})$ and $\boldsymbol{\eta}' := (\eta_1, \eta'_{21})$.  
In our notation, $[\a_1, p]$ gets regularized with $\eta_1$  as before,
whereas $[\a_2,p]$ gets regularized with $\eta_{21}$.
 See (\ref{E:greenshortdef}) for a similar convention.

Using definition (\ref{E:Kdef}) and making dispersive estimates as in 
the direct term proof of Lemma \ref{L:0,0}, we have:
\begin{align*}
        \mathcal{W}_{n_1}&(t_1,t_2; 2,1, b,\bell_{0,b}) \\
        =& \varrho^{2m-b} \int_0^{t_2*} \[d t_{2j}\]_1^3   
                \int_0^{t_2*} \[dt'_{2j}\]_1^3 
                \int_0^{t_{22}*} \[ds_j\]_3^{b-1} \int_0^{t'_{22}*} \[ds'_j\]_3^{b-1} 
                \Big| \int d\fr_{3,b-1} \\
        &       \prod_{j=3}^{b-1} \frac{ (s_j s'_j)^{\ell_j} 
                e^{-i (s_j-s'_j) r_j^2/2}}{(\ell_j!)^2}  
                F(t_{23}, \fr_{3,b-1}^{\bell_{3,b-1}}) 
                \overline{ F(t'_{23}, \fr_{3,b-1}^{\bell_{3,b-1}}) }
                A_{n_1}(t_1, t_{21}, t'_{21}; r_3, r_{b-1})  \Big| \\
        \leq& \int_0^{t_2*} \[d t_{2j}\]_1^3   \int_0^{t_2*} \[dt'_{2j}\]_1^3 
                \int_0^{t_{22}*} \[ds_j\]_3^{b-1} \int_0^{t'_{22}*} \[ds'_j\]_3^{b-1} 
                \prod_{j=3}^{b-1} \frac{ (s_j s'_j)^{\ell_j} }{\< s_j - s'_j\>^{3/2} } \\
         & \times |\!|\!| F(t_{23}, \fr_{3,b-1}^{\bell_{3,b-1}}) 
                \overline{ F(t'_{23}, \fr_{3,b-1}^{\bell_{3,b-1}}) }
                A_{n_1}(t_1, t_{21}, t'_{21}, r_3, r_{b-1}) |\!|\!|_{d\fr_{3,b-1}}.
\end{align*}
We now trivially bound $(\ell_j)^{-1} \leq 1$ to get
 \footnote{More careful analysis should allow us to make use of the $(\ell_j !)^{-1}$ 
factors by arguing as in the proof of the Direct estimate of Lemma \ref{L:0,0}.  
This would yield a factor of $m!^{-1}$ instead of $b!^{-1}$ but since we do not 
need the former, we opt for the cruder estimate.} :
\begin{align*}
        \mathcal{W}_{n_1}&(t_1,t_2; 2,1, b,\bell_{0,b}) \\
         \leq& \varrho^{2m-b} \int_0^{t_2*} \[d t_{2j}\]_1^3   
                \int_0^{t_2*} \[dt'_{2j}\]_1^3  \frac{ t_2^{2\| \bell_{3,b-1}\|+ b-4}}{(b-4)! 
                \<t_{22} - t'_{22}\>^{3/2}  }   \\
        & \times|\!|\!| F(t_{23}, \fr_{3,b-1}^{\bell_{3,b-1}}) 
                \overline{ F(t'_{23}, \fr_{3,b-1}^{\bell_{3,b-1}}) }
                A_{n_1}(t_1, t_{21}, t'_{21}, r_3, r_{b-1}) |\!|\!|_{d\fr_{3,b-1}} .
\end{align*}
where $\| \bell_{3,b-1} \| := \sum_{j=3}^{b-1} \ell_j $. 
 Repeated use of Lemma \ref{L:tdivfdecay} and
(\ref{Ndef}) imply that 
\begin{align*}
        |\!|\!| F(t_{23}, &\fr_{3,b-1}^{\bell_{3,b-1}}) 
                \overline{ F(t'_{23}, \fr_{3,b-1}^{\bell_{3,b-1}}) }
                A_{n_1}(t_1, t_{21}, t'_{21}, r_3, r_{b-1}) 
 |\!|\!|_{d\fr_{3,b-1}}  \\
        \leq& \frac{ (M\lambda_0)^{\| 
 \bell_{3,b-1}\| + b-4}}{\< t_{23}\>^{3/2} \<t'_{23}\>^{3/2} }
                \sup_{ \stackrel{0\leq \xi_3, \xi_{b-1}\leq 2}{r_3,r_{b-1}}} 
                \Big|  \frac{\<r_{b-1}\>^{60}}{\<r_3 \>^{48} } 
                \grad_{r_{b-1}}^{\xi_{b-1}} \grad_{r_3}^{\xi_3} 
                A_{n_1}(t_1, t_{21}, t'_{21}, r_3, r_{b-1})\Big|  \\
        &\times \sum_{\xi_{3,b-1} \in \{0,2\}^{b-3}} 
      N^{\xi_{3,b-1}}_{d\fr_{3,b-1}} 
                \Big( \frac{\<r_3\>^{48}}{\<r_{b-1}\>^{60}} \prod_{j=4}^{b-1} 
                \frac{1}{\<r_{j-1}-r_j\>^{52}} \Big)\\
        \leq& \frac{(M\lambda_0)^{\| \bell_{3,b-1}\| 
   + b-4}}{\< t_{23}\>^{3/2} \<t'_{23}\>^{3/2}}
                \sup_{\stackrel{ 0\leq \xi_3, \xi_{b-1} 
         \leq 2  }{r_3, r_{b-1}}}
                \Big|  \frac{\<r_{b-1}\>^{60}}{\<r_3 \>^{48} } 
                \grad_{r_{b-1}}^{\xi_{b-1}} \grad_{r_3}^{\xi_3} 
                A_{n_1}(t_1, t_{21}, t'_{21}, r_3, r_{b-1})\Big| \; .
\end{align*}
Given the form of $A_{n_1}$, the derivatives on $r_3$ and $r_{b-1}$ 
pass onto only the potentials, which, by Lemma \ref{L:Blemma} 
are sufficiently smooth.  Hence, after taking derivatives and 
then absolute values into the integrals, we can perform estimates 
as in the sections \ref{sss:2,1smalln1} and \ref{sss:2,1bign1} to show that:
\begin{align*}
        \sup_{\xi_3, \xi_{b-1} \in \{0,1,2\}}& \Big|  \frac{\<r_{b-1}\>^{60}}{\<r_3 \>^{48} } 
                \grad_{r_{b-1}}^{\xi_{b-1}} \grad_{r_3}^{\xi_3} 
                A_{n_1}(t_1, t_{21}, t'_{21}, r_3, r_{b-1})\Big|  \\
        \leq& (M\lambda_0)^{n_1+3 + \|\bell_{0,1}\| + \ell_b}
                 t_1 t_2^{2\|\bell_{0,1}\|+2 \ell_b +1} (\log t)^{\O(1)} .
\end{align*}Combining these estimates yields:
\begin{align*}  
        \mathcal{W}_{n_1}&(t_1,t_2; 2,1, b,\bell_{0,b}) \\
        \leq&\varrho^{2m-b} \int_0^{t_2*} \[d t_{2j}\]_1^3   
                \int_0^{t_2*} \[dt'_{2j}\]_1^3 
        \frac{(M\lambda_0)^{m+n_1} t_1 t_2^{2m-b-3}}
                {(b-4)! \<t_{22} -t'_{22} \>^{3/2} \<t_{23}\>^{3/2}
 \<t'_{23}\>^{3/2} } \\
        \leq&  \frac{(M\lambda_0)^{m+n_1} \varrho (\varrho t_1) 
(\varrho t_2)^{2m-b-2} (\log t)^{\O(1)} }
                {(b-4)!}  \, .
\end{align*}The case of $2< \b < b-1$ is handled in a similar way.  
We start by applying Proposition \ref{P:mixedK} and (\ref{E:defF})
 on both component kernels in the time-divided kernel.  We then bound:
\begin{align}
        \mathcal{W}_{n_1}&(t_1,t_2; 2,1, b,\bell_{0,b})\notag  \\
        \leq& \varrho^{2m-b}\int_{0}^{t_1*} \[dt_{1j}\]_1^3 
\int_0^{t_1*}\[ dt'_{1j}\]_1^3 \int_{0}^{t_2*}\[dt_{2j}\]_1^3
                \int_0^{t_2*} \[dt'_{2j}\]_1^3 \int d\fr_{2,\b-1} 
d\fr_{\b+1,b-1} 
                \notag  \\
         &\times K(t_{12}; \fr_{2,\b-1}^{\bell_{2,\b-1}}) 
                \overline{ K(t'_{12}; \fr_{2,\b-1}^{\bell_{2,\b-1}})}
                K(t_{22}; \fr_{\b+1, b-1}^{\bell_{\b+1, b-1}}) 
                \overline{ K(t'_{22}; \fr_{\b+1,b-1}^{\bell_{\b+1,b-1}})} 
\notag \\
        & \times F(t_{13};\fr_{2, \b-1}^{\bell_{2, \b-1}}) 
                \overline{F(t'_{13};\fr_{2, \b-1}^{\bell_{2, \b-1}})}
                F(t_{23};\fr_{\b+1, b-1}^{\bell_{\b+1, b-1}})
                \overline{F(t'_{23};\fr_{\b+1, b-1}^{\bell_{\b+1, b-1}})} \notag \\
        &\times A_{n_1}(t_{11},t_{21}, t'_{11}, t'_{21}; r_2,r_{\b-1},r_{\b+1}, r_{b-1})\; ,
                \label{E:(2,1)bigb2}
\end{align}
where        
\begin{align*}
        A_{n_1}(&t_{11},t_{21}, t'_{11}, t'_{21}; r_2,r_{\b-1},r_{\b+1}, r_{b-1}) \\
        :=& \int dp_0 d\fr_{0,1} d\fr'_{0,1} dr_{\b} dr_b \,
   | \psi_0(r_b)|^2 \int d\balpha d\balpha'
                e^{-i [\balpha\cdot (t_{11}, t_{21}) - \balpha'\cdot(t'_{11},t'_{21}) ]}
                 \\
        &\times e^{ \boldsymbol{\eta} \cdot (t_{11},t_{21}) 
                + \boldsymbol{\eta}' \cdot (t'_{11}, t'_{21}) } \frac{1}{[\a_1, p_0]}\frac{1}{ \overline{[\a'_1, p_0]}} 
                \frac{ B(\a_1, r_\b, r_\b)^{\ell_{\b1}} }{[\a_1, r_\b]^{\ell_{\b1}+1}}
                \frac{ \overline{ B(\a'_1, r_\b, r_\b)}^{\ell_{\b1} }}
                {  \overline{[\a'_1,r_\b]}^{\ell_{\b1}+1}} \\
        &\times 
                \prod_{j=0}^{1}  
\frac{ B(\a_1,r_j, r_j)^{\ell_j}}{[\a_1, r_j]^{\ell_j +1} }  
                \frac{ \overline{ B(\a'_1,r_j, r_j)}^{\ell_j}}
         { \overline{[\a'_1, r_j]}^{\ell_j+1}}
                \prod_{J= \b2, b} \frac{ B(\a_2, R_J, R_J)^{\ell_J}}{ [\a_2, R_J]^{\ell_J+1} } 
                \frac{\overline{ B(\a'_2, R_J, R_J)}^{\ell_J}}
                {  \overline{[\a'_2,R_J]}^{\ell_J+1}} \\
        &\times B(\a_1,\fr_{0,2}) B(\a_1, r_{\b-1},r_\b) 
                \overline{ B(\a'_1,\fr'_{0,1},r_2) B(\a'_1, r_{\b-1},r_\b)} \\
        &\times B(\a_2, r_\b, r_{\b+1}) B(\a_2,r_{b-1},r_b)
                 \overline{ B(\a'_2, r_\b, r_{\b+1} )   
                B(\a'_2,r_{b-1},r_b)} \\
        &\times \Big[    \int d\fu_{0,n_1-2} d\fu'_{0,n_1-2} \,
                \Delta_1(\fu_{0,n_1-2}, \fu'_{0,n_1-2}, \fr_{0,1},\fr'_{0,1} )  \\
        &\qquad \times B(\a_1; p_0, \fu_{0,n_1-2}, r_0)
                \overline{ B(\a'_1; p_0, \fu'_{0,n_1-2} , \rr_0)}   
                \prod_{j=0}^{n_1-2} \frac{1}{[\a_1,u_j]}
 \frac{1}{  \overline{[\a'_1, \uu_j]}}  \Big]
\end{align*}
with $\boldsymbol{\eta} := (\eta_{11}, \eta_{21})$ and $\balpha := (\a_1, \a_2)$.  
We now proceed as in the small $\beta$ case by estimating (\ref{E:(2,1)bigb2})
 using dispersive estimates.  We get:
\begin{align*}
        \mathcal{W}_{n_1}&(t_1,t_2; 2,1, b,\bell_{0,b})\\
        \leq&\varrho^{2m-b} \int_{0}^{t_1*} \[dt_{1j}\]_1^3 
                \int_0^{t_1*}\[ dt'_{1j}\]_1^3 \int_{0}^{t_2*}\[dt_{2j}\]_1^3
                \int_0^{t_2*} \[dt'_{2j}\]_1^3 \\
        &\times \frac{ C^b t^{b-5 +2 ( \| \bell_{2,\b-1} \| + \| \bell_{\b+1,b-1}\|)} }
                {(b-5)! \<t_{12}-t'_{12}\>^{3/2} \< t_{22} - t'_{22} \>^{3/2}} \\
        &\times |\!|\!| F(t_{13};\fr_{2, \b-1}^{\bell_{2, \b-1}}) 
                \overline{F(t'_{13};\fr_{2, \b-1}^{\bell_{2, \b-1}})}
                F(t_{23};\fr_{\b+1, b-1}^{\bell_{\b+1, b-1}})
                \overline{F(t'_{23};\fr_{\b+1, b-1}^{\bell_{\b+1, b-1}})} \notag \\
        &\qquad \times A_{n_1}(t_{11},t_{21}, t'_{11}, t'_{21}; r_2,r_{\b-1},
                r_{\b+1}, r_{b-1})|\!|\!|_{d\fr_{2,\b-1}, d\fr_{\b+1,b-1}} \; .
\end{align*}
Repeated use of Lemma \ref{L:tdivfdecay} gives:
\begin{align*}
        |\!|\!| F(&t_{13};\fr_{2, \b-1}^{\bell_{2, \b-1}}) 
                \overline{F(t'_{13};\fr_{2, \b-1}^{\bell_{2, \b-1}})}
                F(t_{23};\fr_{\b+1, b-1}^{\bell_{\b+1, b-1}})
                \overline{F(t'_{23};\fr_{\b+1, b-1}^{\bell_{\b+1, b-1}})}  \\
        &\quad \times A_{n_1}( r_2,r_{\b-1},r_{\b+1}, r_{b-1})
    |\!|\!|_{d\fr_{2,\b-1}, d\fr_{\b+1,b-1}} 
        \leq \frac{ (M\lambda_0)^{b-5 +\| \bell_{2,\b-1}\| + \| \bell_{\b+1,b-1}\|}}{
                \<t_{13}\>^{3/2} \<t'_{13}\>^{3/2}\<t_{23}\>^{3/2}\<t'_{23}\>^{3/2}} \\
        &\times \sup \Big| 
                \frac{\<r_{\b-1}\>^{40}\<r_{b-1}\>^{60}}{\<r_2\>^{36} \<r_{\b+1}\>^{48}} 
                \grad_{r_2}^{\xi_2} \grad_{r_{\b-1}}^{\xi_{\b-1}} 
                \grad_{r_{\b+1}}^{\xi_{\b+1}} \grad_{r_{b-1}}^{\xi_{b-1}} 
                A_{n_1}( r_2,r_{\b-1},r_{\b+1}, r_{b-1}) \Big| \; ,
\end{align*}
where the supremum is over $0\leq \xi_2,\xi_{\b-1},\xi_{\b+1},\xi_{b-1}\leq 2 $ 
and $r_2,r_{\b-1},r_{\b+1}, r_{b-1} $.  Again, we can check that the derivatives
 on $A_{n_1}$ only affect potential terms and we can estimate the last factor 
following the techniques shown in  the sections \ref{sss:2,1smalln1} and 
\ref{sss:2,1bign1} to yield:
\begin{align*}
        \mathcal{W}_{n_1}(t_1,t_2; 2,1, b,\bell_{0,b}) 
        \leq& \frac{ (M\lambda_0)^{n_1+m} \varrho (\varrho t_1) (\varrho t_2)^{2m-b-2} 
                (\log t)^{\O(1)}}{(b-5)!}   \; .
\end{align*}
Finally, when $b-1 \leq \b \leq b$, we apply (\ref{E:KF}) to $\mathcal{K}(t_2)$, 
while using Proposition \ref{P:mixedK} to write:
\begin{align*}
        \mathcal{K}(t_1&; p_0,\fu_{0,n_1-2}, \fr_{0,\b}^{\bell_{0,\b}})  \\
        =& \frac{1}{2\pi} \int_0^{t_1*} \[dt_{1j}\]_1^3 
                K(t_{12}; \fr_{2,\b-1}^{\bell_{2,\b-1}}) F(t_{13}; 
   \fr_{2,\b-1}^{\bell_{2,\b-1}})
                \int d\a_1 \,e^{-i\a_1 t_{11}}  e^{\eta_{11} t_{11}}   \\
        &\times \frac{ B(\a_1, p_0,\fu_{0,n_1-2},r_0)}{[\a_1,p_0]} 
                \prod_{j=0}^{n_1-2} \frac{1}{[\a_1,u_j]} \prod_{J=0,1,\b1} 
                \frac{B(\a_1, R_J,R_J)^{\ell_J}}{[\a_1,R_J]^{\ell_J+1} }        
                B(\a_1, \fr_{0,2}, \fr_{\b-1,\b}) \; ,
\end{align*}The rest of the estimate is handled as in 
the previous cases for $\beta$.  

We conclude that in all cases in this subsection
\begin{align}
        \mathcal{W}_{n_1}(t_1,t_2; 2,1, b,\bell_{0,b}) \leq&
                \frac{(M\lambda_0)^{n_1+m} \varrho (\varrho t_1)
  (\varrho t_2)^{2m-b-2} (\log t)^{\O(1)}}{(b-4)!} \; .
\label{West2}
\end{align}

\subsubsection{Term (I), case
  $(\g_1,\g_2)\neq (2,1)$, $2\leq n_1\leq 8$}\label{sss:neq2,1smalln1}
Returning to the pairing functions, (\ref{E:pairrecdeltaI}), observe that our 
assumption on $(\g_1,\g_2)$ imply that there is at least one nontrivial relation 
in $\Delta_{2,b}(\fr_{0,b},\fr'_{0,b})$, a trivial relation being of the form
 $\delta(\rr_j = r_j)$.   This will allow us to avoid the use of the costly
 $L^{\infty}$ estimate (\ref{E:inftynodecay}) and obtain a bound which is 
smaller by a factor of $\varrho$ compared to the $(\g_1,\g_2) =(2,1)$ case. 
 The main mechanism is utilizing estimates like (\ref{E:crossingDelta1}).  

We apply (\ref{E:KF}) to obtain (\ref{E:(2,1)}) and (\ref{E:pairrecsmalln1}).
  Suppose first that $\g_2< \b<\g_1$, which implies that 
$\Delta_{2,b}(\fr_{0,b},\fr'_{0,b})$ contains the nontrivial relation 
$\delta(\rr_{\b} = r_\b - r_{\g_2}+\rr_{\g_2})$.  Once again we choose
 $k_1= k'_1 =0 $, $k_2 = k'_2 = \g_2$ and $k_3=k'_3 = \b$.  

We begin as in the case of $(\g_1,\g_2) = (2,1), \, b\leq 4$ by performing 
estimates (\ref{E:inftyell}) and (\ref{E:crossingDelta1}),  the latter performed
 with $\sigma:= 0$ and $\a'(\sigma):= \a'_1$.  We then bound the integration in
 $\fu_{n_1-2}$ and $\fu'_{n_1-2}$ as before and    apply (\ref{E:inftynodecay}):
\begin{align}
        \frac{1}{| \a'_1, p_0| } \prod_{j=1;\neq \g_2}^{\b-1} \frac{1}{|\a'_1,r_j|}
 \prod_{j=\b+1}^{b-1} 
                \frac{1}{|\a'_2, r_j|} \leq t_1 t^{b-3}  \, .
\end{align}This allows us to integrate over $\rr_j$ for $j\neq 0, \g_2, \b$ which 
removes the corresponding delta functions.  We then estimate the 
integrals of $r_j$, $j \neq \g_2,\b, b$ by applying (\ref{E:intnodecay}) and use the bound:
\begin{align}
        \sup_{\a'_2, r'_{\b}} 
        \int &\frac{d\a'_1}{\<\a'_1\>} \int 
\frac{dr_{\g_2} d\rr_{\g_2} d\rr_{\b} }{\<r_{\g_2}\>^4 
                \< \rr_{\g_2} \>^4 \<\rr_\b\>^4} 
                \frac{\delta( \rr_\b = r_\b - r_{\g_2} + \rr_{\g_2})
                     \<\a'_2\> }{ |\a_1,r_{\g_2}| \,
                | \a'_1 , \rr_{\g_2}| \, | \a'_1, \rr_{\b} | \, 
               | \a'_2, \rr_\b | }  \notag \\
        \leq& C(\log t)^2 \Big(\int \frac{d\rr_\b}{\< \rr_\b\>^4} 
\frac{\<\a'_2\> }{ | r_\b - \rr_\b | \,  
                 | \a'_2, \rr_\b | }   \Big) \Big( \sup_{\rr_\b} \int \frac{d\a'_1}{\<\a'_1\>} 
                 \frac{1}{| \a'_1, \rr_\b|}\Big)   \notag \\
        \leq& C (\log t)^4 \, .
        \label{E:alphacrossing}
\end{align}
After applying (\ref{E:alphaint}) to integrate $\a_1$, we use either (\ref{E:intadecay})
  or (\ref{E:intnodecay}) on $|\a_2, r_\b|^{-1}$, depending on whether the factor 
$\<\a_2\>/\<r_{\b}\>^4$ was used in (\ref{E:inftyell}).  The remaining terms are 
handled by (\ref{E:alphaint}).  

Now suppose that $\b=\g_2$.  By assumption, either $\g_2 \geq 2$ or $\g_1 - \g_2 \geq2$,
 which implies that $\Delta_{2,b}$ contains either $\delta(\rr_1 = r_1 -r_0 + \rr_0)$ 
or $\delta(\rr_{\g_2+1} = r_{\g_2 +1} - r_{\g_2} + \rr_{\g_2})$, respectively.  
The first case is handled by applying (\ref{E:crossingDelta1}) with $\sigma=0$ while 
applying the same type of estimate in integrating $r_0, r_1$ and $\rr_1$.  We rest of
 the estimates are trivial.  When $\g_1 - \g_2 \geq 2$, we will need to apply 
(\ref{E:crossingDelta1}) with $\sigma =0$ and then apply an estimate of the form 
(\ref{E:alphacrossing}) to handle $r'_{\g_2}, r_{\g_2+1}, \rr_{\g_2+1}$ and the 
integral in $\a_1'$.  The rest of the estimates are trivial.

The case where $ 0 \leq \b < \g_2$ follows analogously
 as the reader can check.  Finally, the case where $\b\geq \g_1$ requires 
two estimates of the form (\ref{E:crossingDelta1}) - one to handle $\Delta_1$ 
and the other to handle the nontrivial relation in $\Delta_2$.  We omit the 
details, leaving them as an exercise.  The result is:
\begin{align}
        \mathcal{W}_{n_1}(t_1,t_2; \g_1,\g_2, b,\bell_{0,b}) \leq m! (M\lambda_0)^{n_1+m} 
        \varrho^2 \< T\>^{2m-b-2} (\log t)^{m + \O(1)}\, .
\label{West3}
\end{align}

\subsubsection{Term (I), case  $(\g_1,\g_2)\neq (2,1)$, $n_1> 8$}

We form the Born series term as in the corresponding case as in section \ref{sss:2,1bign1}. 
 This eliminates the paring relation 
$\Delta_1(\fu_{0,n_1-2}, \fu'_{0,n_1-2}, r_0,\rr_0,r_{\g_2},\rr_{\g_2})$ 
and makes $r_0 , \,\rr_0, \, r_{\g_2}$ and $\rr_{\g_2}$
 free variables allowing their factors to be estimated by (\ref{E:intadecay}) 
or to participate in estimates of the form (\ref{E:alphacrossing}).  Either way,
 we avoid the costly estimate (\ref{E:inftydecay}).  Again, the condition 
$(\g_1, \g_2) \neq (2,1)$ implies that there is at least one nontrivial relation 
in $\Delta_2$.  This is exploited as in section \ref{sss:neq2,1smalln1} using
 estimates such as (\ref{E:alphacrossing}).  The details are omitted 
and the result is:
\begin{align}
        \mathcal{W}_{n_1}(t_1,t_2; \g_1,\g_2, b,\bell_{0,b}) \leq m! (M\lambda_0)^{n_1+m} 
        \varrho^2 \< T\>^{2m-b-2} (\log t)^{m + \O(1)}\, .
\label{West4}
\end{align}

\subsubsection{Summary of the estimates of the term (I)}

Summarizing (\ref{West}), (\ref{West1}) (\ref{West2}), (\ref{West3}) and (\ref{West4}),
we have shown that for all cases of $(\g_1, \g_2)$, $b$, and $\bell_{0,b}$, that
\begin{align*}
        \mathcal{W}_{n_1}(t_1,t_2; \g_1,\g_2, b,\bell_{0,b}) \leq& (M\lambda_0)^{n_1+m} 
        \Big[ \frac{\varrho}{n} \<T\>^{2m-b-1} \big( \chi_{b\leq4} + 
\frac{\chi_{b>4}}{(b-5)!}\big)\\
        & +
        m!      \varrho^2 \< T\>^{2m-b-2} (\log t)^{m + \O(1)}\Big],
\end{align*}
which, after summation in (\ref{summ}), yields  Lemma \ref{L:pairrec}
for the term (I) in case $n_1\ge 2$.

\bigskip

 Now we turn to the term (II).

\subsubsection{Term (II), $n_1\ge 2$}
To get explicit expressions, we will first treat the case where $\a_{\k_1} \in B$ 
but $\a_{\k_2}\notin B$.  Suppose $\g_1$ is defined so that $b(\g_1) = \k_1$ and 
$\g_2$ is defined so that $b(\g_2) < \k_2 < b(\g_2+1)$.  As before, taking the 
expectation in (\ref{E:pairrec12}) involves the random phase.  Explicitly:
\begin{align} 
        \bE_{B}&\Big[ \bE_{A\setminus B} 
                \chi(\mathcal{A}(n_1)\oplus A; p_0,\fu_{0,n_1-2}, \fr_{0,m})  
                \overline{      \bE_{A\setminus B} \chi(\mathcal{A}(n_1)
                \oplus A; p_0,\fu'_{0,n_1-2}, \fr'_{0,m})}  \Big] \notag \\
        =& |\Lambda|^{-(2m-b)} 
                \delta\Big[ \Sigma_{k=1}^{[n_1/2]} (u_{n_1-2k-1}-u_{n_1-2k}) +
 (r_{\k_2-1} -    
                r_{\k_2})\Big]   \notag \\
        &\times \delta\Big[ \Sigma_{k=1}^{[n_1/2]} (u'_{n_1-2k-1}-u'_{n_1-2k}) +
 (\rr_{\k_2-1} -        
                \rr_{\k_2})\Big]   \notag \\
        &\times \delta \Big[ \Sigma_{j=0}^{n_1-2} (-1)^{n_1-j} (u_j -\uu_j) 
                -(r_0-\rr_0) +(r_{b(\g_1)} - \rr_{b(\g_1})) \Big] \notag  \\
        &\times \prod_{j=1;\neq \g_1}^{b}
                \delta[ (r_{b(j)-1}- r_{b(j)}) - (\rr_{b(j)-1} -\rr_{b(j)})] \notag  \\
        &\times \prod_{j=0;\neq\g_2}^b\Big( \prod_{k=b(j)+1}^{b(j+1)-1}
                \delta(r_{j-1}-r_j)\delta(\rr_{j-1}-\rr_j) \Big) \notag  \\
        &\times \prod_{j=b(\g_2)+1}^{\k_2-1}
                \delta(r_{j}- r_{j-1}) 
                \prod_{j=\k_2+1}^{b(\g_2+1)-1} \delta(r_{j}- r_{j-1})   \;,
\label{E:pairrecdeltaII}
\end{align}
where $[ \cdot ]$ is the least integer function and $u_{-1} = \uu_{-1} := p_0$.

We now integrate $\fr_{m} \setminus \{ [r_{b(j)}]_0^b , r_{\k_2} \}$ 
and their prime counterparts.  Of the variables left, relabel $r_{b(j)} \to r_j$ 
and $\rr_{b(j)} \to \rr_j$, $1\leq j \leq b$, and $r_{\k_2} \to \hat{r}_{\g_2}$, 
$\rr_{\k_2} \to \hat\rr_{\g_2}$.  One can check that we can rewrite our pairing relations as:
\begin{align*}
        &\delta\Big[ \Sigma_{k=1}^{[n_1/2]} (u_{n_1-2k-1}-u_{n_1-2k}) + (r_{\g_2} -     
                \hat{r}_{\g_2})\Big]  \\
        &\times \delta\Big[ \Sigma_{k=1}^{[n_1/2]} (\uu_{n_1-2k-1}-\uu_{n_1-2k}) 
        + (r_{\g_2} -r_0+\rr_0-         
                \hat{\rr}_{\g_2})\Big]  \\      
        &\times \prod_{j=1}^{\g_2} \delta[\rr_j = r_j -(r_0-\rr_0)]
                \prod_{j=\g_2+1}^{\g_1-1} 
\delta[\rr_{j} = r_j -(\hat{r}_{\g_2} -\hat\rr_{\g_2})]
                \prod_{j=\g_1}^b \delta( \rr_j = r_j).
\end{align*}We now proceed as for the term (I).  For $2\leq n_1 \leq 8$, we use
 (\ref{E:KF}) and exploit the pairing relations.  Unlike the simplified relations
 (\ref{E:pairrecdeltaI}), we have two separate relations:  one involving $\fu_{0,n_1-2}$ 
and the other one involving $\fu'_{0,n_1-2}$.  Each of these will be involved in an estimate
 of the form (\ref{E:crossingDelta1}) or (\ref{E:alphacrossing}).  The net effect is that 
we gain a factor of $\varrho^2$.  Similar argument is valid in the case
 when $\a_{\k_1} \notin B$ and $\a_{\k_2} \in B$.  

When $\a_{\k_1}, \a_{\k_2} \notin B$ we can simplify the pairing relations so that we 
have one separate pairing relation for each group $\fu_{0,n_1-2}$ and $\fu'_{0,n_1-2}$, 
and no other relations involving these variables, making it similar to the other cases in
 (II).  We then exploit these relations performing estimates as in (\ref{E:crossingDelta1}) 
or (\ref{E:alphacrossing}) twice, while avoiding the costly $L^{\infty}$ estimate
 (\ref{E:inftynodecay}).  This gains a factor of $\varrho^2$.  The details of these 
calculations are left to the reader but the conclusion is  that for $2\leq n_1 \leq 8$, 
we have:
\begin{align*}
        \mbox{(II)} \leq m! (M\lambda_0)^{n_1+m} \<T\>^{2m-2}\varrho^2 (\log t)^{m+\O(1)}.
\end{align*}When $n_1>8$, we form the two-obstacle Born series term as in (I). 
 Since we have one separate relation for $\fu_{0,n_1-2}$ and one for
$\fu'_{0,n_1-2}$, the formula (\ref{E:defR}) will introduce:
\begin{align*}
        \scriptB_{\nu,n_1}(\a_1,p_0,r_0) \overline{ \scriptB_{\nu',n_1}(\a_1', p_0,\rr_0)}\;,
\end{align*}
where we also have an additional integration over $\nu'$.  However, by
 Lemma \ref{L:Rest}, we have sufficient decay in $\nu'$ to handle the integration. 
 We can deduce the same bound as in the case of the small $n_1$ and leave this to 
the reader to check.

\subsection{Proof of Lemma \ref{L:pairrec} for $n_1=1$}

The case $n_1=1$ requires a separate treatment, but the methods
are analogous to the ones in the previous section for $n_1\ge 2$.

We  start with the following estimate:
\begin{align*}
        \bE \| \sum_{\k=2}^m \sum_{A:|A|=m} 
  \scriptU^{\circ;\k}_{[1], m_1,m_2,A}(t_1,t_2)\psi_0 \|^2 
                \leq \, \mbox{(I)} + \mbox{(II)}
\end{align*}for
\begin{align*}
        \mbox{(I)} :=& \sum_{\stackrel{B: |B|=b}{0\leq b\leq m}}                
                \sum_{\stackrel{A: |A|=m}{B\prec A}}  \sum_{ \a_{\k} \in B}      
                C(N,m,b)        \,
                \bE_B \| \bE_{A\setminus B} \,
                \scriptU^{\circ; \k}_{[1] , m_1,m_2,A}(t_1,t_2)\psi_0\|^2 \\
        \mbox{(II)} :=& \sum_{\stackrel{B: |B|=b}{0\leq b\leq m}}               
                \sum_{\stackrel{A: |A|=m}{B\prec A}}  
                \sum_{\a_{\k} \notin B} C(N,m,b) \,
                \bE_B \| \bE_{A\setminus B} \,
                \scriptU^{\circ; \k}_{[1], m_1,m_2,A}(t_1,t_2)\psi_0 \|^2\;,
\end{align*}
where $C(N,m,b) = \binom{N-m}{m-b} m! (m-1)$.  

We first treat term (I).  Define $\g$ so that $b(\g) = \k$.  
Computing the expectations, we have:
\begin{align*}
        \mbox{(I)} \leq& \sum_{b=2}^m \sum_{\bell} \sum_{\k=2}^b b! \, 
                \mathcal{W}_1(t_1, t_2; \g , b , \bell_{0,b}) ,
\end{align*}for
\begin{align*}
        \mathcal{W}_1(t_1, t_2; \g , b , \bell_{0,b}) :=&
         \varrho^{2m-b}
                \Big| \int dp_0 d\fr_{0,b} d\fr'_{0,b} \, \psi(r_b)
  \overline{\psi_0(\rr_b)}
                \Delta(\fr_{0,b}, \fr'_{0,b})  \\
        &\times \mathcal{K}(t_1, t_2; p_0, \fr_{0,b}^{\bell_{0,b}})
                \overline{\mathcal{K}(t_1, t_2; p_0, {\fr'}_{0,b}^{\bell_{0,b}})} \Big| .
\end{align*}The pairing relations are given by:
\begin{align}
        \Delta(\fr_{0,b}, \fr'_{0,b}) =&
                \prod_{j=1}^{\g-1} \delta[ \rr_j = r_j - (r_0-\rr_0)] 
  \prod_{j=\g}^b \delta(\rr_j=r_j)\; .
                \label{E:relationsn1=1}
\end{align}
We will treat the $\gamma =2$, $\gamma>2$ and $\gamma=1$ cases separately.

\subsubsection{Term (I), case  $\g=2$, $n_1=1$}
For $\g = 2$, we have the nontrivial relation $\rr_1= r_1 -(r_0-\rr_0)$.  
When $b$ is small, this allows us to perform estimates such as 
(\ref{E:crossingDelta1}) or (\ref{E:alphacrossing}).  This gains a power of 
$\varrho$.  When $b > 4$, we will follow the beginning of Section \ref{sss:2,1b>4}  
by applying Proposition \ref{P:mixedK} and (\ref{E:KF}) to split our kernels.  
We then perform time dependent estimates which produce a factor of $(b!)^{-1}$ as before. 
 The details can be gathered from previous estimates.  The result is:
\begin{align*}
        \mbox{(I: $\gamma=2$)} \leq& C (M\lambda_0)^{m+1} \< T \>^m 
                \varrho \frac{ \<T\>^{2m-1}}{n}  .
\end{align*}

\subsubsection{Term (I), case  $\g>2$, $n_1=1$}
Returning to (\ref{E:relationsn1=1}), the condition $\g>2$ gives us at least two 
nontrivial relations.  When $\b \geq \g$, we exploit the relations $\rr_1 = r_1 -r_0 +\rr_0$ 
and $\rr_2 = r_2 - r_0 + \rr_0$.  Using bounds such as (\ref{E:crossingDelta1}) 
and (\ref{E:alphacrossing}) we avoid the use of $L^{\infty}$ estimates 
(which produce powers of time) for $\rr_1$ and $\rr_2$ as well as $\rr_0$, 
which proves the lemma in this case.  For $\b= 0$ we need the following inequality:
\begin{align}
        \int& \frac{d\a_1 d\a'_1}{\<\a_1\>\<\a'_1\>} \int
 \frac{dr_0 d\rr_0}{\<r_0\>^8 \<\rr_0\>^8}
                \frac{\<\a_2\>\<\a'_2\>}{ |\a_1, r_0| \, |\a_2, r_0|\, 
  |\a'_1, \rr_0| \, |\a'_2, \rr_0|} \notag \\
        &\times  \int \frac{dr_1 d\rr_1}{\<r_1\>^4 \<\rr_1\>^4} 
                \frac{\delta(\rr_1=r_1-r_0+\rr_0) }{|\a_2, r_1| \, |\a'_2, \rr_1 |}
                \int  \frac{dr_2 d\rr_2}{\<r_2\>^4 \<\rr_2\>^4} 
\frac{\delta(\rr_2=r_2-r_0+\rr_0)}
                {|\a_2, r_2| \, |\a'_2, \rr_2 |} \notag \\
        \leq& C (\log t)^6 \int \frac{dr_0 d\rr_0}{\<r_0\>^4 \<\rr_0\>^4 \< r_0 - \rr_0\>^4} 
                \frac{\<\a_2\> \<\a'_2\>}{ |\a_2,r_0| \, |\a'_2, \rr_0|}
 \frac{1}{|r_0 -\rr_0|^2+\eta^2}
                             \notag \\
        \leq& C (\log t)^9 \, ,
        \label{E:doublealphacrossing}
\end{align}where the first inequality uses (\ref{E:alphaint}) twice as well as an 
estimate similar those in the proof of Proposition \ref{P:crossing}.  
The remaining cases of $1\leq \b < \g_1$ will follow in the same way (after a 
change of variables).  The rest of the estimate follows from previous ones.  
 It follows that
\begin{align*}
        \mbox{(I:  $\g > 2$)} \leq m! (M\lambda_0)^{m+1}  \varrho^2\<T\>^{2m-2} 
 (\log t)^{m+ \O(1)} .
\end{align*}
\subsubsection{Term (I), case  $\g=1$, $n_1=1$}\label{sec:g1n1}
The case of $\g=1$ requires a separate argument.  The pairing relations force $r_j=\rr_j$ 
for $1\leq j \leq b$.  Note that the  constraint $\a_{b(\gamma)} \neq \a_1$ forces
 $\ell_0>0$ in this case.  Separating the internal and external kernels as in the 
direct estimate in Lemma \ref{L:0,0}, we need to consider:
\begin{align*}
        &\int_0^{t_1*} \[dt_{1j}\]_1^2  \int_0^{t_2*} \[dt_{2j}\]_1^2
                \int_0^{t_1*} \[dt'_{1j}\]_1^2  \int_0^{t_2*} \[dt'_{2j}\]_1^2 
                \int dp_0 dr_0 dr'_0 d\fr_b | \psi_0(r_b)|^2     \\
        &\times
                K_{m_1,m_2}(t_{11},t_{21};p_0, r_0^{\ell_0}, \fr_b^{\bell_b}) 
                \overline{ K_{m_1,m_2}(t_{11}',t_{21}';p_0, \rr_0^{\ell_0},\fr_b^{\bell_b})}
                \\ 
        &\times F(t_{12}, t_{22}; p_0,r_0^{\ell_0}, \fr_b^{\bell_b}) 
                \overline{ F(t'_{12}, t'_{22}; p_0,\rr_0^{\ell_0}, \fr_b^{\bell_b})} \; ,
\end{align*}where
\begin{align*}
        F(t_{12}, t_{22}; p_0, \fr_{0,b}^{\bell_{0,b}}):=& \sum_{k_1,\ldots, k_m=0}^{\infty} 
                \int d\fq_{J_{m,\bk}}
                K(t_{12}, \fq_{J_1}) K(t_{22}, \fq_{J_2}) 
                L(p_0,r_0^{\ell_0},\fr_b^{\bell_b}, \fq_{J_{m,\bk}})\; ,
\end{align*}
and $J_1, J_2$ are defined in (\ref{Jdef}).  We now show that:   
\begin{align}
        \Big| \int& dp_0 dr_0 dr'_0 d\fr_b\,  |\psi_0(r_b)|^2
                K(t_{11},t_{21};p_0,r_0^{\ell_0},\fr_b^{\bell_b})
  \overline{ K(t_{11}',t_{21}'; p_0,
                \rr_0^{\ell_0},\fr_b^{\bell_b} )}
                 \notag \\ 
        &\qquad \qquad \times  F(t_{12}, t_{22}; p_0,r_0^{\ell_0}, \fr_b^{\bell_b}) 
                \overline{ F(t'_{12}, t'_{22}; p_0,\rr_0^{\ell_0}, \fr_b^{\bell_b} )} 
 \Big| \notag \\
        \leq& C^m t^{m-b}  \frac{ t_1^{m_1-1}}{(m_1-1)!} \frac{t_2^{m_2}}{m_2!} 
        \| \psi_0 \|_{30, 0}^2 \notag \\
        &\times \sup_{r_b} |\!|\!|F(t_{12}, t_{22}; p_0,r_0^{\ell_0}, \fr_b^{\bell_b}) 
                \overline{ F(t'_{12}, t'_{22}; p_0,\rr_0^{\ell_0}, \fr_b^{\bell_b })} 
                \<r_b\>^{-60} |\!|\!|_{dp_0 dr'_0 d\fr_{0,b-1}}  \; .
                \label{E:g=1mainbound}
\end{align}
We assume first that $\b\notin \{0,b\}$.  This implies that $m_1\geq2$ in 
(\ref{E:g=1mainbound}) and consequently yields a factor of $1/n$.  
The proof of (\ref{E:g=1mainbound}) begins with an identity
similar to
 (\ref{E:timedivfreesplit}).  However, we replace $\< \sigma_0 - \sigma'_0\>^{-3/2}$ 
in the latter with $\< \sigma_0\>^{-3/2} \<\sigma'_0\>^{-3/2}$.  Integration of 
the $d\sigma'_J$ variables yield:   
\begin{align*}
        \int_0^{t_{11}*} & \[d\sigma_J\]_{-1}^{\b1}  \int_0^{t'_{11}*} 
                 \[d\sigma'_J\]_{-1}^{\b1} \int_0^{t_{21}*} \[ d\sigma_J\]_{\b2}^b 
 \int_0^{t'_{21}*}
                 \[d\sigma'_J\]_{\b2}^b \\
        &\frac{1}{\<\sigma_{-1}-\sigma_{-1}'\>^{3/2}}   
                \frac{ (\sigma_0 \sigma'_0)^{\ell_0}}{\< \sigma_0\>^{3/2} 
                \<\sigma'_0\>^{3/2} (\ell_0!)^2}          \prod_{j=1; j\neq \b}^{b-1} 
                \frac{  (\sigma_j \sigma'_j)^{\ell_j}}{\< \sigma_j - \sigma'_j\>^{3/2}
(\ell_j!)^2} \\
        & \frac{ (\sigma_{\b1} \sigma'_{\b1})^{\ell_{\b1}}
                 ( \sigma_{\b2} \sigma'_{\b2})^{\ell_{\b2}}}
                 { \< (\sigma_{\b1}-\sigma'_{\b1}) + (\sigma_{\b2}-\sigma'_{\b2})\>^{3/2} 
                 (\ell_{\b1}!)^2 (\ell_{\b2}!)^2} 
\frac{(\sigma_b\sigma'_b)^{\ell_b}}{(\ell_b!)^2}  \\
        \leq& C^{b+1} \frac{  t^{m-b-1/2}}{(\ell_0! \cdots \ell_{\b1}!\ell_{\b2}!\cdots
                \ell_b!)^2} \int_0^{t_{11}*}  \[d\sigma_J \]_{-1}^{\b1}
                \int_0^{t_{21}*}  \[d\sigma_J\]_{\b2}^b 
                \frac{ \sigma_0^{\ell_0-1}}{\<\sigma_0\>^{1/2}} \sigma_1^{\ell_1}
                \cdots \sigma_b^{\ell_b} \; .
\end{align*}
Finally, we estimate the last integral by integrating by parts as in the direct 
estimate of Lemma \ref{L:0,0} and use:
\begin{align*}
        \int_0^{t_{11}}d\sigma_0 \frac{ (t_{11}- \sigma_0)^{  \ell_1 + 
\cdots + \ell_{\b1}+ \b}}
                {(\ell_1 + \cdots + \ell_{\b1} + \b)!}  
\frac{\sigma_0^{\ell_0 -1}}{\<\sigma_0\>^{1/2}
                \ell_0!} \leq C^{m_1} \frac{t_{11}^{m_1- 1/2}}{m_1!} \;,
\end{align*}
where $\ell_0 + \cdots + \ell_{\b1} = m_1-\b$.  Putting this 
together we obtain (\ref{E:g=1mainbound}).  We can now argue as in Lemma 
\ref{L:tdivfdecay} to show that,
\begin{align*}
        \sup_{r_b} |\!|\!|  F(t_{12}, t_{22}&; p_0,r_0^{\ell_0}, \fr_b^{\bell_b}) 
                \overline{ F(t'_{12}, t'_{22}; p_0,\rr_0^{\ell_0}, 
\fr_b^{\bell_b})}\<r_b\>^{-60}
                 \|_{dp_0 d\rr_0 d\fr_{0,b-1}} \\
        \leq& \frac{(M\lambda_0)^m}{\<t_{12}\>^{3/2}\<t_{22}\>^{3/2}
                \<t'_{12}\>^{3/2}\<t'_{22}\>^{3/2}}.
\end{align*}The cases of $\b \in \{0,b\}$ are handled similarly and are left to the reader.
Consequently, we get the estimate:
\begin{align*}
        \mbox{(I: $\gamma=1$)} \leq& C (M\lambda_0)^{m+1} \< T \>^m 
                \varrho \frac{ \<T\>^{2m-1}}{n} .
\end{align*}

\subsubsection{Term (II), case $n_1=1$}

Moving onto (II), if $\g$ is chosen so that $b(\g)< \k < b(\g +1)$, where 
$b(0) := 0$, then the case where $\g =0$ is analogous to the $\g=1$ case for the term 
(I) and the case of $\g>0$ is analogous to the $\g>1$ case for the term (I).  
The former is handled with the time division
 and the latter by using (\ref{E:KF}) and making use of two nontrivial pairing
 relations and yields a factor of $\varrho^2$.  One can check that
\begin{align*}
        \mbox{(II)} \leq  (M\lambda_0)^{m+1} \<T\>^{m} \Big[ \varrho
 \frac{\<T\>^{m-1}}{n} + m! \varrho^2 \<T\>^{m-2} (\log t)^{m+ \O(1)} \Big]\; .
\end{align*}            
Putting all of estimates in  this section together, we complete the proof of the Lemma
\ref{L:pairrec}. $\;\; \Box$

\subsection{Proof of Lemma \ref{L:pairrecplus1}}

Here we have a new collision in the time interval $[t_2, t_1 + t_2)$ which will
 provide an extra factor of $\varrho t_1$ and hence a factor of $1/n$. 
 The amputation of propagator essentially yields an extra factor of $\varrho$ 
compared to Lemma \ref{L:pairrec}.    

To compute the expectation, we will use a similar argument as in  Lemma \ref{L:bcd}.
 Starting with the case of $n_1 \geq 2$, we have  
\begin{align*}
        \bE \Big\|&\sum_{\k_1=2}^m \sum_{\k_2=1}^{\k_1-1} 
                \sum_{\a_0,A:|A|=m}^{\mathrm{no\, rec}} 
                \scriptUt^{\k_1,\k_2}_{[n_1], m_1,m_2;\a_0, A}(s,t_2)\psi_0 \Big\|^2 \\
        =& \sum_{\stackrel{1\leq \k_2<\k_1\leq m}{1\leq \k_2'<\k_1'\leq m}}
                 \sum_{\stackrel{B: |B|=b}{0\leq b\leq m+1}}
                \sum_{\sigma\in\operatorname{S}(b)}     
                \sum_{(\a_0 \oplus A, \a'_0 \oplus A')}          \\
        &\qquad 
                \bE \Big(
                \scriptUt^{\k_1,\k_2}_{[n_1],m_1,m_2;\a_0 A}(s ,t_2)\psi_0 , \,
                \scriptUt^{\k'_1,\k'_2}_{[n_1], m_1,m_2; \a'_0, A'}(s,t_2)\psi_0 \Big)\; ,
\end{align*}
where the sum on $(\a_0 \oplus A, \a'_0\oplus A)$ is short for summing over ordered 
sets $A,\,A'$ of size $m$ and $\a_0,\, \a'_0$ such that $\a_0\oplus A$ and 
$\a_0'\oplus A'$ have no repeating elements, $(\a_0\oplus A )\cap (\a_0'\oplus A') = 
B$, $B\prec (\a_0 \oplus A) $ and $\sigma(B) \prec (\a'_0 \oplus A')$. 
 We now use independence of our obstacles, the Schwarz inequality and symmetry to get:
\begin{align*}
        \bE\| &\scriptUt^{\k_1,\k_2}_{1,[n_1],m_1,m_2}(s,t_2)\psi_0 \|^2 \\
        \leq& \sum_{\stackrel{B: |B|=b}{0\leq b\leq m+1}}       
                \sum_{\stackrel{\a_0,A: |A|=m}{B\prec (\a_0\oplus A)}}   
                \sum_{1\leq\k_2<\k_1<m}  C_{N,m,b}  \,
                \bE_B \Big\| \bE_{(\a_0\oplus A)\setminus B} \,
                \scriptUt^{\k_1,\k_2}_{\a_0,[n_1],m_1,m_2, A}(s,t_2)\psi_0 \Big\|^2  \\
        :=& \, \mbox{(I)} + \mbox{(II)}         \, ,
\end{align*}for
\begin{align*}
        \mbox{(I)} :=& \sum_{\stackrel{B: |B|=b}{0\leq b\leq m+1}}              
                \sum_{\stackrel{\a_0,A: |A|=m}{B\prec(\a_0\oplus A)}}^{\mathrm{no\,rec}} 
                \sum_{\stackrel{1\leq \k_2 <\k_1 \leq m}{\{\a_{\k_1},\a_{\k_2} \} 
      \subseteq B} }        
                C_{N,m,b}       \,
                \bE_B \Big\| \bE_{(\a_0\oplus A)\setminus B} \,
                \scriptUt^{\k_1,\k_2}_{\a_0 ,[n_1],m_1,m_2,A}(s,t_2)\psi_0\Big\|^2 \\
        \mbox{(II)} :=& \sum_{\stackrel{B: |B|=b}{0\leq b\leq m+1}}             
                \sum_{\stackrel{\a_0,A: |A|=m}{B\prec(\a_0\oplus A)}}^{\mathrm{no\,rec}}  
                \sum_{\stackrel{1\leq \k_2 <\k_1 \leq m}{\{\a_{\k_1},\a_{\k_2} \} 
    \nsubseteq B} }       
                C_{N,m,b}       \, \bE_B \Big\| \bE_{(\a_0\oplus A)\setminus B} \,
                \scriptUt^{\k_1,\k_2}_{\a_0 ,[n_1],m_1,m_2,A}(s,t_2)\psi_0\Big\|^2 
\end{align*}and $C_{N,m,b} := \binom{N-m-1}{m+1-b} \frac{(m+1)! (m-1) (m-2)}{2} $. 

 We first treat the term (I).  
Define $(\g_1,\g_2)$ such that $\k_1 = b(\g_1)$ and $\k_2 = b(\g_2)$.  
 By considering separately the cases $\a_0\in B$ and $\a_0 \notin B$, we
 can compute the expectations and calculate 
the combinatorics as in Lemma \ref{L:bcd}, to get bound:
\begin{align*}
        \mbox{(I)} \leq& \sum_{b=2}^{m+1} \sum_{\bell_{0,b}} \sum_{1\leq \g_2<\g_1\leq b} b!
                \varrho^{2m-b} \\
        & \Big\{ \varrho \Big| \int dp_0 dp_1 d\fu_{0,n_1-2} d\fu'_{0,n_1-2} d\fr_{0,b} 
                d\fr'_{0,b} \psi_0(r_b) \overline{\psi_0(\rr_b)} 
                \Delta(\fu_{0,n_1-2}, \fu'_{0,n_1-2}, \fr_{0,b}, \fr'_{0,b}) \\ 
        &\times | \hV(p_0-p_1) |^2 
                \mathcal{K}(s,t_2; p_1, \fu_{0,n_1-2}, \fr_{0,b}^{\bell_{0,b}})
                \overline{\mathcal{K}(s,t_2; p_1, \fu'_{0,n_1-2}, 
 {\fr'}_{0,b}^{\bell_{0,b}})} \Big| \\
        +&  \varrho^{2} \Big| \int dp_0 d\fu_{0,n_1-2} d\fu'_{0,n_1-2} d\fr_{0,b} 
                d\fr'_{0,b} \psi_0(r_b) \overline{\psi_0(\rr_b)} 
                \Delta(\fu_{0,n_1-2}, \fu'_{0,n_1-2}, \fr_{0,b}, \fr'_{0,b}) \\
        & \times | \hV(0)|^2\mathcal{K}(s,t_2; p_0, \fu_{0,n_1-2}, \fr_{0,b}^{\bell_{0,b}})
                \overline{\mathcal{K}(s,t_2; p_0 , \fu'_{0,n_1-2},
   {\fr'}_{0,b}^{\bell_{0,b}})} 
                \Big| \Big\}.
\end{align*}The first term arises from cases in which $\a_0 \in B$ and the
 second when $\a_0 \notin B$.  We will only bound the first term, the second will be smaller
 by a factor of $\varrho$ when treated in the same way. 
Our pairing relations are:
\begin{align*}
        \Delta(&\fu_{0,n_1-2},\fu'_{0,n_1-2},\fr_{0,b},\fr'_{0,b}) = \\
        &\delta\Big[ -\Sigma_{j=0}^{n_1-2} (-1)^{n_1-j} (u_j -\uu_j) 
                -(r_0-\rr_0) +(r_{\g_2} - \rr_{\g_2}) \Big] \notag \\
        &\times \prod_{j=1}^{\g_2-1} \delta[ \rr_{j} = r_j -(r_{0} - \rr_{0}) ] 
                \prod_{j=\g_2+1}^{\g_1-1} \delta[ \rr_{j} = r_j -(r_{\g_2} - \rr_{\g_2}) ] 
                \prod_{j=\g_1}^b \delta(\rr_j = r_j).           
\end{align*}
This is essentially identical to equation 
 (\ref{E:pairrecdeltaI}).  The extra integration in $p_0$ is handled through the
 decay of $\hV(p_0-p_1)$ after appealing to Lemma \ref{L:Blemma}.  Consequently, 
by following the proof of Lemma \ref{L:pairrec}, it is easy to verify the bound:
\begin{align*}
        \mbox{(I)} \leq& (M\lambda_0)^{n_1+m} \varrho^2 \<T\>^{m-2} (\log t)^{\O(1)}
                \Big[\frac{\<T\>}{n} + m! \varrho (\log t)^m\Big] \, .
\end{align*}

Aside from the extra factor of $\varrho$ from the amputation, the case of
 (II) is analogous to the one in Lemma \ref{L:pairrec}, as is the case of $n_1 =1$. 
This completes the proof of Lemma \ref{L:pairrecplus1}. $\;\;\Box$


\subsection{Proof of Lemma \ref{L:nested2rec}}

The amputation gains a factor of $\varrho$.  The pairing relations to follow will
 show that we will be able to use estimates such as (\ref{E:crossingDelta1}) to gain 
another factor of $\varrho$.  The last factor of $\varrho$ is obtained through either
 time division
 estimates like those in Section \ref{sec:g1n1}
or by utilizing another non-trivial pairing relation.

Again, we calculate $ \bE_B \| \bE_{A\setminus B}\, 
\scriptUt^{\k_1,\k_2}_{[2],m_1,m_2,A}\psi_0\|^2$
 as in Lemma \ref{L:pairrec} to get:
\begin{align*}
         \bE \Big\| \sum_{2\leq \k_1<\k_2\leq m} 
                \scriptUt^{\k_1,\k_2}_{[ 2], m_1,m_2}(s,t_2)\psi_0 \Big\|^2 
                \leq& \mbox{ (I)} +  \mbox{(II)}
\end{align*}where
\begin{align*}
        \mbox{(I)} :=& \sum_{\stackrel{B: |B|=b}{0\leq b\leq m}}                
                \sum_{\stackrel{A: |A|=m}{B\prec A}}  \sum_{ \stackrel{1\leq \k_1<\k_2\leq m}
                {\{\a_{\k_1},\a_{\k_2} \}  \subseteq B}          }
                C(N,m,b)        \,
                \bE_B \| \bE_{A\setminus B} \,
                \scriptUt^{\k_1,\k_2}_{[2], m_1,m_2;A}(s,t_2)\psi_0\|^2   \\
        \mbox{(II)} :=& \sum_{\stackrel{B: |B|=b}{0\leq b\leq m}}               
                \sum_{\stackrel{A: |A|=m}{B\prec A}}  
                \sum_{\stackrel{1\leq \k_1<\k_2 \leq m}
                {\{\a_{\k_1},\a_{\k_2} \}  \nsubseteq B}} C(N,m,b) \,
                \bE_B \| \bE_{A\setminus B} \,
                \scriptUt^{\k_1,\k_2}_{[2], m_1,m_2;A}(s,t_2)\psi_0 \|^2
\end{align*}and $C(N,m,b) := \binom{N-m}{m-b} \frac{m! (m-1) (m-2)}{2} $.  

Starting with the term(I), 
define $(\g_1,\g_2)$ so that $ (b(\g_1),b(\g_2)) = ( \k_1, \k_2)$.  We have
\begin{align*}
        \mbox{(I)} \leq& \sum_{b=2}^m \sum_{\bell_{0,b}} 
                \sum_{ \substack{1\leq \gamma_1<\gamma_2\leq b\\ b(\gamma_1)\ge 2}}
                b! \, \varrho^{2m-b} \Big| \int dp_0 du_0  du'_{0} d\fr_{0,b} d\fr'_{0,b} \,
                \psi_0(r_b) \overline{\psi_0(\rr_b) }\\  
        &\times 
                \Delta(u_0,u'_{0}, \fr_{0,b}, \fr'_{0,b})   \hV(p_0-u_0)
 \overline{\hV(p_0-u'_0)} 
                \mathcal{K}(s, t_2; u_0, \fr_{0,b}^{\bell_{0,b}})
                \overline{ \mathcal{K}(s,t_2;  u'_0, {\fr_{0,b}'}^{\bell_{0,b}})} \Big|\; ,
\end{align*}
where $\Delta(u_0, u'_0,\fr_{0,b},\fr'_{0,b})$ are the pairing relations.   

\subsubsection{Term (I), case  $\g_1 = 1$} \label{sss:n1=1g=1}

The condition of $b(\g_1) \geq 2$ forces $\ell_0 >0$, and
\begin{align*}
        \Delta(u_0&, u'_0, \fr_{0,b},\fr'_{0,b})  \\
        =&\delta[ (u_0- u'_0) - (r_1-\rr_1)]  
 \prod_{j=2}^{\g_2-1} \delta(\rr_j = r_j - r_1 +\rr_1)
                \prod_{j=\g_2}^b \delta(r_j - \rr_j) \, .
\end{align*}Since none of these relations actually depends on $r_0$ or $\rr_0$, 
we use Proposition \ref{P:mixedK} to split our kernels to isolate the momenta 
$r_0$ and $\rr_0$ so that we can estimate them separately with time division.  
Recalling (\ref{E:defbeta}), we have, for $\b>0$:
\begin{align*}
        \mathcal{K}(s &, t_2; u_0, \fr_{0,b}^{\bell_{0,b}}) \\
        =&\int_{0}^{t_1*} \[dt_{1j}\]_1^2  
\mathcal{K}(t_{11}, r_0^{\ell_0})    
                \frac{e^{\boldsymbol{\eta} \cdot (t_{12} , t_2) }}{4\pi^2}
                \int d\balpha  e^{-i \balpha \cdot (t_{12},t_2)} 
                \frac{B(\a_1, u_0, r_0)}{ [\a_1, u_0] } \\
        &\times \prod_{J=1}^{\b1} \frac{B(\a_1,R_J,R_J)^{\ell_J}}{[\a_1, R_J]^{\ell_J+1}}
                \prod_{J=\b1}^b  \frac{B(\a_2,R_J,R_J)^{\ell_J}}{[\a_2, R_J]^{\ell_J +1}}
                B(\a_1,\fr_{0,\b}) B(\a_2, \fr_{\b,b}) \, ,
\end{align*}
where $\balpha=(\a_1, \a_2)$ and $\boldsymbol{\eta} = (\eta_{12}, \eta_2)$.
The propagator $[\a_1, p]$ is regularized with $\eta_{12}=\eta(t_{12})$,
whereas $[\a_2, p]$ is regularized with $\eta_2=\eta(t_2)$.
We have an analogous expansion for 
$\overline{ \mathcal{K}(s , t_2; u'_0, {\fr'}_{0,b}^{\bell_{0,b}}) }$
with primed variables, $\balpha'=(\a_1,', \a_2')$ and 
$\boldsymbol{\eta} = (\eta_{12}', \eta_2)$.

The factors in the $\balpha$ and $\balpha'$ integration are estimated using
 the same techniques as in Lemma \ref{L:pairrec}.  The first relation
 in $\Delta(u_0,u'_0,\ldots)$ and the $du_0, du_0'$ integration
will allow us to avoid the $L^{\infty}$-estimate in one of the  resolvents
with $r'_1$ by applying Proposition \ref{P:crossing} as in (\ref{E:crossingDelta1}).  
Consequently, we get a contribution of $t^{b-2 +2 \|\bell_{1,b}\|}$
and we gain effectively a factor $\varrho$.
 Recalling (\ref{E:basicdisp}), the bound:
\begin{align*}
                \Big| \int dr_0 \, 
  \mathcal{K}(t_{11}; r_0^{ \ell_0}) f (r_0) \Big| \leq 
                        (M\lambda_0)^{2\ell_0 + 1} \< t_{11}\>^{\ell_0 -3/2} 
                        |\!|\!| f |\!|\!|_{dr_0}
\end{align*}
handles the remaining terms. Since $\ell_0\ge 1$, after $dt_{11}$
integration we gain a factor $\varrho^{1/2}$ compared to
the trivial estimate and a similar gain comes from $\overline{\mathcal{K}}$.
Finally an additional  $\varrho$ factor comes from the amputation
and this gives the result of the Lemma.
The $\b=0$ case is handled the same way. 
 This time, we will need to apply Proposition \ref{P:mixedK} twice since $r_0$ will 
appear in both $\mathcal{K}(t_1)$ and $\mathcal{K}(t_2)$.  The details are left to the reader.

\subsubsection{Term (I), case  $\g_1 > 1$}
Our pairing relations are:
\begin{align*}
        \Delta(u_0,\uu_0, \fr_{0,b}, \fr'_{0,b}) =&\delta[ (u_0 - u'_0 ) - 
(r_{\g_1} -\rr_{\g_1}) ]
                \prod_{j=1}^{\g_1-1}\delta( \rr_j = r_j -r_0+\rr_0) \\
        &\quad          
               \times  \prod_{j=\g_1+1}^{\g_2-1} \delta(\rr_j =r_j -r_{\g_1} + \rr_{\g_1})
                \prod_{j=\g_2}^b \delta(\rr_j=r_j)\;,
\end{align*}
where the first product is non-empty.  After applying (\ref{E:KF})
 we can perform estimates as in Lemma \ref{L:pairrec} to exploit two nontrivial
 relations.  The calculations are similar and we conclude the Lemma for
the term  (I).  

\subsubsection{Term (II)}

For the term (II) we apply (\ref{E:KF}) and perform the usual
 estimates which exploit nontrivial pairing relations as in 
the estimates for (II) in Lemma \ref{L:pairrec}.    

The case of $\k_1< \k_2 \leq b(1)$ requires time division arguments. 
 The key estimates in these cases are 
\begin{align*}
        \int dp_0& du_0 du'_0dr'_0  d\fr_{0,b} \, f(p_0, u_0,u'_0, \rr_0  ,\fr_{0,b}) \\
        &\times 
                K(s_1, s_2;  u_0, r_0^{\ell_{01}},u_0^{\ell_{02}}, \fr_{1,b}^{\bell_{1,b}}) 
                \overline{ K(s'_1,s'_2; u'_0, \rr_0^{\ell_{01}},\uu_0^{\ell_{02}}, 
                \fr_{1,b}^{\bell_{1,b}})} \\
        \leq& C t^{2m-b -3} |\!|\! | f|\!|\!|_{dp_0 du_0 du'_0dr'_0  d\fr_{0,b} } ,
\end{align*}
when $b(1)=\kappa_1$ 
 and
\begin{align*}
        \int dp_0& du_0 du'_0dr'_0  d\fr_{0,b} \, f(p_0, u_0,u'_0, \rr_0  ,\fr_{0,b}) \\
        &\times 
                K(s_1, s_2;  u_0, r_0^{\ell_{01}},u_0^{\ell_{02}}, p_0^{\ell_{03}},
 \fr_{1,b}^{\bell_{1,b}}) 
                \overline{ K(s'_1,s'_2; u'_0, \rr_0^{\ell_{01}},\uu_0^{\ell_{02}}, 
                p_0^{\ell_{03}} , \fr_{1,b}^{\bell_{1,b}})} \\
        \leq& C t^{2m-b -4} |\!|\! | f|\!|\!|_{dp_0 du_0 du'_0dr'_0  d\fr_{0,b} }  ,
\end{align*}
when $b(1)> \kappa_1$. In the first estimate
$\ell_{01} + \ell_{02}  +1 =\ell_0$ and $\ell_{01}>0$ and
in the second one $\ell_{01}+\ell_{02} + \ell_{03} + 2 = \ell_0$ 
with $\ell_{01}>0$.
  Note that the summation over $\k_1$ and $\k_2$ in the definition of (II)
in effect  also sums over possible $\ell_{01}, \ell_{02}$ and $\ell_{03}$.  
The reader can verify that applying these estimates and  following 
the estimates of the direct terms in Lemma \ref{L:0,0} one obtains  
 Lemma \ref{L:nested2rec}. $\;\;\;\Box$

\subsection{Proof of Lemma \ref{L:pingpong2rec}}

Starting from (\ref{E:pingpongdef}), we make the usual decomposition:
\begin{align*}
        \Big\|  &\sum_{\stackrel{\k_3=1}{\k_3\neq\k_1,\k_2}}^m 
                \sum_{1\leq \k_2< \k_1=m} 
                \sum_{A: |A|=m}^{\mathrm{no\,rec}} 
  \scriptUt^{\k_1,\k_2,\k_3}_{*,[n_1],m_1,m_2;A}
                (s,t_2)\psi_0 \Big\|^2  \leq \mbox{ (I)} + \mbox{ (II)} ,
\end{align*}where
\begin{align*}
        \mbox{(I)}  :=& \sum_{\stackrel{B: |B|=b}{0\leq b\leq m}}               
                \sum_{\stackrel{A: |A|=m}{B\prec A}}  
                \sum_{\stackrel{\k_1, \k_2, \k_3}{ \{\a_{\k_1},\a_{\k_2},\a_{\k_3} \} 
   \subseteq B}}     
                C(N,m,b)        \,
                \bE_B \| \bE_{A\setminus B} \,
                 \scriptUt^{\k_1,\k_2,\k_3}_{*,[n_1],m_1,m_2;A} (s,t_2)\psi_0\|^2   \\
        \mbox{(II)} :=& \sum_{\stackrel{B: |B|=b}{0\leq b\leq m}}               
                \sum_{\stackrel{A: |A|=m}{B\prec A}}  
                \sum_{ \stackrel{ \k_1,\k_2,\k_3}{\{\a_{\k_1},\a_{\k_2},\a_{\k_3} \}  
                \nsubseteq B} }C(N,m,b) \,
                \bE_B \| \bE_{A\setminus B} \,
                \scriptUt^{\k_1,\k_2,\k_3}_{*,[n_1],m_1,m_2;A} (s,t_2)\psi_0 \|^2 \, ,
\end{align*}
$C(N,m,b) := \binom{N-m}{m-b} \frac{m! (m-1) (m-2)^2}{2} $ and the sums
 on $\k_1,\k_2,\k_3$ are restricted to $1\leq \k_2 < \k_1 \leq m$ and $\k_3 \neq \k_1,\k_2$.  

As before, the first term is the leading order term.  Defining $(\g_1,\g_2,\g_3)$ 
such that $( b(\g_1), b(\g_2), b(\g_3)) = ( \k_1, \k_2, \k_3)$, we have:
\begin{align*}   
        \mbox{(I)} \leq& \sum_{b=2}^m \sum_{\bell_{0,b}} 
                \sum_{1\leq \gamma_2<\gamma_1\leq b}\sum_{\g_3\neq \g_1,\g_2}
                b! \, \varrho^{2m-b} \Big| \int d\fp_{0,1} d\pp_1 d\fu_{0,n_1-2}
    d\fu'_{0,n_1-2} 
                d\fr_{0,b} d\fr'_{0,b} \\  
        &\times \psi_0(r_b) \overline{\psi_0(\rr_b) }
                \Delta(p_1,\pp_1,\fu_{0,n_1-2},\fu'_{0,n_1-2}, \fr_{0,b}, \fr'_{0,b})  \\
        &\times \mathcal{K}(t_1, t_2; p_1, \fu_{0,n_1-2}, \fr_{0,b}^{\bell_{0,b}})
                \overline{ \mathcal{K}(t_1,t_2;\pp_1, \fu'_{0,n_1-2},
  {\fr_{0,b}'}^{\bell_{0,b}})} \Big|.
\end{align*}
The pairing functions can be simplified to always yield two nontrivial 
relations: one of which involves the variables 
$(\fu_{0,n_1-2}, \fu'_{0,n_1-2}, \fr_{0,b},\fr'_{0,b})$ and the other one involves
$(p_1, \pp_1, \fr_{0,b}, \fr'_{0,b})$.   As in the estimates of section
 \ref{sss:neq2,1smalln1} in Lemma
 \ref{L:pairrec}, we exploit these two nontrivial relations and obtain two factors 
of $\varrho$ (one from each nontrivial relation).  This, combined with an extra 
$\varrho$ resulting from the amputation, will yield the correct estimate.  
In particular, we apply estimates such as (\ref{E:crossingDelta1}), 
(\ref{E:alphacrossing}) or (\ref{E:doublealphacrossing}) on 
three of the variables $(\rr_0, \rr_{g_1}, \rr_{g_2}, \rr_{\b})$,  where $g_1$ 
and $g_2$ are the first and second smallest element
of $(\g_1, \g_2, \g_3)$.  The rest of the factors are handled as in previous proofs.  

When $n_1>8$, we take the additional step of rewriting the pairing relation 
involving $\fu_{0,n_1-2}$ and $\fu'_{0,n_1-2}$ in position space, applying 
(\ref{E:defR}) and forming the Born series term.  The second nontrivial relation
 (which involves $p_1$ and $\pp_1$) is used in bounds of the form
 (\ref{E:crossingDelta1}), (\ref{E:alphacrossing}) or (\ref{E:doublealphacrossing}).
  The result is that we avoid $L^{\infty}$-estimates on three of 
the propagators containing the  variables $(\rr_0, \rr_{g_1}, \rr_{g_2}, \rr_{\b})$.  

The terms in (II) are handled in the same way as in Lemma \ref{L:pairrec}. 
 We isolate two nontrivial relations, one involving $\fu_{0,n_1-2}$ and 
the other $\fu'_{0,n_1-2}$.  When $n_1 \leq 8$ we use these relations in
 estimates such as (\ref{E:crossingDelta1}), (\ref{E:alphacrossing}) or
 (\ref{E:doublealphacrossing}) and when $n_1>8$ we form the Born series
 terms.  See the discussion in Lemma \ref{L:pairrec}. 
 The details are left to the reader. This completes the proof of Lemma   
\ref{L:pingpong2rec}. $\;\;\;\Box$

\section{Estimates on the propagator}\label{sec:est}

Recall that $\|f\|_{n,n'} := \| \langle x \rangle^n \langle\grad_x\rangle^{n'} f(x)\|_2$. 
 In what follows, we will treat functions $f$ which are possibly dependent on the parameter
 $\a \in \mathbb{R}$.  In this case, we abuse the notation to write
 $ \| f_\a \|_{n,n'} := 
\sup_{\a}  \| \langle x \rangle^n \langle\grad_x\rangle^{n'} f_{\a}(x)\|_2$.

For $0<\eta < 1 $, and $\a \in \mathbb{R}$, define the following operator:
\begin{align}
       B_\eta(\a,u)f (p) =
   B(\a,u)f (p) &:=  \int_{\mathbb{R}^3}dq \frac{\hV(p-q) f(q)}{\a - (q+u)^2/2 + i\eta}
           \label{E:defoperatorB}.
\end{align}
We will usually suppress the dependence in $\eta$ unless it becomes critical.

\begin{lemma} \label{L:operatorBest}Let $N >0$ and $n,n' \leq N$.  If
 $\| \hV\|_{N+2,N+2} \leq \lambda_0 \ll1$, then there exists a constant 
$C$ depending only on $N$ (and implicitly on the dimension $d=3$) such that:
\begin{align*}
     \sup_{\eta, u,\alpha}   | \< p \>^n \< \grad_p\>^{n'} B_\eta(\a,u)f (p) | 
   \leq& C \lambda_0 \| f\|_{n,2} \\
        \sup_{u,\alpha} | \< p \>^n \< \grad_p\>^{n'} \partial_{\a} 
\big( B_\eta(\a,u)f_\a  \big) (p) | \leq& 
                C \lambda_0 \Big(\frac{ \| f_\a \|_{n,2} }{| \a + i\eta|^{1/2}} + 
\| \partial_\a f_\a \|_{n,2} \Big).
\end{align*}
We also have the same bounds for $\| B_\eta(\a,u)f \|_{n,n'}$ and 
$\| \partial_\a B_\eta(\a,u)f \|_{n,n'}$, respectively.  
\end{lemma}
\begin{proof}
A direct calculation of the Fourier Transform of the Yukawa potential yields the identities:
\begin{align*}
        \mathcal{F}^{-1}   (\a - (\cdot+u)^2/2 + i\eta)^{-1} (x) &= 
                C \frac{e^{i|x|\sqrt{\a +i\eta}- ixu} }{|x|} := G_u (x)
 := G(x) \\
         B(\a,u) f(p) &= \int dx \, e^{ixp} V_0(x)\int dy \, G(x-y) \check{f}(y) \; .
\end{align*} 
Here we consider the branch of the square root with positive imaginary part
and we omitted $\eta$ from the notations.
Hence we can make use of the bound:
\begin{align*}
        | \< p\>^n  \<\grad_p\>^{n'} B(\a,u) f (p) | \leq 
        \Big\| \< \grad_x \>^n \< x\>^{n'} V_0(x) \int dy \, 
G (x-y) \check{f}(y) \Big\|_{L^1(dx)}
\end{align*} Using the Leibniz rule, we have:
\begin{align*}
        \Big\| (\grad_x )^n & \< x\>^{n'} V_0(x) \int dy \, 
   G (x-y) \check{f}(y) \Big\|_{L^1(dx)} \\
        \leq& \sum_{j'+j = n} \Big\| (\grad_x)^{j}(\< x\>^{n'} V_0(x) ) \int dy \,
 G(x-y) (\grad_y)^{j'}      
                \check{f}(y) \Big\|_{L^1(dx)} \\
        \leq&  \sum_{j'+j = n} \| \hV \|_{n, n'+2} \| \<x\>^{-2} \<y\>^{-2} 
G(x-y)\|_{L^2(dxdy)}
                \| \<y\>^2 \< \grad_y\>^{j'} \check{f} \|_{L^2(dy)} .
\end{align*}
The integral involving $G(x-y)$ can be bounded uniformly since we know 
that $\eta \leq 1$.  This proves the first inequality.

The second inequality follows in the same way as the first and uses the bound 
$| \partial_\a G_{\a, u}(x) | \leq C  | \a + i \eta|^{-1/2}$ .  The last two bounds
 follow in the same manner.  
\end{proof}

\begin{lemma} \label{L:Blemma}
Define $B_\eta(\a,p,r)$ as in (\ref{E:defB}), let $N>0$ and $n,n' \leq N$.  
If $\|\hV\|_{N+2,N+2} \leq \lambda_0 \ll 1$, then there exists a constant $M$ 
depending only on $N$ such that:
\begin{align*}
        | \< p_1-p_2\>^n \<\grad_{p_1}\>^{n'} \<\grad_{p_2}\>^{n'} B_\eta(\a, p_1,p_2) | 
                \leq& M \lambda_0 \\
        | \< p_1-p_2\>^n \<\grad_{p_1}\>^{n'} \<\grad_{p_2}\>^{n'} \partial_\a 
B_\eta(\a, p_1,p_2) |                 \leq&  \frac{M \lambda_0}{ |\a+i\eta|^{1/2}}  \; .
\end{align*}
\end{lemma}

\begin{proof} 
We omit $\eta$ from the notation. For a fixed $k$, consider
\begin{align*}
        B(k; \a,p_1,p_2) :=  \int d\fq_{k} \hV(p_1-q_1) \frac{\hV(q_1-q_2)}{\a- q_1^2/2 +i\eta}
                \cdots \frac{\hV(q_k-p_2)}{\a -q_k^2/2 +i\eta} \; ,
\end{align*}where the $k=0 $ term is defined to be $\hV(p_1-p_2)$.  
This implies:
\begin{align*}
        \<\grad_{p_1}\>^{n'} &\<\grad_{p_2}\>^{n'}  B(k; \a,p_1,p_2) \\
        =&  
                \int d\fq_{k} (\<\grad\>^{n'} \hV)(p_1-q_1) 
                \frac{\hV(q_1-q_2)}{\a- q_1^2/2 +i\eta}
                \cdots \frac{(\<\grad\>^{n'} \hV)(q_k-p_2)}{\a -q_k^2/2 +i\eta} \\
        =&  \int d\fq_{k} (\<\grad\>^{n'} \hV)(p_1-p_2-q_1) \\
        &\times \frac{\hV(q_1-q_2)}{\a- (q_1+p_2)^2/2 +i\eta}
                \cdots \frac{(\<\grad\>^{n'} \hV)(q_k)}{\a -(q_k+p_2)^2/2 +i\eta}  .
\end{align*}
By writing $p = p_1-p_2$ and $u= p_2$, it suffices to bound:
\begin{align*}  
        \<p\>^n \<\grad_p\>^{n' } B(\a,u)^k ( \<\grad\>^{n'}\hV )  (p) = 
                \<p\>^n \<\grad_p\>^{n'} \underbrace{ B(\a,u)\circ \cdots \circ B(\a,u)}_{k}
                ( \<\grad\>^{n'}\hV )  (p)  .
\end{align*}
We can now apply the previous lemma to get:
\begin{align*}
        | \<p\>^n \<\grad_p\>^{n' } B(\a,u)^k ( \<\grad\>^{n'}\hV )  (p)| \leq C\lambda_0 
                \| B(\a,u)^{k-1} ( \<\grad\>^{n'}\hV ) \|_{n,2} \; ,
\end{align*}
and inductively, using the bounds on $ \| B(\a,u)f \|_{n,n'}$, we obtain
\begin{align*}
        | \<p\>^n \<\grad_p\>^{n' } B(\a,u)^k 
 ( \<\grad\>^{n'}\hV )  (p)| \leq( C\lambda_0)^{k+1} .
\end{align*}
By definition, $B(\a,p_1,p_2) = \sum_{k=0}^{\infty} B(k; \a, p_1, p_2)$. 
 Summing the last bound yields the first estimate.  The second estimate follows from 
the Leibniz rule and is proven in an analogous way.

\end{proof}

\subsection{Summing for Pair Recollisions}

This section provides the estimates on the two-obstacle Born series term mentioned
 in Section \ref{ss:error3}.  For $n_1\geq2$, recall (\ref{E:defR}):
\begin{align}
        \scriptB_{n_1, \nu }(\a,p_0,r_0) :=& 
                \int d \fu_{0,n_1-2}  B(\a, p_0,u_0)
                \frac{B(\a,u_0,u_1) e^{i\nu u_0}}{\a-u_0^2/2 + i\eta}
                \frac{B(\a, u_1, u_2)
   e^{-i\nu u_1}}{\a - u_1^2/2 + i\eta} \notag \\
        &\times \cdots\times
                \frac{B(\a,u_{n_1-2},r_0)
    \exp{\big[ (-1)^{n_1-2}i \nu u_{n_1-2}  \big] }  }{\a - u_{n_1-2}^2/2+i\eta}  \; ,
\end{align} 
where $0< \eta \leq 1$, and $B$ is defined in (\ref{E:defB}).  
The dependence on $\eta$ is omitted from the notation as the estimates
below are uniform for $0< \eta\leq 1$.
\begin{lemma} \label{L:Rest}Let $n_1\geq 2$ and $N>0$.  Then there exists a 
constant $M$ depending only on $N$ such that for $n \leq N$, we have:
\begin{align}
        | \<p_1-p_2\>^{n} \scriptB_{n_1,\nu}(\a,p_1,p_2) |       \leq& 
                (M \lambda_0)^{n_1}\<\nu \>^{-(n_1-1)/2}\; ,
                \label{E:Rest}
\end{align}where $ \| \hV \|_{N+3,N+3} \leq \lambda_0   $. 
\end{lemma}
\begin{proof}
Define the following operator (compare to (\ref{E:defoperatorB})):
\begin{align}
        \scriptB_{\nu}(\a,u)f(p) :=& \int dq  \frac{\hV(p-q) e^{i \nu (q+u)}}
                {\a - (q+u)^2/2 +i\eta}
                f(q) \; .\label{E:defoperatorR}      
\end{align} We claim that for $n$, $n' \leq N$:
\begin{align}
        |  \<p \>^n \< \grad_p\>^{n'} \scriptB_{\nu}(\a,u)f (p) | \leq
                 C \lambda_0 \<\nu \>^{-1/2}  \|f\|_{n,3}  . \label{E:operatorRest}
\end{align}
To show this, we proceed as in Lemma \ref{L:operatorBest}, 
except that we  estimate:
\begin{align*}
        | \<p \>^n &\< \grad_p\>^{n'} \scriptB_{\nu}(\a,u)f(p) |  \\ 
        \leq& \sum_{j+j'=n} \Big\| (\grad_x)^j  \<x\>^{n'} V_0(x) 
                \int dy \, G_{\nu,u} (x-y) (\grad_y)^{j'}\check{f}(y)
  \Big \|_{L^1(dx)} \\
        \leq& \sum_{j+j'=n} \|\hV\|_{n,n'+3} 
        \Big\|\<x\>^{-3} \<y\>^{-3} G_{\nu,u} (x-y) \Big\|_{L^2(dx dy)} 
                 \| \<y\>^3  (\grad_y)^{j'}\check{f}(y) \|_2 ,
\end{align*} where $G_{\nu,u}(x) = \frac{1}{|x+\nu|}
 e^{i |x + \nu | \sqrt{\a+i\eta} - i ux }$.  It is easy to see that    
\begin{align*}
        \|\<x\>^{-3} \<y\>^{-3} G_{\nu,u} (x-y) \|_{L^2(dx dy)} \leq C\<\nu \>^{-1/2} ,
\end{align*}which justifies (\ref{E:operatorRest}).

For the proof of Lemma \ref{L:Rest} we  write $p= p_1-p_2$, $u = p_2$ 
and estimate
\begin{align*}
        \Big| \<p\>^n (\scriptB^{k_0}_{0}(\a,u)&\circ \scriptB_{\nu}(\a,u)\circ \\
        & \circ \scriptB_{0}^{k_1}(\a,u)\circ\cdots 
                \circ \scriptB_{(-1)^{n_1-2} \nu}(\a,u)\circ
                \scriptB_{0}^{k_{n_1-1}}(\a,u) \circ \hV) (p) \Big|  
\end{align*}by applying (\ref{E:operatorRest}) repeatedly as in Lemma \ref{L:Blemma}. 
 We then sum over $k_0, \ldots, k_{n_1-1}$ to complete the proof
of Lemma  \ref{L:Rest}.
\end{proof}

\section{Wigner Transform of Main Term}

\subsection{Renormalization}

Recall (\ref{E:defB}) and define:
\begin{align}
        \phi_A&(t,p_0)
                :=  \int d\fp_m \, \psi_0(p_m) \chi(A; \fp_{0,m}) K(t; \fp_{0,m}) 
\prod_{j=1}^m 
                B(p_m^2/2,p_{j-1},p_j) 
        .\label{E:defphi}
\end{align} 
We note that $B=B_\eta$ and all quantities derived from it
depend on $\eta:=\eta(t)$ (see (\ref{E:defeta})) throughout the whole
section but this fact will be omitted from the notation.
At the end we will use that the necessary estimates on $B$ are uniform in $\eta$.

As usual, we define $\phi^{\mathrm{no\,rec}}_m := 
\sum_{A: |A|=m}^{\mathrm{no\, rec}} \phi_A$ and recall the definition
of $\psi_m =   \psi^{\mathrm{no\,rec}}_m$  from 
(\ref{E:defpsim}). 
We will suppress the "no rec" notation in the following section.

We will first estimate the error of replacing $\psi_m$ with $\phi_m$.
\begin{lemma} \label{L:phiminuspsi}  We have
\begin{align*}
        \bE \| \psi_m(t) - \phi_m(t) \|^2 \leq C (M \lambda_0)^m m! \varrho^{1/2}
         (\varrho t)^{m-1/2} (\log t)^{m+\O(1)} \; .
\end{align*}
This implies  that for fixed $m$ and our scaling $\varrho = \varrho_0 \ep$,
 $t=T\ep^{-1}$ we obtain
\begin{align*}
        \lim_{\ep \to 0} \bE \| \psi_m(t) - \phi_m(t) \|^2  = 0 \; .
\end{align*}
\end{lemma}     

\begin{proof}
 We begin by 
appealing to Lemma \ref{L:Kidentity} to write:
\begin{align*}
        \phi_A(t;p_0) =& \frac{ ie^{\eta t}}{2\pi } \int d\fp_m\, \chi(A; \fp_{0,m} )
                \psi_0(p_m)  \int \frac{ d\a \, e^{-i \a t}}{\a -p_0^2/2 + i \eta} 
                \prod_{j=1}^m \frac{B(p_{j-1},p_j)}{\a- p_j^2/2 +i \eta }  \,
\end{align*}and applying (\ref{E:KF}) to $\psi_A$.  To compute 
$\bE \| \sum_{A: |A|=m}^{\mathrm{no\, rec}} (\psi_A - \phi_A )\|^2$ 
we appeal to Lemma \ref{L:bcd} with 
\begin{align*}
        G(\fp_{0,m}) =&  \frac{ i e^{\eta t} }{2\pi }   \psi_0(p_m) \int d\a 
                \frac{e^{-i\a t} }{\a - p_0^2/2+i \eta} \sum_{k=1}^m  \Big(
                \prod_{j=1}^{k-1} \frac{B(\a, p_{j-1}, p_j)}{\a-p_j^2/2 + i \eta } \\
        &\times
                \frac{B(\a,p_{k-1},p_k) - B(p_m^2/2, p_{k-1}, p_k) }{ \a - p_k^2/2 + i \eta } 
                \prod_{k+1}^m \frac{B(p_m^2/2, p_{j-1},p_j)}{\a-p_j^2/2 + i \eta }\Big).
\end{align*}We now proceed by estimating
\begin{align*}
        \sum_{b=0}^m \sum_{\sigma \in \operatorname{S}(b) }\sum_{\ell,\ell'} \varrho^{2m-b} 
                \Big| \int d\fp_{0,b} d\fp'_b  G(p_0, \fp_{b}^{\bell_b}) 
                \overline{G(p_0, {\fp'_b}^{\bell'_b})} 
                \Delta_\sigma(p_0, \fp_b, \fp'_b) \Big| 
\end{align*}using Proposition \ref{P:crossingtools} as in our previous crossing estimates.  
This time, we do not exploit any structure of the pairings and crudely estimate the integrands
with variables
$\fp'_{b-1}$ in $L^{\infty}$.  However, we eliminate a factor of $t^{1/2}$ 
as a result of the bound 
\begin{align*}
        |B(\a',p'_{j-1},p'_j) - B(\pp_m^2/2 ,p'_{j-1},p'_j) | \leq C(M\lambda_0) t^{1/2} 
                \frac{|\a' - \pp_m^2/2|}{\<p'_{j-1}-p'_j \>^{30}}.
\end{align*}which follows trivially from Lemma \ref{L:Blemma}.  Indeed the numerator 
will cancel its corresponding singular factor $ | \a' -\pp_m^2/2 + i\eta |^{-1} $ 
and consequently we 
 eliminate a total factor of $t^{1/2}$.   
\end{proof}

For $v\in \mathbb{R}^3$ define:
\begin{align}
        T_v(p,q) :=& B\big( (p_m+v)^2/2, p-v, q-v \big) \notag \\
        T_v(p) :=& T_v(p,p)
\label{E:defT}
\end{align}
where we have suppressed the dependence on $p_m$ in our notation. 
 When $v=0$, we will also conveniently drop the subscript $v$ altogether. 
Moreover, these quantities also depend on $\eta$.
 We now define the {\it renormalized}
operator kernels:
\begin{align}
        B^{\mathrm{ren}}(p_{j-1}, p_j) :=& T(p_{j-1},p_j) - \frac{1}{|\Lambda|} 
                T(p_j)\delta(p_{j-1}-p_j) ,\label{E:defBren}  \\
        K^{\mathrm{ren}}(t; \fp_{0,m}) :=& (-i)^m \int_0^{t*} \prod_{j=0}^m ds_j \, 
                e^{-i s_j (p_j^2/2 + \varrho T(p_j)   )} .
                \label{E:defren}
\end{align}
We recall that the momenta are on a discrete lattice (see (\ref{E:deltadef}) and 
(\ref{E:contconv})) before we take $L\to \infty$.  The benefit of renormalization is that 
\begin{align}
        \bE_{\a} B^{\mathrm{ren}}(p_{j-1},p_j)e^{ix_\a (p_{j-1} -p_j)} = 0 . 
\label{E:normconseq}
\end{align}
With the notation $B^{\mathrm{ren}}(\fp_{0,m}) :=  
 \prod_{j=1}^m B^{\mathrm{ren}}(p_{j-1}, p_j)$, and 
with a similar definition for $T(\fp_{0,m})$, we define the renormalized wave
 function with less than $m$ external collisions to be:
\begin{align}
        \phi_{< m}^{\mathrm{ren}}(t; p_0) := \sum_{m=0}^{m-1} 
                \sum_{A:|A|=m}^{\mathrm{no\,rec}}
                \int d\fp_m K^{\mathrm{ren}}(t;\fp_{0,m}) B^{\mathrm{ren}}(\fp_{0,m}) 
                \chi(A; \fp_{0,m}) \psi_0(p_m). 
        \label{E:defrenphi}
\end{align}


\begin{lemma} \label{L:renerror}For $\varrho = \varrho_0 \ep$, $t= T\ep^{-1}$  we get:
\begin{align*}
         \lim_{m \to \infty}\lim_{\ep\to 0} \limsup_{L\to \infty}
                \bE \| \phi^{\mathrm{ren}}_{<m}(t) - \phi_{< m}(t) \|^2 = 0 \; .
\end{align*}
\end{lemma}
\begin{proof}
Using the definition of $B^{\mathrm{ren}}(p_{j-1}, p_j)$, one can verify that:
\begin{align}
        \sum_{n=0}^{m-1} &\sum_{A: |A|=n}^{\mathrm{no\,rec}} 
                \int d\fp_n K(t; \fp_{0,n}) \chi(A;\fp_{0,n}) 
                \psi_0(p_n)  T(\fp_{0,n})  \notag \\
        =&\sum_{b=0}^{m-1} \sum_{A: |A| =b}^{\mathrm{no\,rec}}
                 \sum_{\stackrel{ \bell_{0,b}}{ \| \bell_{0,b} \| <m-b} }
                 \Big[ \int d\fp_b K(t;\fp_{0,b}^{\bell_{0,b}}) 
                 \chi(A;\fp_{0,b}) B^{\mathrm{ren}}(\fp_{0,b})
                 \psi_0(p_b) \notag \\
        &\times
                \prod_{j=0}^b \(\varrho T(p_j)\)^{\ell_j} \Big] + 
\O\Big(\frac{1}{| \Lambda |}\Big).
                \label{E:resumphi}
\end{align}where once again $\| \bell_{0,b} \| := \sum_{j=0}^b \ell_j$.  Also, the identity
\begin{align*}
        K(t;\fp_{0,b}^{\bell_{0,b}}) = (-i)^b\int_0^{t*} \[ ds_j\]_0^b 
                \prod_{j=0}^b \frac{(-is_j)^{\ell_j}}{\ell_j!}
                e^{-i s_j p_j^2/2}  
\end{align*}implies the relation:
\begin{align}
        \sum_{\ell_0,\ldots,\ell_b =0}^\infty \Big[ K(t;\fp_{0,b}^{\bell_{0,b}})    
                \prod_{j=0}^b 
                \(\varrho T(p_j)\)^{\ell_j}\Big] = K^{\mathrm{ren}}(t;\fp_{0,b}).
                \label{E:Kren} 
\end{align}With (\ref{E:Kren}) and (\ref{E:resumphi}) in hand, it suffices bound:
\begin{align}
        \bE \Big\| \sum_{b=0}^{m-1} \sum_{A: |A| =b}^{\mathrm{no\,rec}} 
                \sum_{\stackrel{\bell_{0,b}}{\| \bell_{0,b}\|  \geq m -b}}
                \int d\fp_b& \psi_0(p_b)  B^{\mathrm{ren}}(\fp_{0,b}) \chi(A;\fp_{0,b})
                K(t; \fp_{0,b}^{\bell_{0,b}}) \prod_{j=0}^b 
\( \varrho T(p_j)\)^{\ell_j}\Big\|^2 \; .
                \label{E:renerror} 
\end{align} We expand the $L^2$-norm:
\begin{align}
        \sum_{b,b' = 0}^{m-1} &\sum_{\stackrel{A,A'}{|A|=b;|A'|=b'}}^{\mathrm{no\,rec}}
                \sum_{\bell_{0,b} ,\bell'_{0,b'}} 
                \int d\fp_{0,b} d\fp'_{b'} \psi_0(p_b)  
                \overline{\psi_0(p'_{b'}) } \prod_{j=0}^b \(\varrho T(p_j)\)^{\ell_j}
                \prod_{j = 0}^{b'} \( \varrho \overline{T(p'_{j})}\)^{\ell'_{j}}  \notag \\
        & \times\bE \[ \chi(A; p_0, \fp_b) \overline{ \chi(A'; p_0, \fp'_{b'})} \] 
                K(t; \fp_{0,b}^{\bell_{0,b}}) 
                \overline{ K(t; p_0^{\ell'_0}, {\fp'_{b'}}^{\bell'_{b'} })} 
                B^{\mathrm{ren}}(\fp_{0,b}) 
                \overline{B^{\mathrm{ren}}(p_0,\fp'_{b'})} \notag  \\
        \leq& \sum_{b=0}^{m-1} \sum_{\sigma \in \operatorname{S}(b)} 
                \sum_{\bell_{0,b},\bell'_{0,b}} 
                \varrho^{b}\Big|  \int d\fp_{0,b} d\fp'_b | \psi_0(p_b)|^2 
                B^{\mathrm{ren}}(\fp_{0,b}) \overline{B^{\mathrm{ren}}(p_0,\fp'_{b})}\notag \\
        & \times  
                K(t;\fp_{0,b}^{\bell_{0,b} }) 
\overline{ K(t; p_0^{\ell'_0}, {\fp'_{b}}^{\bell'_{b}}) }
                \Delta_{\sigma}(\fp_{0,b}, \fp'_b)\prod_{j=0}^b \(\varrho T(p_j)\)^{\ell_j}
                \( \varrho \overline{T(p'_{j})}\)^{\ell'_{j}}  \Big| \; , \label{E:renexp} 
\end{align} 
where after taking expectations, the property (\ref{E:normconseq}) of
$B^{\mathrm{ren}}$ forces the existence of a permutation $\sigma$ such that
$ A'=\sigma(A)$  and
  $\Delta_{\sigma}(\fp_{0,b},\fp'_b)$ contains the pairing relations as usual.
We then distinguish between direct terms ($\sigma = \operatorname{Id}$)
 and crossing terms ($\sigma \neq \operatorname{Id}$).  For the direct terms, 
we will use the Schwarz inequality:  
\begin{align*}
        \mbox{(Direct)}\leq& \sum_{b=0}^{m-1} \sum_{m^*=m}^{\infty}  
                \sum_{\stackrel{\bell_{0,b}}{   \|\bell_{0,b}\| = m^*-b}}
                \binom{m^*}{b} \varrho^b 
                \int d\fp_{0,b} |\psi_0(p_b)|^2 |B^{\mathrm{ren}}(\fp_{0,b})|^2 \\
        &\times
                \prod_{j=0}^b |\varrho T(p_j)|^{2\ell_j} | K(t;\fp_{0,b}^{\bell_{0,b} })|^2\; .
\end{align*}
$B^{\mathrm{ren}}$ can be estimated from 
(\ref{E:defT}),  (\ref{E:defBren}) and Lemma \ref{L:Blemma}, and we have:
\begin{align*}
        \mbox{(Direct)}\leq& \sum_{b=0}^{m-1} \sum_{m^*=m}^{\infty}  
                \sum_{\stackrel{\bell_{0,b}}{   \|\bell_{0,b}\| = m^*-b}} \binom{m^*}{b}
 \varrho^{2m^*-b}
                \lambda_0^{2m^*} \sup_{p_b} \int d\fp_{0,b} 
  | K (t; \fp_{0,b}^{\bell_{0,b}})|^2  \\
        &\times \frac{1}{\<p_b\>^{60}}  
                \prod_{j=1}^b\frac{1}{\<p_{j-1}-p_j\>^{60}} \; .
\end{align*}We now use dispersive estimates as in the estimate of the 
direct terms of Lemma \ref{L:0,0} to get:
\begin{align*}
        \mbox{(Direct)}\leq&  \sum_{b=0}^{m-1} \sum_{m^*=m}^{\infty}  
                C^{m^*} \lambda_0^{2m^*}  \frac{T^{2m^*-b}}{(m^*-b)! b!}
                \leq \sum_{m^*=m}^{\infty} \frac{(C T)^{m^*}}{m^*!} 
\end{align*} which vanishes as we take $m\to \infty$.           

To handle the crossing terms, $\sigma\neq \mbox{Id}$ on the right hand side 
of (\ref{E:renexp}),
 we begin to proceed as in the direct estimate.  
An application of the Schwarz inequality and symmetry gives:
\begin{align*}
        \mbox{(Crossing)} \leq& \sum_{b=2}^{m-1} \sum_{\sigma\neq Id}\sum_{m^*=m}^{\infty}
         \sum_{\stackrel{\bell_{0,b}}{  \|\bell_{0,b}\| = m^*-b}} \binom{m^*}{b} \varrho^b
                \int d\fp_{0,b} d\fp'_b  |\psi_0(p_b)|^2   \\
        &\times         
                \Delta_{\sigma}(\fp_{0,b}, \fp'_b)                      
                |B^{\mathrm{ren}}(\fp_{0,b})|^2  
                \prod_{j=0}^b |\varrho T(p_j)|^{2\ell_j} | K(t; \fp_{0,b}^{\bell_{0,b}})|^2.
\end{align*}
The conjugate momenta are then integrated out using the pairing functions. 
 Following the steps of the direct estimate, we get
\begin{align*}
        \mbox{(Crossing)} \leq& \sum_{b=2}^{m-1} m! \sum_{m^*=m}^{\infty}  
                  \frac{ (CT)^{m^*}}{m^*!} 
\end{align*} where the $m!$ is due to estimating the number of permutations $\sigma$. 
 This crude bound implies that we can appeal to the dominated convergence theorem in 
order to pass our limit in $\ep$ through the infinite sum.  To get the limit, we 
return to (\ref{E:renexp}) and expand $K$ using Lemma \ref{L:Kidentity}.  As in Lemma
 \ref{L:0,0}, the non-triviality of $\sigma$ will imply that we can gain 
a factor $\varrho$ compared to the crude bound.   
Finally, the dominated convergence theorem yields:
\begin{align*}  
        \lim_{\ep\to 0} \mbox{(Crossing)} \leq& \sum_{b=0}^{m-1}  m!
                \sum_{m^*=m}^{\infty}\lim_{\ep\to0} \varrho \, C^{m^*} T^{2m^*-1} = 0
\end{align*}
for every fixed $m$.
\end{proof}
%
%

\subsection{Computation of Wigner Transform}

 The rescaled Husimi function associated with $\psi^{\ep}_{\omega, T\ep^{-1}}$ 
(see (\ref{E:husimi2})) can be written as:
\begin{align*}
        H^{(\ep,\mu)}_{\psi}(X_0, V_0) =& \Big( W^{\ep}_{\psi} *_{X_0} G^{\ep^{\mu}}*_{V_0} 
                G^{\ep^{\mu}}\Big) (X_0, V_0) \\
        =& \int dx dw_0\, W^{\ep}_{\psi} (x,w_0) G^{\ep^{\mu}}(x-X_0) G^{\ep^{\mu}}(w_0-V_0) ,
\end{align*} 
where $G^{\ell}$ is the Gaussian function
with scaling given in (\ref{E:gaussian}) and 
$W^{\ep}_\psi (x,w_0) := \ep^{-3} W_{\psi}(x/ \ep, w_0)$ is the rescaled Wigner transform.

Recall our decomposition (\ref{E:1stdecomp}).  It has the disadvantage that our 
threshold $m_0$ is dependent on $\ep$. 
 To cure this, fix $M^*>0$ and write:
\begin{align*}
        e^{itH}\psi_0 = \sum_{m=0}^{M^*-1}
        \psi_{m}^{\mathrm{no\,rec}}(t) + \sum_{m=M^*}^{m_0-1} 
                   \psi_{m}^{\mathrm{no\,rec}}(t) 
 + \Psi_{m_0}^{\mathrm{error}}(t)  \,.
\end{align*}
According to Lemma \ref{L:errorest},
the last term vanishes in the $\ep \to 0$ limit when we set $m_0= m_0(\ep)$ 
as in  (\ref{E:m0choice}).  We have the bound:      
\begin{align*}
        \bE \| \psi_m^{\mathrm{no\,rec}}(T\ep^{-1}) \|^2 \leq 
(M\lambda_0)^m \<T\>^m \Big[ \frac{T^m}{m!} + m! \varrho T^{m-1} (\log T\ep^{-1})^{m+5}\Big] 
\end{align*}
which is essentially the same estimate as Lemma \ref{L:0,0} except we do 
not have time division and hence it is easier to prove.  This implies:
\begin{align}
        \lim_{M^*\to \infty} \lim_{\ep \to 0} \lim_{L\to\infty} \bE \Big\|
 \sum_{m=M^*}^{m_0-1} 
                \psi_{m}^{\mathrm{no\,rec}}(t) \Big\|^2  =0   \, . \label{E:cuttail}
\end{align}

We ultimately need to prove that for any given  bounded and
 continuous function $J$ on $\mathbb{R}^6$ and any fixed $0<T<\infty$
we have
\begin{align*}
        \lim_{\ep\to0} \lim_{L\to \infty} \Bigg|
        \int dX_0 dV_0 \, J(X_0,V_0)   
        \Big[ \bE H^{(\ep,\mu)}_{\psi(T\ep^{-1} )} (X_0, V_0) 
      - F_T(X_0,V_0) \Big]\Bigg| =0 \; ,
\end{align*}
where $F_T$ is the solution of the Boltzmann equation (\ref{E:boltz}).
Recall that the Husimi function defines a probability density
on $\mathbb{R}^6$. Since the Boltzmann equation preserves
positivity and the $L^1(\mathbb{R}^6)$-norm,  and
 $\| F_0\|_1=1$, the solution
 $F_T(X_0,V_0)$ is also a probability density. 
Therefore we have to prove weak convergence of probability measures.
It is well known that it is sufficient to test such convergence 
for smooth, compactly supported testfunctions. For the rest
of the section, we thus fix a function  $J\in \mathcal{S}(\mathbb{R}^6)$.

Using (\ref{E:husimi2}) and an argument nearly identical
 to the one justifying (2.10) in \cite{EY1}, we have:
\begin{align}
        \Bigg| \int dX_0 &dV_0 \, J(X_0, V_0) 
        \Big[\bE H^{(\ep,\mu)}_{\psi(T\ep^{-1}) }(X_0, V_0) 
        - \bE H^{(\ep,\mu)}_{\psi_1(T\ep^{-1})} (X_0, V_0)\Big] \Bigg|
        \notag \\
        \leq& C \Big( \sup_{\ep<1} \int \sup_{V_0}  
\big| \widehat{J_\ep}(\xi, V_0) \big| d\xi \Big)
        \sqrt{ \bE \| \psi_1 \|^2 \bE \| \psi_2\|^2 },
\label{psipsi}
\end{align}
for any decomposition $\psi =: \psi_1 + \psi_2$ and
with  $\widehat{J_{\ep}}(\xi,V_0):= \ep^{-3} \widehat{J}(\xi\ep^{-1}, V_0)$.
We recall the definition (\ref{E:defrenphi}) and we set
$$
\phi_t=\phi(t) := \phi_{<M^*}^{\mathrm{ ren}}(t)
$$
 for brevity.
 We apply the estimate (\ref{psipsi}) with $\psi=\psi_t$ and
$\psi_1= \phi(t)$. Then 
Lemmas \ref{L:errorest}, \ref{L:phiminuspsi}, 
\ref{L:renerror} and equations (\ref{E:1stdecomp}), (\ref{E:cuttail}) 
imply that it suffices to show that 
\begin{align}
        \lim_{M^*\to \infty} \lim_{\ep\to 0} \lim_{L\to \infty} 
 \bE H^{(\ep,\mu)}_{\phi(T\ep^{-1}) }(X_0, V_0) = F_T(X_0, V_0) 
\label{E:testS}
\end{align}
in $\mathcal{S'}$.

An application of the Fourier inversion theorem gives us the 
following identity: 
\begin{align*}
        H^{(\ep,\mu)}_{\phi}(X_0, V_0) = \int dp dw_0 \, e^{ipX_0} 
                \widehat{W}_{\phi}(\ep p, w_0) \frac{1}{\ep^{3\mu}} G^{\ep^{-\mu}}(p) 
                G^{\ep^\mu}(w_0-V_0) \; ,
\end{align*}
 where 
\begin{align*} 
        \widehat{W}_{\phi}(\xi , w_0) = \overline{\phi_t\Big(w_0 - 
          \frac{\xi}{2} \Big)} 
                \phi_t\Big(w_0+ \frac{\xi}{2}\Big) 
\end{align*}
is the Fourier transform of the Wigner function in the first variable.
Using (\ref{E:defrenphi}), we have:
\begin{align*}
        \widehat{W}_{\phi}(\ep p, w_0) =& \sum_{m,n=0}^{M^*-1}
\sum_{A: |A|=m}     \sum_{A':  |A'|=n}  
                \int d\fp_{m} d\fp'_n \, \chi\Big(A; w_0+\frac{\ep p}{2},\fp_m\Big)
                \overline{\chi\Big(A'; w_0-\frac{\ep p}{2}, \fp'_n\Big)} 
                  \\
        &\times   K^{\mathrm{ren}}\Big(t; w_0 + \frac{ \ep p}{2} , \fp_m\Big) 
                \overline{K^{\mathrm{ren}}\Big(t; w_0 - \frac{\ep p}{2},  \fp'_n\Big)}    \\
        &\times \psi_0^{\ep}(p_m) \overline{\psi_0^{\ep}(\pp_n)}        
                B^{\mathrm{ren}}\Big(w_0+ \frac{\ep p}{2} , \fp_m\Big)
                \overline{ B^{\mathrm{ren}}\Big(w_0- \frac{\ep  p}{2}, \fp'_n\Big)}.
\end{align*}
We next take the expectation of this expression, 
and we use that the renormalization forces $n=m$  and $A' = \sigma (A)$
 for $\sigma \in \operatorname{S}(n)$
(see (\ref{E:normconseq})).
 Rename variables $w_j := p_j - \frac{\ep p}{2}$ and $w'_j := \pp_j +\frac{\ep p}{2}$ 
and define the following:
\begin{align*}
        B^{\mathrm{ren}}_p(\fw_{0,m}) :=& \prod_{j=1}^m 
       B^{\mathrm{ren}}( w_{j-1}+p, w_j +p) \\
   B^{\mathrm{ren}}(  w_{j-1}+p, w_j +p):=&
   B\Big( \frac{(w_m+ p)^2}{2}, w_{j-1}+p, w_j +p\Big) \\&
   -\frac{1}{|\Lambda|}
   B\Big( \frac{(w_m+ p)^2}{2}, w_{j}+p, w_j +p\Big)\delta(w_{j-1}-w_j) \\
        K^{\mathrm{ren}}(t; \fw_{0,m} \pm p) :=& 
                K^{\mathrm{ren}}(t; w_0 \pm p, \ldots ,w_m \pm p) \; .    
\end{align*}
Using Lemma  \ref{L:bcd}, the limit (\ref{rhorho})
and a change of variables $p\to 2p$, we obtain that
\begin{align}
       \lim_{L\to\infty} &
       \int dX_0  dV_0 \, J(X_0,V_0) \bE H_{\phi}^{(\ep,\mu)}
   (X_0, V_0)  \label{E:expanded} \\
        =&\sum_{m=0}^{M^*-1} \! \sum_{\sigma\in \operatorname{S}(m)} \! \!
        2^3        \varrho^m
      \!  \int dX_0 dV_0 \, J(X_0,V_0) \int dp dw_0 \, e^{ 2 i pX_0}  
                \frac{1}{\ep^{3\mu}} G^{\ep^{-\mu}}(2 p) G^{\ep^\mu}(w_0-V_0) \notag \\ 
        &\times \int d\fw_m d\fw'_m \, 
                \prod_{j=1}^m \delta[( w_{j-1}-w_j ) - 
(w'_{\sigma(j)-1} - w'_{\sigma(j)})] \notag  \\
        &\times
                K^{\mathrm{ren}}(t; \fw_{0,m} +  \ep p) 
                \overline{ K^{\mathrm{ren}}(t; \fw_{0,m}-\ep p)}
                B^{\mathrm{ren}}_{\ep p}(\fw_{0,m}) \overline
 {B^{\mathrm{ren}}_{-\ep p}(\fw_{0,m})}
                \widehat{W}_{\psi^{\ep}_0}(2 \ep p, w_m) \; . \notag
\end{align}
As before, we can show that cross terms, which arise from terms in 
which $\sigma \neq \operatorname{Id}$, are smaller by a factor 
of $t^{-1}= \operatorname{O}(\ep)$ and vanish in the $\ep \to 0$ 
limit (recall we are taking $\ep \to 0$ for a fixed $M^*$).  
The proof of this statement is nearly identical to estimate of 
the crossing terms in Lemma \ref{L:0,0} except without the 
time-division and using $K^{\mathrm{ren}}$ in place of $K$.  

One can prove a representation for the renormalized kernel (which differs by a small 
perturbation in the dispersion relation from the free kernel)
 analogous to Lemma \ref{L:Kidentity} and subsequently, 
estimates mirroring those in Proposition \ref{P:crossingtools} and in Lemma \ref{L:0,0}. 
 The key observation
is that in the necessary
estimates  the renormalization can be removed from the propagators using the
bound
$$
    \Big|\frac{1}{ \a - (p_j^2/2 + \varrho T(p_j) ) + i\eta }\Big|
    \leq \frac{C}{ | \a - p_j^2/2 + i\eta|}
$$
that follows from
\begin{align*}
        \Big|& \frac{1}{ \a - (p_j^2/2 + \varrho T(p_j) ) + i\eta }
 - \frac{1}{ \a - p_j^2/2 + i\eta} \Big| \\
        &\qquad \leq
        \frac{ \varrho | T(p_j)| }{ | \a- ( p_j^2/2 + \varrho T(p_j) ) + i\eta| 
\, | \a - p_j^2/2 + i\eta|} \\
        &\qquad \leq \frac{C}{ | \a - p_j^2/2 + i\eta|} ,
\end{align*}
using $\varrho \eta^{-1} \leq C$.  Consequently, 
we are left with only the direct terms ($\sigma=\mbox{Id}$)
 after the $\ep \to 0$ limit. 

Our next step is to replace 
$$
B_{\pm \ep p}^{\mathrm{ren}}(\fw_{0,m}) =
\prod_{j=1}^m B^{\mathrm{ren}}\Big( \frac{ (w_m\pm\ep p)^2}{2},
w_{j-1}\pm \ep p,  w_j\pm \ep p\Big)
$$
 with 
$$
 T_{\pm \ep p}(\fw_{0,m}) = \prod_{j=1}^m B\Big( \frac{ (w_m\pm\ep p)^2}{2},
w_{j-1}\pm \ep p,  w_j\pm \ep p\Big)
$$ 
in (\ref{E:expanded}).  
That is, we remove the renormalization on the potential part.  
By definition, we have:
\begin{align*}
        B^{\mathrm{ren}}\Big( \frac{(w_m\pm \ep p)^2}{2},
        &w_{j-1} \pm \ep p,  w_j \pm \ep p\Big) 
                = B\Big( \frac{(w_m\pm\ep p)^2}{2},
                w_{j-1} \pm\ep  p ,w_j \pm \ep p\Big)  \\
        & -\frac{1}{|\Lambda|} \delta(w_{j-1} - w_j) 
        B\Big(\frac{ (w_m\pm \ep p)^2}{2}, w_j \pm \ep p, w_j \pm \ep p\Big) \; .
\end{align*}
{F}rom this we can show that for a fixed $M^*$, that the error
 associated with replacing $B^{\mathrm{ren}}_{\ep p}(\fw_{0,m} )$ 
with $T_{\ep p}(\fw_{0,m})$ will be of order $|\Lambda|^{-1} $ and 
hence it vanishes in the $L \to \infty$ limit.  It is important to note 
here that our momenta are on the discrete lattice and hence we have
 the identity:  $ (\delta (p) / |\Lambda|)^2 = \delta(p)/ |\Lambda|$. 
In particular, higher powers of the delta functions that  may arise 
in the product  are harmless.     

The free-evolution portion 
in (\ref{E:expanded}) can be written, using our scaling $t= T\ep^{-1}$ 
and $\varrho= \varrho_0\ep $, as:
\begin{align*}
        K^{\mathrm{ren}}(t; &\fw_{0,m}+ \ep p)
\overline{K^{\mathrm{ren}}(t; \fw_{0,m}-\ep p)} \\
        =& \int_0^{t*} \[ds_j\]_0^m  \int_0^{t*}\[ds'_j\]_0^m \prod_{j=0}^m 
                e^{-i s_j [ (w_j+ \ep p)^2/2 + \varrho T_{\ep p}(w_j) ] } 
              \;  e^{i s'_j [  (w_j-\ep p)^2/2 + \varrho \overline{T_{-\ep p}(w_j)} ]} \\
        =& \frac{2^{m}}{\ep^{m} } \int_0^{T*} \[da_j \]_0^m \prod_{j=0}^m 
                e^{-i a_j [ 2w_j p + \varrho_0 \scriptT( w_j;\ep p )   ]} 
                \Big[ \prod_{j=0}^m \int_{- a_j / \ep  }^{  a_j / \ep } db_j 
                e^{- i b_j \Omega(w_j; \ep p)}\Big] \delta( \Sigma b_j) \; ,
\end{align*}where 
\begin{align*}
        \scriptT(w_j; \ep p) :=&  T_{\ep p}(w_j) - \overline{T_{-\ep p}(w_j )} \\ 
        \Omega(w_j; \ep p) :=&  w_j^2 + \varrho \big(T_{\ep p} (w_j )
                +\overline{T_{-\ep p} (w_j)}\big) \; ,
\end{align*}
and we introduced $a_j =\frac{\ep}{2}(s_j+ s_j')$, $b_j = s_j - s_k'$.
We also define:
\begin{align}
        M_\ep(\boldsymbol{a}_{0,m} ; \fw_{0,m}, \ep p) :=&
                \Big[ \prod_{j=0}^m \int_{- a_j  / \ep }^{ a_j / \ep } db_j 
                e^{-i b_j \Omega(w_j; \ep p)}\Big] \delta( \Sigma b_j ) 
 \label{E:defM}    \\
        =& \int_{\mathbb{R}^m} \[db_j\]_{0}^{m-1} 
\chi_{\boldsymbol{a}_{0,m}\ep^{-1}}( b) 
                \prod_{j=0}^{m-1} e^{-i b_j [ \Omega(w_j ; \ep p) - \Omega(w_m ; \ep p)]}
   \; ,           \notag
\end{align}
where
\begin{align*}
        \chi_{\boldsymbol{a}_{0,m}\ep^{-1}}(b) := 
                \chi \Big( -a_m\ep^{-1} \leq \sum_{j=0}^{m-1} b_j \leq a_m 
                \ep^{-1}\Big)
                \prod_{j=0}^{m-1} \chi(-a_j \ep^{-1} \leq b_j \leq a_j\ep^{-1}).
\end{align*}
Neglecting terms which vanish in the $L\to\infty$ and
$\ep \to 0$ limits, we have:
\begin{align}
        \int dX_0 & dV_0 \, J(X_0,V_0) \bE H_{\phi_M}^{(\ep,\mu)}
 (X_0, V_0) \notag \\
        =& 2^3 \sum_{m=0}^{M^*-1} (2\varrho_0)^m 
        \int dX_0 dV_0  \, J(X_0,V_0)
                \int dp \, d\fw_{0,m} 
                e^{2 i p X_0 }    \frac{1}{\ep^{3\mu}} G^{\ep^{-\mu}}(2p)     
                \notag \\
        &\quad \times  G^{\ep^{\mu}}(w_0-V_0)  \widehat{W_{\psi_0^{\ep}} }(2\ep p, w_m)  
                T_{\ep p} (\fw_{0,m}) \overline{ T_{-\ep p} (\fw_{0,m})}  \notag \\
        &\quad \times \int_0^{T*} \[da_j\]_0^m  \prod_{j=0}^m 
                e^{-i a_j [ 2w_j p+ \varrho_0 \mathcal{T}(w_j;\ep p)  ]} 
                M_{\ep}(\boldsymbol{a}_{0,m} ; \fw_{0,m}, \ep p) \; .
        \label{E:husimiexpanded}
\end{align}
We would now like to argue that $p\sim \ep^{-\mu}$ and therefore we can replace 
instances of $\ep p$ with $0$ since we are taking $\ep \to 0$.
We also would like to remove the restriction
$\chi_{\boldsymbol{a}_{0,m}\ep^{-1}}$ from (\ref{E:defM}) to
extend to $db_j$ integration and pick up the onshell delta function.
To justify this rigorously, we need the following
Proposition.

\begin{proposition} \label{P:M}Let $a_j \leq T$ for $j=0, \ldots, m$, $M_{\ep}$ 
be defined in (\ref{E:defM}) and $f \in \mathcal{S}(d\fw_{0,m})$.  We have:
\begin{align*}
        \sup_{p}\Big| \int d\fw_{0,m} \, f(\fw_{0,m}) M_{\ep}(\boldsymbol{a}_{0,m} 
                ; \fw_{0,m}, \ep p) \Big| \leq&
                C(T) \int dw_m |\!|\!| f |\!|\!|_{d\fw_{0,m-1}}  \, ,
\end{align*}where $C(T)$ is independent of $\ep$.  
Moreover, for each fixed values $a_j >0$, we have:
\begin{align*}
         \int d\fw_{0,m} \, f(\fw_{0,m}) M_{\ep}(\boldsymbol{a}_{0,m}; \fw_{0,m}, \ep p) \to
         \int d\fw_{0,m} \, f(\fw_{0,m}) \prod_{j=0}^{m-1} 2\pi \delta( w_j^2 - w_m^2),
\end{align*} as $\ep \to 0$.
The same limit holds if $f=f_\ep$ depends on $\ep$, but
$\int dw_m \,  |\!|\!| f_\ep |\!|\!|_{d\fw_{0,m-1}} $
is uniformly bounded. 
\end{proposition}
This result says that $M_\ep$ creates the "on-shell" condition in the limit.
\begin{proof}
For $\ep>0$ we can use Fubini to write:
\begin{align*}  
        \int d\fw_{0,m}& \, f(\fw_{0,m}) M_{\ep}(\boldsymbol{a}_{0,m} ; \fw_{0,m}, \ep p) \\
        =&
                \int_{\mathbb{R}^m} \[db_j\]_0^{m-1} 
                \chi_{\boldsymbol{a}_{0,m}\ep^{-1}}(b) \int d\fw_{0,m}\, 
                \prod_{j=0}^{m-1} e^{-i b_j w_j^2} \\
        &\times f(\fw_{0,m})
                \prod_{j=0}^{m-1} 
                e^{- i b_j [ -w_m^2 + \varrho(T_{\ep p}(w_j)+\overline{T_{-\ep p}(w_j)}
                -T_{\ep p}(w_m ) - \overline{T_{-\ep p}(w_m )}]}   .
\end{align*}
We now apply the dispersive estimate iteratively, (\ref{E:basicdisp}) to get the bound:
\begin{align*}
        \Big| \int& d\fw_{0,m} \, f(\fw_{0,m}) 
                M_{\ep}(\boldsymbol{a}_{0,m} ; \fw_{0,m}, \ep p) \Big| \\
        \leq&C \int_{\mathbb{R}^m} \[db_j\]_0^{m-1} 
                \chi_{\boldsymbol{a}_{0,m}\ep^{-1}}(b)  \prod_{j=1}^{m-1}\<b_j\>^{-3/2}  \\ 
        &\times
         \int dw_m \Big|\!\Big|\!\Big| f(\fw_{0,m}) \prod_{j=0}^{m-1} 
                e^{- i b_j [ -w_m^2 + \varrho(T_{\ep p}(w_j)+\overline{T_{-\ep p}(w_j)}
                -T_{\ep p}(w_m ) - \overline{T_{-\ep p}(w_m )}]}
 \Big|\!\Big|\!\Big|_{d\fw_{0,m-1}}.
\end{align*}A simple computation bounds the factor in the triple norm by 
$|\!|\!| f |\!|\!|_{d\fw_{0,m-1}} \sum_{j=0}^{m-1} \<\varrho b_j\>^2 $ 
which is independent of $p$.  The condition of $|b_j| \leq a_j \ep^{-1}$ and
 $\varrho = \varrho_0 \ep$ allows us to complete the proof of the first statement
of Proposition \ref{P:M}.

For the second statement, define 
\begin{align*}
        \widetilde{M}_{\ep} ( \boldsymbol{a}_{0,m} ; \fw_{0,m})
                :=& \int_{\mathbb{R}^m} \[db_j\]_{0}^{m-1} 
\chi_{ \boldsymbol{a}_{0,m}\ep^{-1}}( b) 
                \prod_{j=0}^{m-1} e^{-i b_j (w_j^2 - w_m^2 )} \\
        M(  \boldsymbol{a}_{0,m} ; \fw_{0,m}) :=&
                \int_{\mathbb{R}^m} \[db_j\]_{0}^{m-1}\prod_{j=0}^{m-1}
 e^{-i b_j (w_j^2 - w_m^2 )} = \prod_{j=0}^{m-1} 2\pi \delta(w_j^2-w_m^2) .
\end{align*}
Then
\begin{align*} 
        \Big|\int &d\fw_{0,m} \, f(\fw_{0,m}) 
                \Big(M_{\ep}(\boldsymbol{a}_{0,m} ; \fw_{0,m}, \ep p)  
                - \widetilde{M}_{\ep}(\boldsymbol{a}_{0,m}; \fw_{0,m}) \Big)  \Big| \\
        \leq& \Big| \int \[db_j \]_{j=0}^{m-1} 
\chi_{\boldsymbol{a}_{0,m}\ep^{-1}}(b) \prod_{j=0}^{m-1} \<b_j\>^{-3/2} 
                \int d w_m \\
        &\times \Big|\!\Big|\!\Big| f(\fw_{0,m})\Big( 1- \prod_{j=0}^{m-1} 
                e^{-ib_j  \varrho[T_{-\ep p}(w_j)+\overline{T_{\ep p}(w_j)}
                -T_{-\ep p}(w_m ) - \overline{T_{\ep p}(w_m )}]} \Big)
  \Big|\!\Big|\!\Big|_{d\fw_{0,m-1}}  \\
         \leq& C(T)  \ep^{1/2}|\log\ep|
         \int dw_m  \, |\!|\!| f|\!|\!|_{d \fw_{0,m-1} },
\end{align*}
since the factor in the triple norm is bounded by $\sum_j |\varrho b_j|$
for $|b_j|\leq a_j\ep^{-1}$.
Finally,
\begin{align*}
        \Big| \int d\fw_{0,m}& \, f(\fw_{0,m}) \big( \widetilde{M}_{\ep}(\boldsymbol{a}_{0,m}; 
                        \fw_{0,m}) -M(\boldsymbol{a}_{0,m}; \fw_{0,m}) \big) \Big| \\
        \leq& C \int dw_m \,  |\!|\!| f|\!|\!|_{d \fw_{0,m-1}} \int \[db_j\]_0^{m-1} 
\big( 1 - \chi_{\boldsymbol{a}_{0,m}\ep^{-1}}(b) \big) \prod_{j=0}^{m-1} \< b_j \>^{-3/2}.
\end{align*}
When $a_j >0$, $j = 0, \ldots, m$,  the $db_j$ integral goes to
zero as $\ep\to0$ and we prove the proposition.  
\end{proof}

We now show that we can make the the following
replacement
\begin{align*} 
        T_{\ep p}(\fw_{0,m}) \overline{ T_{-\ep p}(\fw_{0,m})}
  e^{-i \varrho_0 \Sigma a_j \mathcal{T}(w_j; \ep p)}  \to
        |T(\fw_{0,m})|^2 e^{-i \varrho_0 \Sigma a_j \mathcal{T}(w_j; 0)} ,
\end{align*} in (\ref{E:husimiexpanded}).  By Proposition \ref{P:M}, it 
 suffices to estimate:
\begin{align}\label{TT}
         \int dw_m \, \Big|\!\Big|\!\Big| &
         \int dp dX_0 dV_0  \, J(X_0,V_0) e^{2 i p X_0 }    
                \frac{1}{\ep^{3\mu}} G^{\ep^{-\mu}}(2p) G^{\ep^{\mu}}(w_0-V_0) 
                \widehat{W_{\psi_0^{\ep}} }(2\ep p, w_m) \notag \\
        &\times \int_0^{T*}  \[da_j\]_0^m 
                \Big[ |T(\fw_{0,m})|^2 
                e^{-i \varrho_0 \Sigma a_j \mathcal{T}(w_j; 0)}
 \notag \\
  & \qquad -T_{\ep p}(\fw_{0,m})
                \overline{T_{-\ep p}(\fw_{0,m})} 
                e^{-i \varrho_0 \Sigma a_j \mathcal{T}(w_j; \ep p)} \Big]
               \Big|\!\Big|\!\Big|_{d\fw_{0,m-1}} .
\end{align}
Using Lemma \ref{L:Blemma} and the smoothness of $J$ in the $V_0$
variable,
 the expression in the square brackets can be shown to be order 
$\eta^{-1/2}\ep |p|\sim \ep^{1/2}| p|$. Since $|p|\lesssim \ep^{-\mu}$
apart from exponentially small terms, we see that (\ref{TT}) vanishes
if $\mu<\frac{1}{2}$.

 Summarizing these replacements, we are left with, to leading order,
\begin{align*}
        \int dX_0 & dV_0 \, J(X_0,V_0) \,  
   \bE H_{\phi}^{(\ep,\mu)} (X_0, V_0) \\
        =&2^3 \sum_{m=0}^{M^*-1} (2\varrho_0)^m \int dX_0 dV_0 \,  J(X_0,V_0)
                \int dp \, d\fw_{0,m} 
                e^{2 i p X_0 }  \\
&\times  \frac{1}{\ep^{3\mu}} G^{\ep^{-\mu}}(2p) G^{\ep^{\mu}}(w_0-V_0)             
\widehat{W_{\psi_0^{\ep}} }(2\ep p, w_m)  
                |T(\fw_{0,m})|^2 e^{-i \varrho_0\Sigma \mathcal{T}(w_j;0)  }\\
             &\times    \int_0^{T*} \[da_j\]_0^m  
                e^{- i \Sigma 2 a_j  w_j p} 
                \prod_{j=0}^{m-1}2\pi  \delta ( w_j^2 - w_m^2) \; .
\end{align*}
We now use Fourier inversion to write:
\begin{align*}
        2^3 \int dp & \, e^{2 i p X_0 } 
  \frac{1}{\ep^{3\mu}} G^{\ep^{-\mu}}(2p)  
                \widehat{W_{\psi_0^{\ep}} }(2\ep p, w_m)e^{- i \Sigma 2 a_j  w_j p} \\
        =& \int dx \, G^{\ep^{\mu}}\Big(X_0 -x - \Sigma_0^{m} a_j w_j \Big) 
                W_{\psi_0^{\ep}}^{\ep}(x,w_m) \; .
\end{align*}
Using our explicit form for the initial wave function,
\begin{align*}
        \psi_0^{\ep} (x) := \ep^{3/2} h(\ep x) e^{iu_0 x}
\end{align*}
it can be verified that: 
\begin{align*}
        \lim_{\ep \to 0} \int dX_0 & dV_0 \, J(X_0,V_0)\,  \bE 
                H_{\phi}^{(\ep,\mu)} (X_0, V_0) \\
        =&\sum_{m=0}^{M^*-1} \varrho_0^m \int dX_0 dV_0 \, J(X_0,V_0)
                \int_0^{T*} \[da_j\]_0^m \int d\fw_{0,m} \delta(V_0-w_0) \\      
        &\times F_0(X_0- \Sigma_{0}^{m} a_j w_j, w_m) e^{2T\varrho_0 
                \operatorname{Im}T(w_0)} \prod_{j=1}^m 4\pi 
                |T(w_{j-1}, w_j)|^2  \delta(w_j^2-w_m^2) \; ,
\end{align*}where $F_0(X, V) =| h(X)|^2 \delta (V-u_0)$ is the 
initial condition to the Boltzmann equation.  
Taking $M^* \to \infty$ we proved the limit
 (\ref{E:testS}), where $F_T$ is the solution
to the Boltzmann equation (written in the iterated time integration form)
with collision kernel $4\pi\varrho_0 |T(w_{j-1}, w_j)|^2$.

Finally, we have to identify the collision kernel.
 We recall that by definition
$$
 T(w_{j-1}, w_j) = B_\eta\Big(\frac{w_m^2}{2}, w_{j-1}, w_j\Big)\; .
$$
We also recall that the dependence of  
$B$ on $\eta$ is controlled by Lemma \ref{L:operatorBest}, and that
 $\lim_{\eta\to 0+0}B_\eta$ is identified with the scattering T-matrix
$T_{\mathrm{scat}}$
in (\ref{tscat}). We therefore have
\begin{align*}
        \prod_{j=1}^{m} | T(w_{j-1}, w_j) |^2 \delta(w_j^2 - w_m^2) = 
                \prod_{j=1}^m 
 | T_{\mathrm{scat}}(w_{j-1}, w_j) |^2 \delta(w_{j-1}^2 - w_j^2) \; .
\end{align*}
Defining 
$$
\sigma(U,V): =4\pi |T_{\mathrm{scat}}(U,V)|^2 \delta(U^2 - V^2) \; ,
$$
 and applying the optical theorem to get 
$$
 \operatorname{Im}T_{\mathrm{scat}}(V,V)
 = - \frac{\sigma}{2} = -\frac{1}{2} \int dU \,  \sigma(U,V) \, ,
$$
 we conclude that:
\begin{align}
       \bE  H^{(\ep,\mu)}_{\psi_{T/\ep}^\ep}(X_0, V_0) \to F_T(X_0, V_0)
\label{conv}
\end{align}
as $\ep \to 0$ in $\mathcal{S}'(\mathbb{R}^6)$,
 where $F_T(X_0,V_0)$ solves the Boltzmann equation with
 collision kernel $\Sigma(U,V) := \varrho_0 \sigma(U,V)$.    
This completes the proof of the Main Theorem. $\;\;\Box$

\subsubsection*{Acknowledgements}
This work began as a joint project with H.-T. Yau and many of the ideas here have 
been developed in collaboration with him (see the conference proceeding announcement
 \cite{EY2}).  We would like to thank him for his support and advice, without which 
this work would not have been possible.


\end{document}